\documentclass{article}%
\usepackage{amsmath}
\usepackage{amsfonts}
\usepackage{amssymb}
\usepackage{graphicx}%
\providecommand{\U}[1]{\protect\rule{.1in}{.1in}}
\newtheorem{theorem}{Theorem}

\begin{document}

\title{Modular localization and the holistic structure of causal quantum theory, a
historical perspective \\{\small Dedicated to the memory of J\"{u}rgen Ehlers (1929-2008) }\\{\small Studies in History and Philosophy of Modern Physics 49, (2015) 109}}
\author{Bert Schroer\\present address: CBPF, Rua Dr. Xavier Sigaud 150, \\22290-180 Rio de Janeiro, Brazil\\permanent address: Institut f\"{u}r Theoretische Physik\\FU-Berlin, Arnimallee 14, 14195 Berlin, Germany}
\date{February 2014}
\maketitle
\tableofcontents

\begin{abstract}
Recent insights into the conceptual structure of localization in QFT ("modular
localization") led to clarifications of old unsolved problems. The oldest one
is the Einstein-Jordan conundrum which led Jordan in 1925 to the discovery of
quantum field theory. This comparison of fluctuations in subsystems of heat
bath systems (Einstein) with those resulting from the restriction of the QFT
vacuum state to an open subvolume (Jordan) leads to a perfect analogy; the
globally pure vacuum state becomes upon local restriction a strongly impure
KMS state. This phenomenon of localization-caused thermal behavior as well as
the vacuum-polarization clouds at the causal boundary of the localization
region places localization in QFT into a sharp contrast with quantum mechanics
and justifies the attribute "holstic". In fact it positions the E-J
Gedankenexperiment into the same conceptual category as the cosmological
constant problem and the Unruh Gedankenexperiment. The holistic structure of
QFT resulting from "modular localization" also leads to a revision of the
conceptual origin of the crucial crossing property which entered particle
theory at the time of the bootstrap S-matrix approach but suffered from
incorrect use in the S-matrix settings of the dual model and string theory.

The new holistic point of view, which strengthens the autonomous aspect of
QFT, also comes with new messages for gauge theory by exposing the clash
between Hilbert space structure and localization and presenting alternative
solutions based on the use of stringlocal fields in Hilbert space. Among other
things this leads to a radical reformulation of the Englert-Higgs symmetry
breaking mechanism.

\end{abstract}

\section{Preface}

The subject of this paper grew out of many discussions about Jordan's
discovery of quantum field theory (QFT) which I had with the late J\"{u}rgen
Ehlers. They focussed in particular on events between the publication of
Jordan's thesis on quantum aspects of statistical quantum mechanics in 1924
\cite{Jo}, and his discovery of QFT around 1925 which was published in one
section of the famous 1926 "Dreim\"{a}nnerarbeit" \cite{Drei} together with
Born and Heisenberg. This paper was in fact the second paper after
Heisenberg's discovery of quantum mechanics (QM). The resistance of Born and
Heisenberg against Jordan's section has its natural explanation in that these
two authors felt that Jordan was burdening the conceptual struggle to
understand the new quantum mechanics with something which may distract from
this project.

I met J\"{u}rgen Ehlers the first time around 1957 at the University of
Hamburg when he was Jordan's assistant and played the leading role in Jordan's
general relativity seminar. Our paths split, after I wrote my diploma thesis
on a topic of particle theory at the time when particle physics moved away
from the university physics institute to the newly constructed high energy
laboratory at DESY. Contacts with Ehlers and the relativity group became less
frequent and ended when both of us took up research associate positions at
different universities in the US.

Only 40 years later, when Ehlers moved to Potsdam/Golm in the 90s as the
founding director of the new Albert Einstein Institute (AEI), we met a second
time. After having done important research on problems of general relativity
and astrophysics he became increasingly interested to understand some of
Jordan's famous early work on quantum field theory about which we knew little
at the time of Jordan's weekly relativity seminar\footnote{After wwII Jordan's
interest was mainly focussed on general relativity and philosophical
implications of quantum theory. Since he never mentioned his early work on
QFT, we remained quite ignorant about it.}. Ehlers was in particular
interested to understand some subtle points in a dispute between Jordan and
Einstein concerning Einstein's use of statistical mechanics fluctuation
arguments for black body radiation \cite{Ein2}. The ensuing dispute around
this purely theoretical argument in favor of the existence of photons has been
more recently referred to as the \textit{Einstein-Jordan conundrum
\cite{Du-Ja}.}

As the terminology reveals, the E-J conundrum was a poorly understood relation
between fluctuations caused by restricting the vacuum state to the observables
in a subvolume in Jordan's newly discovered field quantization and Einstein's
use of statistical mechanics within the old Bohr-Sommerfield quantum setting.
This led him to identify a particle-like component in the fluctuation spectrum
of a black body radiation ensemble (which he termed "Nadelstrahlung") with his
1905 interpretation of the photo-electric effect as a manifestation of the
corpuscular nature of light.

The E-J conundrum has sometimes been seen as an illustration of the
particle-wave dualism of quantum mechanics, but with the hindsight of modern
QFT its real significance points into a much deeper level. This was certainly
Ehler's view when he drew my attention to what he considered its real
significance. Coming from general relativity and cosmology he thought that
this problem is analogous \cite{Mainz} to the problems related to vacuum
polarization used to explain the origin of the cosmological constant in terms
of fluctuations of the quantum field theoretic vacuum. He hoped that with my
experience of 40 years of QFT I could be of some help to obtain a better understanding.

I learned recently through John Stachel that conjectures about possible
connections between thermal aspects of the subvolume fluctuations in QFT as
they occur in the E-J conundrum and Hawking-Unruh problems already existed in
the 80s \cite{Stachel}. In fact it will become clear in the course of the
present work that it indeed can and should be viewed this way.

For some time this problem remained out of my range of interest; I did not
want to loose time on something which would draw me into opaque historical
problems away from my research on new foundational insights into to QFT via
"modular localization"\footnote{Here modular localization stands for an
intrinsic formulation of causal localization which is independent on what
quantum field "coordinatization" one uses in order to describe the particular
model. of QFT.} \cite{AOP}. During a two year stay (2002/2003) in Brazil, a
CNPq supported research project "The Modular Structure of Causal Quantum
Physics" provided the chance to extend this research. Around 2007 I suddenly
realized that the complete understanding of the E-J conundrum can be obtained
with the help of precisely those newly gained insights. One just had to apply
the \textit{principle of modular localization,} which assigns a certain number
of unexpected properties to localized subalgebras. Whereas the global vacuum
state is pure, the restriction to a causally localized subalgebra renders it
impure; in fact its impurity can be described as a thermodynamic KMS state
\cite{Haag} with respect to a "modular Hamiltonian". This is a general result
of the application of the so-called Tomita-Takesaki modular theory of local
operator algebras to the subalgebra which observables localized in a spacetime
region (whose causal completion remains smaller than Minkowski spacetime) generate.

This reduced vacuum state is entangled in a more radical sense than the
entanglement of particle states in Schr\"{o}dinger's QM of particle states
under a binary split of the system into spatial inside/outside subsystems.
Entanglement in quantum mechanics resulting from binary inside/outside splits
of degrees of freedom resulting from the reduction to the inside and the
ensuing loss of of the outside information is a well-known phenomenon; it has
been observed in quantum optical experiments and the results led to a Nobel
prize. But the quantum mechanical "vacuum" (the mathematical reference state
which one needs for the "second quantization" multiparticle description of QM)
remains completely inert against entanglement. In fact \textit{the singular
vacuum entanglement caused by localization in QFT is characteristic for the
enormous conceptual distance between the two quantum theories}. The
terminology E-J "conundrum" refers to the fact that for a long time this
aspect of the vacuum remained outside theoretical comprehension.

In fact most theoretical physicists became for the first time aware of the KMS
nature of the QFT restricted vacuum state in connection with the Unruh's
"Gedankenexperiment" in which the localization region is a spacetime wedge.
This aspect of vacuum entanglement also points at the "fleeting" nature of
this effect; it remains many orders of magnitude below the measured quantum
optical entanglement of quantum mechanical particle states. But even if it
will always remain a "Gedanken" concept \footnote{The situation becomes less
"fleeting" if the horizon of the localization region is an (Unruh
observer-independent) black hole "event horizon".}, it is at the heart of QFT
and follows directly from the \textit{quantum adaptation of the}
Faraday-Maxwell "action at the neighborhood" which Einstein converted into the
Minkowski spacetime \textit{causality principle.} Its quantum counterpart is
of a radically different nature whose physical manifestations are somewhat
unexpected. It will be referred to as \textit{modular localization;} a
terminology which relates its mathematical formulation with its physical
implications. In the present work it will be shown that its conceptual range
is not limited to shed light into dark corners of QFT's history as the before
mentioned E-J conundrum, but it also plays an important role in an ongoing
conceptual reformulation of QFT (which includes gauge theories and the
recently much discussed "Higgs mechanism").

The two components in Einstein's statistical mechanics fluctuation properties
are indeed, as Jordan claimed, also present in the physical vacuum state after
restricting it to the ensemble of observables which are localized in a
subvolume. It is important to not impose boundary restrictions (box
quantization) but remain within the realm of "open systems". Here it is
irrelevant whether Jordan's calculation treated this aspect correctly
\cite{Du-Ja}; many important observations in the history of quantum physics
have been made within doubtful calculations.

When I was about to explain my findings \cite{E-J}\cite{Jor}\cite{cau} in 2008
to Ehlers, I learned that he passed away shortly before my return to Berlin.

The main aim of this paper, which I dedicate to the memory of J\"{u}rgen
Ehlers, is to explain my findings and their relation to the ongoing research
in QFT in more details and a larger context as I did peviously in \cite{E-J}.

I remember that Ehlers, in his capacity as the founding director of the AEI in
Potsdam, took an interest in string theory (ST). He was however annoyed by the
fact that he was unable to bridge the gaps between his understanding of
spacetime properties of gravity and the (sometimes bizarre) claims of members
of the ST group at the AEI; notwithstanding the fact of the enormous amount of
mathematical sophistication and the professional reputation of some of the
protagonists of ST.

The work on modular localization also led me to string-localized fields and
their improved short distance property which promised a radical extension of
renormalization theory to interaction between fields with higher spins. The
reason why I mention this here is that this new concept of string-localized
fields in Hilbert space also revealed that string theory (ST) and its
derivatives (embeddings, dimensional reductions, the AdS-CFT isomorphism) has
no relation to causal localization in spacetime; it is rather the result of a
fundamental misunderstanding on these issues. Hence Ehlers' problems with the
ancient Einstein-Jordan conundrum and his new problems with ST were
interconnected in a curious way. His death in 2008 prevented me from conveying
this insight.

It is the purpose of these notes to explain the constructive \cite{E-J} as
well as critical \cite{on-shell} power in a historical context.

Usually a historical paper revisits the past about already closed subjects;
typical examples are research papers on the discovery and the conceptual
development of QM. In contrast to such subjects, which are closed from a
foundational physical point of view (but sometimes still lead to bitter
philosophical feuds), the situation of the problems addressed in this paper is
very different. Most of them, although some having been present in QFT from
its historical beginnings, were only properly understood recently and have not
yet been addressed by philosophers; In contrast to QM, QFT is still far from
its conceptual closure not to mention its philosophical exploration. The
present paper attempts to give an account of the present situation.

The Einstein-Jordan conundrum was often misunderstood as a confirmation of the
particle-wave duality which, since de Broglie's matter-wave idea and
Schr\"{o}dinger's wave equation, was an integral part of QM. But the E-J
dispute addresses a much deeper issue which, before the appearance of modular
localization concept in QFT, had little chance to be properly understood.

My posthumous thanks for introducing me to a fascinating topic from the
genesis of QFT which, far from being a closed part of history, exerts its
conceptual spell over actual particle theory, naturally go to J\"{u}rgen
Ehlers. The present exploration of the foundational principle of modular
localization did not only change the view about hitherto incompletely
understood problems at the dawn of QFT \cite{E-J}, but also promises to have
an important say about its future \cite{on-shell}.

\section{Introduction}

A dispute between Einstein and Jordan (referred to as the E-J conundrum
\cite{Du-Ja}) led Jordan to propose the first quantum field theoretical model
with the purpose to show that there exists a quantum analog of Einstein's
thermal subvolume fluctuations in open subvolumes (intervalls) of
two-dimensional quantized Maxwell waves in a global vacuum state. For this
pupose Jordan invented the simplest QFT which in modern terminology is the
model generated by a conformal chiral current. A brief sketch of the
pre-history which led to the E-J conundrum may be helpful.

\begin{itemize}
\item Einstein 1917 in \cite{Ein}: calculation of mean square fluctuations in
an open subvolume in statistical mechanics of the thermal black body radiation
shows presence of two components: wave- and particle-like ("Nadelstrahlung")
fluctuation structure which Einstein interpreted as a theoretical evidence for
photons (after his 1905 paper based on the observational support coming from
the photoelectric effect).

\item Jordan in his PhD thesis (1924, \cite{Thesis}) argued that the
particle-like component$~\bar{E}_{\nu}\sim h\nu~$is not needed for attaining equilibrium.

\item Einstein's reaction \cite{Ein2} consisted in a publication in which
Jordan's argument is conceded to be mathematically correct but physically
flawed (the absorption is incorrectly described). However he praised Jordan's
statistical innovations ("Stosszahlansatz").

\item Einstein's paper caused Jordan's radical change of mind; he fully
accepted Einstein's view by demonstrating that he can obtain the same wave-
and particle-like fluctuation components by restricting a "two-dimensional
quantized Maxwell field" (modern terminology: d=1+1 chiral current model) to a
subinterval. In this way he discovered field quantization, probably without
understanding \textit{why} a vacuum in QFT behaves radically different from a
quantum mechanical "no particle state", in particular why the reduced vacuum
shares the kind of impurity with that of a KMS statistical mechanics state.
\end{itemize}

Shortly after this episode Jordan published his first field quantization in a
separate section in the famous 1926 \ "Dreim\"{a}nnerarbeit" \cite{Drei}. Gaps
in Jordan's computation and his somewhat artistic treatments of infinities
caused some ruffling of feathers with his coauthors Born and Heisenberg
\cite{Du-Ja}. From a modern point of view the picture painted in some
historical reviews, namely that this was a typical case of a young
brainstorming innovator set against a scientific establishment (represented by
Born), is not quite correct. Born and Heisenberg had valid reasons to consider
Jordan's fluctuation calculations as incomplete, to put it mildly. Conceding
this does however not lessen Jordan's merits as the protagonsist of QFT .

One reason why this discovery of QFT was not fully embraced at the time was
that, although a free field on its own (staying with its linear properties) is
a rather simple mathematical object, the problem of energy fluctuations in
open subvolumes is anything but simple. To understand why subvolume
fluctuations in the vacuum state of QFT are similar to Einstein's statistical
mechanics thermal fluctuations is a deep conceptual problem which could not
have been solved solely by calculations; especially because before the arrival
of the concept of modular localization such calculations could only have been
done in terms of conceptually uncontrolled approximations. But now it can be
satisfactory answered with the help of a new view of QFT which generically
relates the restriction of the vacuum to the observables of a spacetime
subvolume with thermal properties and vacuum polarization ("split inclusions"
of modular localized algebras \cite{Haag}); this is precisely what "modular
localization" achieves. One may safely assume that Born and Heisenberg
perceived that this new quantum field model of Jordan with infinitely many
oscillator degrees of freedom did not quite fit into their quantum mechanical
project which Heisenberg started a short time before; in particular Jordan's
nonchalant way of handling infinities led to critical comments \cite{Du-Ja}.

Nevertheless Heisenberg, who in comparison to Jordan understood less about
statistical mechanics at the time of the E-J conundrum, later became aware of
vacuum polarization (which is absent in QM) probably still under the influence
of Jordan's fluctuation problem. A letter he wrote to Jordan before he
published his famous vacuum polarization paper mentions a logarithmic
divergence $\lim_{\varepsilon\rightarrow\infty}\log\varepsilon,$%
~with$~\varepsilon~$describing the ~"fuzziness" at the interval ends of
Jordan's one dimensional model \cite{Du-Ja}. Indeed vacuum polarization and
thermal manifestations of vacuum entanglement from causal localization are
opposite sides of the same coin.

One note of caution. Since the terminology "particles" and "waves" played an
important role in the Einstein-Jordan dispute, the reader may think that it
refers (as mentioned before) to the quantum mechanical particle-wave dualismus
(the two equivalent descriptions of QM); in this way its real significance,
namely the thermal aspects of vacuum entanglement through causal localization
of quantum matter, is sometimes overlooked.

The important distinction between the global quantum mechanical nature of
infinitely many oscillators and their holistic role in the implementation of
causal localization in a quantum theory of local fields had to wait almost 5
decades before being understood on a foundational level. For some time QFT was
even suspected to be afflicted by internal inconsistencies which lead to
ultraviolet divergencies (the "ultraviolet catastrophe"). Even after
discovering the covariant renormalized perturbation theory for quantum
electrodynamics (and finding an impressively successful agreement of low order
perturbation with experimental observations) some of these doubts lingered on.
Renormalized perturbation theory remained for a long time a collection of
recipes about how to extract finite time-ordered correlation functions from
the quantization rules starting with classical Lagrangians. What convinced
people despite the weakness was the internal consistency of the finite results.

The quantization parallelism to the classical field theory of Faraday and
Maxwell as embodied in the Lagrangian or functional integral quantization
prevented for a long time an awareness about some radical differences
resulting from quantum causal localization as compared to its classical
counterpart. One manifestation of such a difference was that quantum fields,
in contrast to smooth causally propagating classical functions, were rather
singular operator-valued Schwartz distributions. They require testfunction
smearing in order to attain the status of (generally) unbounded operators with
which one then can construct operator algebras of bounded operators which are
causally localized in spacetime regions. The other surprise was that these
operator algebras have properties which were somewhat unexpected from the
conceptual viewpoint of QM. Causal localization causes the global vacuum state
to become impure upon restriction to a local operator subalgebra
$\mathcal{A(O})$ generated by covariant fields $A(x)$ smeared with
$\mathcal{O}$-supported test functions. These impure "partial" states fulfill
the so-called KMS property \cite{Haag} with respect to a \textit{modular
Hamiltonian} which is intrinsically determined by the pair ($\mathcal{A(O}%
),\Omega_{vac}$) of local algebra and vacuum state vector. In fact all
physical (i.e. finite energy) states restricted to a local algebra behave like
statistical mechanics states.

The mathematical theory of operator algebras which highlights such properties
is the \textit{Tomita-Takesaki modular operator theory} which is omnipresent
in QFT thanks to its causal localization structure. The presentation of QFT in
terms of a net of operator algebras and their properties was proposed by
Rudolf Haag \cite{1957} shortly after Arthur Wightman published his
characterization of covariant fields in terms of properties of their
correlation functions \cite{Wight}. Haag's textbook \cite{Haag} on "local
quantum physics" (LQP), based on an operator-algebraic approach to QFT,
appeared only many decades after he gave a first account of this new
formulation \cite{1957}. The terminology LQP in the present article is used
whenever it is important to remind the reader that the arguments go beyond the
view about QFT which he meets in most textbooks (which are usually restricted
to a formulation of perturbation theory within the setting of Lagrangian
quantization and its functional integral formulation).

The mathematical property which guaranties the applicability of the T-T
modular operator theory, is the so-called \textit{standardness} of the pair
($\mathcal{A(O}),\Omega_{vac}$) i.e. the property that the operator algebra
acts on $\Omega_{vac}$ (more generally on all finite-energy state vectors) in
a cyclic ($\overline{\mathcal{A(O})\Omega_{vac}}=H$) and separating
($\mathcal{A(O})$ \textit{contains no annihilators} of $\Omega_{vac}$) manner.
The cyclicity of the vacuum is closely related to the positivity of the energy
of the representation of the Poincar\'{e} group, whereas the separating
property results from spacelike commutativity of observables and is equivalent
to the fact that the commutant, which contains the algebra of the causal
complement $\mathcal{A(O})^{\prime}\supseteq\mathcal{A(O}^{\prime}),$ acts
also cyclic on $\Omega_{vac}$ as long as the spacelike complement
$\mathcal{O}^{\prime}$\ is non-void. This physicists know under the name of
the the "Reeh-Schlieder property" \cite{Haag}, whereas the operator
algebraists call this the "standardness" of the pair ($\mathcal{A(O}),\Omega
$). This property is not shared by QM and accounts for the significant
differences between these two QT \cite{interface}.

For a structural comparison it is convenient to rewrite (the Schr\"{o}dinger
form of) QM into the Fock space setting of "second quantization" which
converts wave functions into fields. As mentioned before in this reformulation
the newly introduced vacuum remains, as opposed to its active role in QFT,
completely inert with respect to the action of the Schr\"{o}dinger "quantum
field" (no vacuum entanglement leading to vacuum polarization). Instead of the
cyclic action the local algebra at a fixed time\footnote{In LQP such an
algebra at a fixed time $\mathcal{A}(\mathcal{R})$ is defined as the
intersection of all spacetime algebras $\mathcal{A}(\mathcal{O})$ with
$\mathcal{R}\subset\mathcal{O}$.} corresponding e.g. to a spatial region
$\mathcal{R\subset}\mathbb{R}^{3},$ one obtains a subspace and a tensor
factorization of $H$
\begin{align}
H(\mathcal{R})~  &  =~\overline{\mathcal{A}(\mathcal{R})\Omega_{QM}}\subset
H=H(\mathcal{R})\otimes H(\mathcal{R}^{\perp})~\\
\mathcal{A}(\mathcal{R})  &  =B(H(\mathcal{R})),~\mathcal{A\equiv
B(H)=A}(\mathcal{R})\otimes\mathcal{A}(\mathcal{R}^{\perp})\nonumber
\end{align}
$~$of with a factorizing vacuum $\Omega_{QM}.$ This inertness against
entanglement of the quantum mechanical vacuum is very different from the
"vacuum polarizability" of $\Omega_{vac}$ in QFT which is connected to the
lack of tensor factorization (despite the the fact that by defintion the
commutant $A(\mathcal{O})^{\prime}$ contains all operators which commute with
$\mathcal{A(O})$). In terms of structural properties of operator algebras
these remarkable differences in the mathematical structure amount to the
existence of two non-isomorphic factor algebras which are met in QFT: the
global $\mathcal{B(}H\mathcal{)}$ algebra of all bounded operators on a
Hilbert space (the unique type I$_{\infty}$ factor) and the local
\textit{monad} algebras $\mathcal{A(O})$ which are all isomorphic to the
unique hyperfinite type III$_{1}$ factor algebra in the Murray-von
Neumann-Connes classification of factor algebras \cite{Haag}.

The choice of terminology reveals the intention to see the new local quantum
physical view of QFT in analogy to the way Leibnitz understood \textit{reality
in terms of relations between monads}. In this extreme relational view, a
monad by itself is nearly structureless, similar to a point in geometry.
Indeed in the local quantum physical description of QFT, all properties of
quantum matter, including the Poincar\'{e} covariance of its localization in
spacetime and its possible localization-preserving inner symmetries, can be
shown to arise from the abstract (non-geometric) modular positioning of a
finite number of copies (depending on the spacetime dimension) of the monad
within a shared Hilbert space (section 3); the Poincar\'{e} group can be
extracted from the modular groups of the contributing algebras and the concept
of modular inclusions \cite{K-W}.

Together with the thermal KMS property of the locally restricted vacuum, there
is the formation of a vacuum polarization cloud at the causal boundary of
localization which accounts for a \textit{localization entropy, }a special
type of entanglement entropy. By replacing the boundary by a thin shell of
size $\varepsilon~$the localization entropy can be described in terms of a
function of the dimensionless area $\alpha=area/\varepsilon^{2}$ which
diverges in the limit $\varepsilon\rightarrow0.$ This relation between the
increasing sharpness of localization and the increasing localization entropy
is the \textit{substitute of the lost quantum mechanical Heisenberg
uncertainty relation}. The position operator $\mathbf{x}_{op}$ is, as all
quantum mechanical observables, of global nature; it does not belong to the
observables obeying the causal localization principle of LQP but may be used
in the (non-covariant) effective description of wave-function propagation. The
divergence in the sharp localization limit $\varepsilon\rightarrow0$ shows
another aspect in which QFT differs from QM.

The entanglement between the wedge-localized algebra and its opposite (that of
the spacelike separated wedge) is always infinite in the sense that it is not
possible to describe the associated state as density matrix (accounting for
the singular nature of vacuum entanglement); indeed there are no pure states
nor density matrix states on monad algebras; all states are impure in a very
radical way. This is not a disease of QFT (ultraviolet divergency of entropy
at a sharpely defined localization) but rather its conceptual heart; without
it there would be no relativistic QFT. In quantum statistical mechanics this
kind of KMS state is only met in the thermodynamic limit of density matrix
Gibbs states diverge and pass to KMS states on a monad algebra. In this case
the QFT generated by the commutant describes a "shadow world" outside the
localization concept \cite{S-W}. Local algebras $\mathcal{A(O})$ in QFT are
monads and have no density matrix or pure states\footnote{A state is a
normalized linear positive functional on an algebra and only if this algebra
consists of all bounded operators in a Hilbert space $B(H)$, states can be
represented by vectors (modulo phase factors).} at all; every global state
restricted to such an algebra will be rather singular. In fact all physical
(i.e. finite energy) states restrict to singular KMS states (i.e. one which
cannot be written as a density matrix state).

The reduced vacuum state assign a \textit{probability} to the ensemble of
local observables contained in $\mathcal{A(O});$ this is a consequence of the
KMS (statistical mechanics-like) nature of the impure reduced vacuum state.
Unlike the probability interpretation, which Born added to QM and which
Einstein rejected ("God does not throw dice"), the ensemble viewpoint of
probability as in statistical mechanics (which Einstein accepted) is intrinsic
to QFT. KMS states on the ensembles of $\mathcal{O}$-localized observables are
like thermal states of statistical mechanics and not "Gedanken-ensembles" as
in case of Born's assignement of probabilities to individual mechanical
systems of QM which refers to the statistics of repeated measurements.
Einstein had no problems with probability of real ensembles in statistical
mechanics, but it is the at that time unknown modular localization aspect
which permits to recognise the ensemble aspect of local observables.

There have been attempts to improve Jordan's approximations \cite{Du-Ja} since
the subvolume fluctuation problem is not solvable in closed form. The
characterization of the algebra of operators localized in a subvolume is a
\textit{holistic problem}; the enclosure of the subsystem in a quantization
box is not the same as reducing the vacuum to the subvolume algebra. Dealing
with open subsystems is an "holistic" challenge in which the knowledge of the
global oscillators is of not much help. Standard QFT does not provide a clear
mathematical concept in order to characterize the ensemble of operators which
is localized in a subvolume $\mathcal{O}$. On way of doing this would be to
smear the quantum fields with $\mathcal{O}$-supported testfunctions and use
the algebra which they generate. Even then one needs some knowledge about the
"modular Hamiltonian" which is related to the kind of statistical mechanics
associated with the KMS state corresponding to the restricted vacuum. In
certain cases one can guess it in the form of a geometric transformation which
leaves $\mathcal{O}$ invariant. For a noncompact wedge region in Minkowski
spacetime e.g. $W_{3}=\left\{  x;x_{3}>\left\vert x_{0}\right\vert \right\}  $
this would be the wedge-preserving Lorentz subgroup $\Lambda_{W_{3}}(\chi),$
for Jordan's model (a chiral subalgebra on a lightlike interval, see section
4) it is the interval-preserving dilation subgroup of the M\"{o}bius group;
but in the generic case on has to refer to modular theory. What is important
in the historical review is not whether Jordan got this right, but rather that
in his attempt to counter Einstein he invented QFT.

In order to avoid any misunderstandings it should be emphasized that in saying
that the concept of probability enters QFT in a more natural way than in QM,
one is not implying that this is changing the epistemic aspects of the
measurement theory in QT. All the conceptual aspects of entanglement
(including Bell's inequality) remain valid. What QFT adds is a more radical
realization of these phenomena on a much smaller scale; as already mentioned
the scale of localization-caused vacuum entanglement is that of the Unruh
effect and Hawking radiation. The reality of entanglement of particle states
with respect to binary subdivisions in QM is experimentally accessible in
terms of quantum optical arrangements, whereas the KMS impurity of the
spacetime-restricted vacuum (e.g. the Unruh effect) will presumably always
remain experimentally inaccessible (including even high energy nuclear experiments).

Part of the problem is that it is nearly impossible to describe precisely in
terms of existing hardware how a perfect causal localization can be realized;
even for noncompact spacetime regions as Unruh's Rindler wedges, the effect
depends on the state of uniform acceleration of the observer;
observer-independent manifestations appear only in the context of
metric-induced event horizons of black holes. Fortunately foundational
principles do not need to permit \textit{direct} observational verification;
they only have to be conceptually consistent, incorporate the reality which
existed before their inception, and lead to new observable consequences. In
this respect QFT, which only shares with QM the Hilbert space and $\hbar$ but
not the causal locality principle, has been and promises to continue to be the
most inclusive successful physical theory.

One can entertain wonderful dreams of what may have happened if important
concepts would have appeared decades earlier. But in the real world big
conceptual jumps against the prevalent ideas of the time (the Zeitgeist) are
virtually impossible; even for getting from inertial systems in Minkowski
spacetime to General Relativity it took Einstein many years and the same can
be said about the development of QM out of the old semiclassical
Bohr-Sommerfeld ideas. The problem for the case at hand is aggravated by the
fact that, up to the middle of the 60s, there did not even exist a
mathematical framework of operator algebras in which ideas about localization
could have been adequately formulated.

It is interesting to note that modular operator theory and its physical
counterpart of modular localization is the only theory to whose discovery and
development mathematicians (Tomita, Takesaki, Connes) and physicists (Haag,
Hugenholz and Winnink) contributed on par. They first realized this at a 1965
conference in Baton Rouge\footnote{The mathematicians worked on the
generalization of the modularity of Haar measures ("unimodular") in group
representation theory whereas the physicists tried to understand quantum
statistical mechanics directly in the thermodynamic infinite volume limit
(open system statistical mechanics) by using the KMS identity instead of
approaching this limit by tracial Gibbs states.}, with statistical mechanics
of open systems and the role of the KMS property representing the physical
side \cite{Haag}. The study of the relation between modular operator theory
and causal localization in LQP started a decade later \cite{Bi-Wi}, and its
first application consisted in a more profound understanding \cite{Sewell} of
the Unruh Gedankenexperiment \cite{Unruh}. The terminology "modular
localization" is more recent and marks the beginning of a new constructive
strategy in QFT based on the modular aspects of localization of states and
algebras \cite{BGL}\cite{AOP}. In mathematics the theory was the decisive
instrument which led to Connes closure of the Murray-von Neumann project of
classifying von Neumann factor algebras.

The E-J conundrum represents in fact a precursor of the Unruh
Gedankenexperiment and, as the latter, can be fully resolved in terms of the
principle of modular localization. In fact in the special case of Jordan's
chiral current model (the historically first and simplest model of a QFT), the
solution of the E-J conundrum amounts to a unitary \textit{isomorphism}
between a system defined by the vacuum state restricted to the algebra
$\mathcal{A}(I)$ localized in an interval $I$ and an associated global
statistical mechanics system at finite temperature. Such isomorphic relations
are referred to as describing an "inverse Unruh effect", \cite{S-W} and the
Jordan model is the only known illustration. However in both cases the KMS
temperature is not something which one can measure with a thermometer or use
for "egg-boiling" (and there is also no "boiling soup" of
particle/anti-particle pairs) \cite{B-S}.

The attribute "holistic" will be used quite frequently in connection with
modular localization. This terminology has been previously introduces by
Hollands and Wald \cite{Ho-Wa} in connection with their critique of
calculations of the cosmological constant in terms of simply occupying global
energy levels (with a cutoff at the Planck mass). In previous papers
\cite{integrable}, it refers to the intrinsicness of localization which is
connected with the cardinality of phase space degrees of freedom and their
subtle local interplay. This distinguishes physical localization of quantum
matter from mathematical/geometrical concepts. In fact it presents a strong
barrier against attempts of geometrization of QFT and explains why the
Atiyah-Witten attempt of the 70ies to "geometrize" QFT did not lead to the
breakthrough which many people (including the author) hoped for.

The simplest illustration of the meaning of holistic consists in the
refutation of the vernacular: "(free) quantum fields are nothing more than a
collection of oscillators" which often students are told in courses of QM.
Knowing continuous families of oscillators in the form of creation and
annihilation operators $a^{\#}(\mathbf{p})$ does not reveal anything about
free quantum fields and their associated local operator algebras. The free
Schr\"{o}dinger field and a free scalar covariant field share the same global
oscillator creation/annihilation operators
\begin{align}
a_{QM}(\mathbf{x,}t)  &  =\frac{1}{\left(  2\pi\right)  ^{\frac{3}{2}}}\int
e^{i\mathbf{px-}\frac{\mathbf{p}^{2}}{2m}}a(\mathbf{p})d^{3}p,~\left[
a(\mathbf{p}),a^{\ast}(\mathbf{p}^{\prime})\right]  =\delta^{3}(\mathbf{p-p}%
^{\prime})\label{f}\\
A_{QFT}(x)  &  =\frac{1}{\left(  2\pi\right)  ^{\frac{3}{2}}}\int\left(
e^{-ipx}a(\mathbf{p})+e^{ipx}a^{\ast}(\mathbf{p})\right)  \frac{d^{3}p}%
{2\sqrt{\mathbf{p}^{2}+m^{2}}},~p=(\mathbf{p},\sqrt{\mathbf{p}^{2}+m^{2}%
})\nonumber
\end{align}
In both cases the global algebra is the irreducible algebra of all operators
$B(H),$ generated by the shared creation/annihilation operators. But the local
algebras\footnote{Technical points as the connection between fields and the
algebras they generate are not important in the present context and therefore
will be omitted.} generated by testfunction-smearing with finitely supported
Schwartz functions $suppf(\mathbf{x})\subset\mathcal{R}$ of the fields and its
canonical conjugate at a fixed time in a spatial region $\mathcal{R}$ are very
different in both cases. In the relativistic covariant case they are identical
to the algebras $\mathcal{A}(\mathcal{O}_{\mathcal{R}}),$ $\mathcal{O}%
_{\mathcal{R}}=\mathcal{R}^{\prime\prime}$ the causal spacetime completion of
$\mathcal{R}$ (which is also generated by smearing with $\mathcal{O}%
_{\mathcal{R}}$-supported spacetime smearing functions). According to what was
stated before, these algebras are of "monad" type and the $\mathcal{A(O}%
_{\mathcal{R}})$-restricted vacuum state is a KMS state; in the case of the
Schr\"{o}dinger field the associated subalgebra $B(H(\mathcal{R}))$ is of the
same type as the global algebra; the QM vacuum continues to be an inert state
in the "smaller" factor Hilbert space $H(\mathcal{R}).~$

Whereas the global QM algebra is simply the tensor product of its factor
algebras, the relation of the net of local algebras to its $\mathcal{A(O})$
"pieces" is a more holistic relation; although together with its complement it
generates the global algebra $\mathcal{A(O})\vee\mathcal{A(O})^{\prime}=B(H),$
the global algebra $B(H)$ is not a tensor product of the two. The most
surprising property which underlines the terminology "holistic" is the fact
that the full net of local operator algebras which contains all physical
informations can be obtained by "modular tuning" \textit{of a finite number of
copies of a monad} in a shared Hilbert space\footnote{This number n is two for
the simplest case of a chiral algebra, whereas for a net in four spacetime
dimension the correct modular positioning can be achieved in terms of n=7
copies. The emergence of the spacetime symmetries in Minkowski spacetime as
well as possible inner symmetries of quantum matter is a consequence of this
holistic tuning.}; the reader who is interested in the precise formulation and
its proof is referred to \cite{K-W}, see also \cite{interface}. The fact that
the global oscillator variables are the same in both cases (\ref{f}) does not
reveal these fundamental holistic differences of spacetime organization of
quantum matter which have very different physical consequences. The present
quantization formalism (Lagrangian, functional integral) does not shed light
on those properties of QFT which solve the Einstein-Jordan conundrum in a
clear-cut way. If it comes to ensemble properties of localized observables,
the global aspects of generating covariant fields (which have no definite
localization region) on which covariant perturbation theory is founded are of
lesser importance than the local operator algebras $\mathcal{A(O)}$ which are
generated by all smeared fields $A(f)$ with $\sup pf\subset\mathcal{O}$. The
emphasis changes from covariance properties of fields to properties of
relative localization of operator algebras and this change finds its
appropriate mathematical form in the LQP ("local quantum physics") setting of
QFT \cite{Haag}.

It is precisely this holistic aspect which renders any calculation of the
subvolume fluctuation difficult; the simplicity of global oscillators is of no
help here. A calculation in closed form is (even in the absence of
interactions) not possible, and the imposition of covariance, which was the
important step for obtaining the modern form of perturbation theory, also does
not provide guidance. For renormalized perturbation theory one has clear
recipes which were extracted from the imposition of covariance, but this is of
not much help when one wants to find appropriate description of localized
fluctuation in open subsystems. Saying that the global aspects can be
described in terms of oscillators is almost as useless as trying to understand
the holistic structure of a living body in terms of its chemical composition
(in this analogy the chemical substances correspond the the global operators
whereas the nature of live corresponds to the organization of global
oscillators into algebras of local observables).

Although modular localization theory asserts the existence of "modular
Hamiltonians", in its present state it does not provide a generic method to
explicitly construct them. Jordan's chiral model is an exceptional case for
which, similar to the Unruh Gedankenexperiment, an explicit identification of
the modular Hamiltonian in terms of the spacetime symmetries of the model is
possible. Actually one may view Jordan's fluctuation problem as a predecessor
of the Unruh effect in other words: QFT was born with the "thermal"
localization aspects of the E-J conundrum which includes a completely
intrinsic pre-Born notion of ensemble-probability; however the proximity of
its date of birth to that of QM prevented an in-depth understanding of
differences beyond the shared $\hbar~$and the Hilbert space.

This begs the question how, with the understanding of foundational properties
of QFT still being that incomplete, it was possible to achieve the remarkable
progress in renormalized perturbation theory. To phrase it in a more
provocative historical context: how could one arrive at the Standard Model
without having first solved the 1925 Einstein-Jordan conundrum? The answer is
surprisingly simple: to get from the old Wenzel-Heitler formulation of
perturbation theory, in which the vacuum polarization contributions were still
missing, to the Tomonage-Feynman-Schwinger- Dyson perturbation theory for
quantum electrodynamics (QED), one only needed \ to impose covariance and
"exorcise" some ultraviolet divergences by finding plausible recipes. It was
the internal consistency of the result and not its derivation from Lagrangian
quantization which made renormalized perturbation theory successful.

Many years later there were also derivation of these renormalization rules by
starting from invariant free field polynomials (without using Lagangian
quantization\footnote{The free fields do not have to fulfill Euler-Lagrange
equations.}) and invoking spacelike commutativity in an inductive way (the
causal perturbation setting of Epstein and Glaser \cite{E-G}). But such
conceptual refinements (of reducing prescriptions to to an underlying
principle) had little impact on the Zeitgeist; in any case it would not have
helped to obtain the foundational insight into modular localization which is
required in order to solve the E-J conundrum.

This lucky situation of making progress by playfully pushing ahead and working
once way through a yet conceptual incomplete formalism with the help of
consistency checks did not extend much beyond Lagrangian quantization and
renormalized perturbation theory. As will be shown in section 6, it is
precisely this setting which determined the fate of QFT for more than half a
century which is now being replaced by a more general setting based on modular
localization. The latter has not only removed unnecessary restrictions from
renormalization theory, but also led to a different view about on-shell
constructions (section 5). When, in the aftermath of the
Lehmann-Symanzik-Zimmermann (LSZ) scattering theory and the successfull
adaptation of the Kramers-Kronig dispersion relations, the first attempts of
S-matrix based on-shell construction were formulated, the conceptual
difficulties of analytic aspects of on-shell properties were underestimated.
As one knows through more recent progress about modular localization, an
important aspect of the S-matrix, namely its role as a relative modular
invariant of wedge-localization, was missing. As a result, the true nature of
the particle crossing property was misunderstood by identifying it with
Veneziano's dual model crossing which was then passed to string theory (ST).

The correct formulation of the on-shell crossing property within a new
S-matrix-based construction project and the solution of the E-J conundrum are
interconnected via the principle of modular localization. It is the aim of
this paper to show the power of the latter by presenting the solution to these
two problems.

The first attempts to formulate particle physics and obtain an constructive
access outside of quantization and perturbation theory was the S-matrix in
Mandelstam's project \cite{Mandel}. As we know nowadays, and as it will be
explained in detail in the present work, this failed as a result of the
insufficient understood on-shell analytic properties. Their connection to the
causality principle are much more subtle than those to the off-shell
correlation functions. In retrospect it is clear that with the scant
understanding of the central crossing property (and more generally the
conceptual origin of on-shell analyticity properties), there was no chance in
70s for Mandelstam's S-matrix based particle theory project to succeed.

In retrospect it is also clear why this happened precisely when Veneziano's
mathematical construction of a crossing symmetric meromorphic function in two
variables was accepted as a model realization of particle crossing for elastic
scattering amplitudes. It is appropriate in an article, whose intention is to
shed light on still ongoing misunderstandings, to explain their origin in a
historical context.

The importance of the E-J conundrum in the development of QFT can be best
highlighted by \textit{following Galileio's example and imagine a dialog
between Einstein and Jordan about subvolume fluctuations but placing it in the
year 1927, after Max Born added his probability interpretation to Heisenberg's
and Schr\"{o}dinger's quantum mechanics.}

\textbf{Einstein}: Dr. Jordan, I appreciate that you finally accepted my
invitation to come to Berlin and I am very interested to understand why, after
first criticizing my fluctuation calculations in my statistical mechanics
thermal blackbody radiation model, you now claim that you find similar
fluctuation components in your new wave quantization at zero temperature.

\textbf{Jordan}: Thank you Professor Einstein for taking so much interest in
my work. The appearance of such a fluctuation spectrum in my new setting of
quantized waves in a vacuum state is indeed surprising. Although my wave
quantization of 2-dimensional Maxwell waves generalizes Heisenberg's
quantization in some sense, the fluctuation properties obtained by restricting
the vacuum to a subinterval leads to a very different situation from that
expected in his and Born's formulation of QM. It seems that my quantized
Maxwell waves cannot be subsumed into a quantum mechanics of systems with an
infinite number of oscillators.

\textbf{Einstein}: As you remember, I have some grave reservation against
associating a probability to an individual measurement on a quantized
mechanical system which I occasionally expressed in the formulation "the Dear
Lord does not throw dice". But I never had any problem with probability in
statistical mechanics, in fact my calculation of the Nadelstrahlung-component
in the black body fluctuation spectrum, which led me to the particle nature of
light on pure theoretical grounds, is based on the probability of quantum
statistical mechanics. Does the result of your subvolume fluctuation
calculation in the pure ground state of your field quantization mean that this
state appears impure if analyzed in the setting of an open subsysten?

\textbf{Jordan}: Professor Einstein, I am glad that you raised this question.
I have been breaking my head over these unexpected consequences of my new
quantized field theory and I would be dishonest with you, if I claim to
understand these conceptual implications. But since the main difference to
mechanics is the causal propagation, (which was already implicit in the
Nahewirkungsprinzip of Faraday and Maxwell and which you then succeeded to
generalize into your new relativity principle in a Minkowski spacetime), I am
inclined to suspect that the ensemble aspect, which one needs in order to
avoid the assignement of a probability to an individual mechanical system (as
proposed by my adviser Prof. Max Born), has its origin in the quantum
realization of causal localization. Somehow this principle creates a natural
ensemble associated with its causal completion of a localization region,
namely the ensemble of all local observables attached to that spacetime
region. This is in contrast to QM which deals with individual mechanical
systems for which the association to an ensemble is a useful mental construct
for the interpretation of QM. I tried to convince Prof. Born and my colleague
Werner Heisenberg, who despite their initial resistance finally agreed to
permit me to present my idees in a separate section of a joint paper which was
published two years ago. But I was not able to remove their doubts. It would
be very helpful for me to obtain some support from your side.

\textbf{Einstein}: I need some time to think about this new situation. Your
conjecture seems to suggest that your new theory of quantum fields, which is
certainly much more fundamental than Heisenberg's and Schr\"{o}dinger's
quantized mechanics, comes with an intrinsic notion of localized ensembles of
observables and an associated statistical mechanics type of probability. If
one could better understand how the less fundamental global quantum mechanics
can be related as a limiting case to your new fundamental quantum field theory
in such a way that Born's postulated probability is a relict of your local
ensemble probability, this may change my view and perhaps even influence my
quantum physical Weltanschauung. Let us remain in contact and please keep me
informed about future clarifications on the points raised in our conversation.//

In this imagined dialog, which could have radically changed the history of
QFT, I avoided the use of advanced mathematical concepts. of modular
localization for which there was no mathematical support in the 20s. The E-J
conundrum is best understood as a progenitor of an Unruh-like Gedankenexperiment.

The organization of this paper is as follows. In the next section the vacuum
polarization on the boundary of causal localization is derived for the
"partial charge", which is a modern formulation of Heisenberg's original
observation. Section 3 sketches the issue of modular localization and its KMS
property with special emphasis on the difference between a KMS ("Carnot")
temperature and that measured by a thermometer. In section 4 the KMS property
is used for the explicit construction of an isomorphism between the thermal
subvolume (interval in Jordan's chiral model) fluctuations in Jordan's model
with a corresponding statistical mechanics model representing Einstein's side.
Section 5 explains modular localization and its relation with the
Tomita-Takesaki modular operator theory. The ongoing impact of modular
localization on on-shell constructions of QFT, with particular emphasis on the
connection of particle crossing with the KMS identity, is addressed in section 7.

The most important consequence of modular localization for the ongoing
research in particle theory is the generalization of renormalized perturbation
to interactions involving arbitrarily high spin through the use of
string-localized fields in section 6. In the case of spin s=1 it leads to a
much deeper understanding of why gauge theory requires the indefinite metric
Krein space setting and how modular localization allows a formulation which
remains throughout in Hilbert space.

The same ideas which lead to unexpected progress also permit to expose the
misunderstandings which led to the dual model and ST as presented in section
7. In contrast to the stringlocal fields in higher spin QFT the localization
which string theorist attribute to it is that of a chain of quantum mechanical
oscillators (Born's localization) which bears no relation to causal
loxalization in spacetime. Section 8 addresses some old and in the maelstrom
of time lost insights about the connection between the cardinality of phase
space degrees of freedom and causal localization. This includes problems
concerning dimensional changes which came from ST but which can also be
formulated in the setting of QFT. The critique of the Maldacena conjecture,
concerning the nature of the AdS-CFT correspondence, addresses one of those
problems. The concluding remarks in the last section attempt to position the
present situation in particle theory within the historical context and the
expectations about its future.

\section{Vacuum polarization, area law}

In 1934 Heisenberg \cite{1934} finally published his findings about vacuum
polarizations (v. p.) in the context of conserved currents which are quadratic
expressions in free fields. Whereas in QM they lead to well-defined partial
charges associated with a volume V,%
\begin{align}
\partial^{\mu}j_{\mu}  &  =0,~Q_{V}^{clas}(t)=\int_{V}d^{3}xj_{0}%
^{clas}(t,\mathbf{x})\\
Q_{V}^{QM}(t)  &  =\int_{V}d^{3}xj_{0}^{QM}(t,\mathbf{x}),~Q_{V}^{QM}%
(t)\Omega^{QM}=0\nonumber
\end{align}
there are no such sharp defined "partial charges" $Q_{V}$ in QFT, rather one
finds (with $g_{T}$ a finite support smooth interpolation of the delta
function) \cite{Req}
\begin{align}
&  Q(f_{R,\Delta R},g_{T}):=\int j_{0}(\mathbf{x},t)f_{R,\Delta R}%
(\mathbf{x})g_{T}(t)d\mathbf{x}dt,~f_{R,\Delta R}=\binom{1,~\left\Vert
x\right\Vert \leq R}{0,\text{ }\left\Vert x\right\Vert \geq R+\Delta
R}\label{partial}\\
&  lim_{R\rightarrow\infty}Q(f_{R,\Delta R},g_{T})=Q,~\left\Vert Q(f_{R,\Delta
R},g_{T})\Omega\right\Vert =%
%TCIMACRO{\QATOPD{\{}{.}{F_{2}(R,\Delta R)\overset{\Delta R\rightarrow0}{\sim
%}C_{2}ln(\frac{R}{\Delta R})}{F_{n}(R,\Delta R)\overset{\Delta R\rightarrow
%0}{\sim}C_{n}(\frac{R}{\Delta R})^{n-2}}}%
%BeginExpansion
\genfrac{\{}{.}{0pt}{}{F_{2}(R,\Delta R)\overset{\Delta R\rightarrow0}{\sim
}C_{2}ln(\frac{R}{\Delta R})}{F_{n}(R,\Delta R)\overset{\Delta R\rightarrow
0}{\sim}C_{n}(\frac{R}{\Delta R})^{n-2}}%
%EndExpansion
\nonumber
\end{align}
where the logarithmic divergence corresponds to $n=2$.

The \textit{dimensionless} partial charge $Q(f_{R,\Delta R},g_{T})$ depends on
the "thickness" (fuzziness, roughness) $\Delta R=\varepsilon$ of the boundary
and becomes the $f$ and $g$-independent (and hence $t$-independent) conserved)
global charge operator in the large volume limit. The deviation from the case
of QM are caused by v. p.. Whereas the latter fade out in the $R\rightarrow
\infty$ limit, they grow with the dimensionless area $(\frac{R}{\Delta
R})^{n-2}$~for $\Delta R\rightarrow0.$ The simplest calculation is in terms of
the two-point function of conserved current of a zero mass scalar free field.
In the massive case the leading term in the limit $\Delta R\rightarrow0$
remains unchanged. We leave the elementary calculations (not elementary at the
time of Heisenberg) to the reader.

The presence of v. p. causes relativistic quantum fields to be more singular
than Schr\"{o}dinger fields and requires the formulation in terms of Schwartz
distribution theory as used in the above smearing of the current with smooth
finitely supported test function. The LQP setting on the other hand avoids the
direct use of such singular objects in favor of local operator algebras. In
such a description the singular nature of vacuum polarization is not directly
perceived in the individual operators but rather shows up in ensemble
properties of operator algebras. It turns out that under rather general
conditions there exists between two monad algebras a distinguished (by modular
theory) intermediate type $I_{\infty}$ algebra $N$ \cite{Haag}%
\begin{align}
&  \mathcal{A(O}_{\mathcal{R}+\Delta R})\supset N\supset\mathcal{A(O}%
_{\mathcal{R}}),~H\overset{V}{\rightarrow}H(N)\otimes H(N^{\prime}),\text{
}\eta\equiv V(\Omega\otimes\Omega)\\
&  VAB^{\prime}\Omega=A\Omega\otimes B\Omega,~A\in\mathcal{A(O}_{\mathcal{R}%
}),B^{\prime}\in\mathcal{A(O}_{\mathcal{R}+\Delta R}),~VNV^{\ast}%
=B(H)\otimes\mathbf{1}\nonumber
\end{align}
i.e. there exists a unitary operator $V$ which permits to write the full
Hilbert in terms of a tensor product such that $\mathcal{A(O}_{\mathcal{R}%
})\subset N,$ $\mathcal{A(O}_{\mathcal{R}+\Delta R})^{\prime}\subset
N^{\prime}$ where the "split vacuum"$~\eta$ is a state in the original Hilbert
space which corresponds to the tensor product of vacua.

In QM the unitary $V$ would be simply the identity operator expressing the
fact that the vacuum is a auxiliary mathematical state which remains
physically inert under splitting, i.e. the QM vacuum is not entangled under
spatial subdivisions. In QFT it is a state which on $N\otimes N^{\prime}$ is
nontrivially entangled in the sense of quantum information theory. However in
the sharp localization limit $\Delta R\rightarrow0$ the "quantum mechanical"
type $I_{\infty}~$converge towards the monads $A(\mathcal{O}_{R}%
),A(\mathcal{O}_{R}^{\prime})$ which commute but do not tensor-factorize. The
limiting entanglement is of a very singular kind which has no counterpart in
quantum information theory and is characteristic for subalgebras which do not
admit density matrix states as the monad. The situation is analogous to that
encountered in finite temperature statistical mechanics in the thermodynamic
infinite volume limit when the tracial nature (the Gibbs formula) of the state
is lost and only the KMS property remains\footnote{Whereas the thermodynamic
limit monad is approximated from the inside, the split property approximates
the local monad from the outside.}.

The above described nontrivial behavior under splitting leads to a nontrivial
$\Delta R$ dependent \textit{localization entropy} which is consistent with
the KMS impurity of the restricted vacuum. In fact, since the vacuum
polarization happens in a layer of size $\Delta R$ (the "fuzzy" boundary) the
entropy is a function of the dimensionless area%
\begin{align}
&  ~En(R,\Delta R)=split~localization~entropy\label{area}\\
&  En|_{\Delta R\rightarrow0}\simeq ca,~~a=\frac{area}{(\Delta R)^{d-2}%
}~~for~d>2\nonumber
\end{align}
where the second line is the leading order in the sharp localization limit
which one expects if the "polarization clouds", which determine the singular
behavior of smeared fields as Heisenberg's partial charges (\ref{partial}),
are the same as those which appear in the above entropy argument.

Note that in contradistinction to the treatment in the literature where the
connection with the model of local QFT is lost by introducing an imagined and
ill-defined \ momentum space cutoff\footnote{One can cut off integrals but to
cut off a model of QFT is ill defined.}, the implementation of the split
property is a construction \textit{within} the QFT model.

The logarithmic behavior for d=2 split entropy can actually be rigorously
derived \cite{BMS} and is well-known to condensed matter physicists. For
Jordan's chiral current model used in the E-J conundrum, the entropy can be
directly obtained from the isometry with a chiral statistical mechanics model
(section 4). This situation is very special and has been termed "the inverse
Unruh effect \cite{S-W}. For d=1+3 't Hooft has obtaind the area behavior in
terms of the "brickwall picture" \textbf{\cite{brick}, }but a rigorous
derivation, solely based in the split property of modular localization, is not
yet available. Bekenstein's area law results if one relates $\Delta R$ with
the Plank length.

There exists a conjecture that even in the general case there could remain a
weak form of the "inverse Unruh effect" \cite{S-W} in which the spatial volume
factor is replaced by the "volume factor" of a box with two spacelike and one
lightlike direction. In that case the two spacelike extensions would account
for the dimensionless area factor and the lightlike contribution would be (as
in the chiral Jordan model) logarithmic \cite{BMS} so that the net result is a
logarithmically modified area law.

Either behavior of localization-entropy shows that although there are genuine
infinities in QFT; they are limited to sharp localization within a model and
not a predicate of QFT; in case of quantum fields they are controlled in terms
of testfunction smearing. Unlike the misunderstood ultraviolet divergencies in
the old formulation of perturbation theory, they have no relation to the
"ultraviolet catastrophe" i.e. they threaten in no way the consistency of QFT;
to the contrary, they are a direct consequence of its most foundational
modular localization property. In a certain sense the divergence of
thermodynamic infinite volume limit correspond to the infinity obtained in the
sharp boundary limit (increasing sharpness of the boundary) $\varepsilon
\rightarrow0.$

With the notion of "localization temperature" and energy one has to be much
more careful than with the dimensionless localization entropy. When one
naively interprets the Unruh temperature as that measured by a thermometer,
one enters a conceptual mine field. The equality of the thermometer (local)
temperature (related to the zeroth thermodynamic law with the "Carnot
temperature" of the second fundamental law of an KMS equilibrium state is only
correct in an inertial system, but the "egg-boiling local temperature of the
Unruh effect refers to an accelerated observer. In fact the thermometer
\textit{temperature in a vacuum state remains zero}; it is a "local
temperature" which does not depend on the Unruh trajectory \cite{B-S}. The
same holds for other situations described by modular theory (next section);
although there is always a dimensionless modular Hamiltonian and a
dimensionless temperature $\beta=2\pi~$associated with modular KMS states, The
still ongoing hot topic about "firewalls" \cite{Raju} is dangerously close to
the Unruh "cooking temperature" and more investigations about possible
differences between causal horizons (Unruh) and event horizons of black holes
are necessary for clarification.

Another useful conceptual warning in passing from classical fields to quantum
fields is to avoid to attribute a direct physical meaning to fields, but
rather to view them in a similar role as that which coordinates play in the
description of geometry. This is facilitated by the fact that quantum fields
are not directly measured (no experimentalist has measured a nuclear field);
rather the notion of a quantum field serves as a \textit{device to describe
particles} which are related to a particular subset of quantum field i.e. the
same particles can be interpolated by many different fields. It has turned out
that to view fields in their role as coordinatizing or generating local
algebras is the most useful way of keeping track of the differences betweem
description-dependent fields from intrinsic particles. In this way particles
do not correspond to individual fields but rather to local field classes which
carry the same superselection charges. All structural properties of LQP and
the resulting general theorems can be expressed in terms of local nets of
operator algebras, but the present formulation of renormalized perturbation
theory still needs generating fields.

Note that the well known entropy conjecture by Bekenstein, based on equating a
certain area behavior in classical General Relativity with quantum entropy,
results formally from the above area law by equating $\Delta R$ with the
Planck length. Quantum Gravity is often thought of as that still elusive
theory which explains why and how the quanta of gravity can escape the
consequences of modular localization for sharp localization which are
responsible for the singular short distance aspects of causal localization. If
Bekenstein's conjecture really describes quantum aspects of gravity (and not
just quantum matter in curved spacetime), then modular localization cannot be
extended to Quantum Gravity.

As mentioned before the relation between $\Delta R$ and the entropy is
reminiscent of Heisenberg's quantum mechanical uncertainty relation in which
the uncertainty in the position is replaced by the split distance $\Delta R$
within which the vacuum polarizations can attenuate, so that outside the
vacuum returns to play its usual role (if tested with local observables in the
causal complement of $\mathcal{O}_{\mathcal{R}+\Delta R})$.

It should be stressed again that the probability interpretation, which Born
had to add to Heisenberg's and Schr\"{o}dinger's formulation of QM, is
completely intrinsic to LQP. It is a consequence of the "thermal" KMS property
of ensembles of operators contained in a localized algebra $\mathcal{A(O)}$ in
$\mathcal{O}$-restricted physical (finite energy states). As such it is not
different from the statistical mechanics probability, which Einstein used in
his fluctuation arguments in terms of which he challenged the physical content
of Jordan's thesis. It is only with the modern concept of modular localization
and the hindsight of more than eight decades of QFT that one realizes how
close the E-J conundrum came to the intrinsic probability coming from the
quantum formulation of the Faraday-Maxwell-Einstein causal locality principle
in Minkowski spacetime. Einstein's problem was the assignement of a
probability to an individual mechanical system (which requires to
\textit{imagine} it as a member of an ensemble for which the probabilistic
nature is seen in repeated measurements). The fact that probability is
intrinsic to QFT and that the vacuum entanglement of sharp localization is
more singular than that of quantum information theory influences the
discussions around Bell`s inequalities but does not invalidate them. The
effects of the (more radical form of) vacuum entanglement in QFT remain orders
of magnitudes below the quantum mechanical entanglement of particle state
which can be directly measured in terms of quantum optical methods.

A particular radical illustration of the conceptual differences between QFT
and QM is the reconstruction of a net of operator algebras from the relative
modular position of a finite number of copies of the monad \cite{interface}.
For chiral theories \ on the lightray one needs two monads in a shared Hilbert
space in the position of a \textit{modular inclusion}, for d=1+2 this "modular
GPS" construction needs three and in case of d=1+3 seven modular positioned
monads are sufficient to create the full reality of a causal quantum matter
world, including its Poincar\'{e} symmetry (and hence Minkowski spacetime)
from the abstract modular groups \cite{K-W}. This possibility of obtaining
concrete models by modular positioning of a finite number of copies of an
abstract monad (indecomposable constructs without inner structure) in a shard
Hilbert space is the strongest "holistic outing" of QFT; the reader is
encouraged to look at this application of modular theory \cite{K-W}. For d=1+1
chiral models the modular positioning leads to a partial classification of
chiral theories as well as to their explicit construction of large classes of
models (section 5).

Apart from d=1+1 factorizing (integrable) models, where modular properties in
the form of \textit{nuclear modularity} were used for existence proofs of
models \cite{Lech}, QFT has not yet reached the state of maturity where such
holistic properties can be applied for classifications and existence proofs of
families of models and their mathematically controlled approximation. An
extension to curved spacetime would be very interesting; the simplest question
in this direction is the modular construction of the local diffeomorphism
group on the circle in the setting of chiral theories.

\section{Modular localization and its thermal manifestation}

The aim of this section is to present the concept of \textit{modular
localization} which is the backbone of LQP and represents the intrinsic
formulation of causal quantum localization. Since, as mentioned before,
subalgebras $\mathcal{A(O})$ localized in spacetime regions $\mathcal{O}$ with
$\mathcal{O}^{\prime\prime}\subsetneq R^{4}$ are known to act cyclic and
separating on the vacuum (the Reeh-Schlieder property \cite{Haag}), the
"standardness" condition for the validity of the Tomita-Takesaki modular
theory is always fulfilled for local subalgebras. This leads to a uniquely
defined Tomita operator $S_{\mathcal{O}}$ whose properties will be the main
subject of this section.

It has been known for a long time that the algebraic structure underlying free
fields allows a functorial interpretation in which operator subalgebras of the
global algebra $B(H)$ are the functorial images of \textit{subspaces of the
Wigner wave function spaces} ("second quantization"\footnote{Not to be
confused with quantization; to quote a famous saying by Ed Nelson:
"quantization is an art, but second quantization is a functor".}).

Before presenting some mathematical details, it is useful to recall some
philosophical points. LQP avoids the parallelism to classical field theory
which charaterizes the Langrangian quantization approach of QFT and the
closely related functional integral representation. Accepting that QFT is more
fundamental than classical field theory, the content of QFT should reveal
itself in terms of its own principles without the detour of a "quantization
parallelism" to classical field theory.

In contrast to QM, the LQP setting of QFT de-emphasizes individual operators
in QFT in favour of \textit{ensembles of operators} which share the same
spacetime localization region. These ideas also follow more closely the
situation in the laboratory, where the experimentalist measures coincidences
between events in spacetime. All the measured particle properties, including
the nature of spin and internal quantum numbers, were obtained by repetitions
and refinements of observations based on counters which are placed in compact
spatial region and are maintained activated for a limited time. Their detailed
internal structure is generally not known, what matters is their localization
in spacetime and the sensitivity of their response. However without a precise
mathematical backup which matches these physical concepts, LQP remains in the
philosophical realm.

The role of covariant quantum fields in LQP is that of generators of a net of
local operator algebras $\{\mathcal{A(O})\}_{O\in R^{4}}$ which act in a fixed
Hilbert space. In the Wightman setting a field is a covariant operator-valued
distribution $A(x)$ which is globally defined for all $x\in R^{4}.$ From its
global definition one passes to (unbounded) $\mathcal{O}$-localized operators,
formally written as $A(f)=\int A(x)f(x)d^{4}x,$ $suppf\subset\mathcal{O}$,
which according to Wightman's axioms define a system of polynomial (generally
unbounded) operator $^{\ast}$-algebras $\mathcal{P(O}).$ Formally these
unbounded operators can be associated with an aforementioned net of
(mathematically easier manageable) bounded operators forming von Neumann
algebras, which is the starting point of Haag's LQP setting. The advantage is
that one obtains access to the well-developed mathematical theory of operator
algebras (omitting from now on "bounded"). Certain causality aspects allow a
more natural definition and more profound understanding in the LQP setting.
The mathematical details, which allow to pass between Wightman's description
to the algebraic local nets of observables in the LQP setting and vice versa,
are tedious and still technically incomplete \cite{Haag}, but this had little
effect on progress.

Whereas both settings are different formulations of closely related physical
concepts, there is a significant distinction between these settings and
constructions based on Lagrangian or (the closely related) functional integral
based quantization methods. Quantization is not a physical principle; whereas
classical descriptions often help to find a perturbative description
("quantization") of a QFT, there is no general correspondence. The fact that
the less fundamental QM which lacks causal localization and its holistic
consequences, is capable to maintain an almost (up to ordering prescriptions
of operators) unique connection to classical mechanics does not imply that
such a close relation must continue to hold in QFT. The strong link between
classical mechanics and its quantum counterpart finds its best known
expression in the fact that Lagrangian quantization (canonical quantization)
and functional quantization (path integrals) enjoy solid mathematical support
from measure theory but not in QFT.

All this breaks down in interacting QFT with realistic short distance
behavior\footnote{Free field short distance behavior of polynomially coupled
scalar fields is still in the reach of measure-theoretical functional methodes
\cite{Gl-Ja}.}. Apart from d=1+1 integrable models (section 5), for which
rigorous methods of LQP led to existence proofs \cite{Lech}\cite{Le}, there is
of course renormalized perturbation theory; but since perturbative expansions
in the coupling strengths (which are consistent on the level of polynomial
relations) inevitably lead to divergent series, they are not the right objects
for a mathematically controlled approach to QFT. In fact there exists not even
a mathematical argument that they define an asymptotic approximation in the
limit of vanishing coupling to an existing model of QFT, although the use of
low order perturbative results led in many cases to quite spectacular
agreements with observations. Whereas the setting of QM has reached its
closure a long time ago, the conceptual/mathematical flanks of QFT remain open.

The \textit{causal perturbation} setting of Epstein and Glaser \cite{E-G}
avoids the ultraviolet divergencies of the Lagrangian or functional setting by
implementing causal locality in terms of time-ordered products in an inductive
way. A specific model is defined in terms of its free field content and the
starting point is a first order interaction density in form of a
Lorentz-invariant (scalar) Wick-polynomial. The scaling degree of the
interaction density is determined in terms of the scaling degrees of the
participating fields and their derivatives. If the scaling degree of the
interaction defining first order polynomial in terms of free fields does not
surpass $d_{s.d.}^{int}=4$ one obtains a renormalizable model in which the
short distance dimensions of quantum fields remain bounded, independent of the
number of iterative steps (order of perturbation).

The problem with this setting is its limitation with respect to the spin of
pointlike free fields in a Hilbert space setting. The short distance dimension
of massive pointlike free fields in Hilbert space increases with spin as
$d_{s.d.}=s+1.$ Hence a $m>0,~s=1$ Proca potential with $d_{s.d.}=2$ does not
admit any renormalizable interaction in Hilbert space and the infrared
divergencies of its m=0 limit are well-known additional obstacles of
perturbation theory. Wigner's 1939 classification of particles in terms of
positive energy representations led to a clear statement about the field
content of covariant $(m=0,s\geq1)$ representations: there are covariant
pointlike field strengths\footnote{Massive pointlike potentials and their
associated field strengths have the same $d_{s.d.}=s+1,$ but whereas the zero
mass limit of field strengths exists, that of potentials does not.} but no
covariant pointlike potentials. This is the famous \textit{clash between
Hilbert space positivity and pointlike localization}. The conventional way out
is that of keeping the pointlike structure and allowing indefinite metric
so-called Krein spaces instead of Hilbert spaces.

This problem is not present in classical Maxwell theory; in that case the use
of vectorpotentials contains a redundancy which affects the connection of
Cauchy data and their causal propagation and is conveniently taken care of in
terms of the concept of gauge transformations and gauge invariance (the return
to field strengths and currents). Lagrangian quantization and functional
integral presriptions for gauge theories lead out of the Hilbert space; in
fact pointlike interaction-free massless vector potentials are well known to
require a Krein space formulation (the Gupta-Bleuler formalism and its BRST
extension). Since the Hilbert space setting is the foundational pillar of QT,
\textit{quantum gauge theory} in the presence of interactions of massive or
massless vectormesons is an undesired but inevitable compromise which is
suggested by Lagrangian quantization. Since classical field theory does not
not know anything about Hilbert space positivity, there is a serious obstacle
to quantization for interacting $s\geq1$ interactions and gauge theory is a
compromise which only describes the vacuum sector which is generated by the
subalgebra of gauge invariant pointlike local fields acting on the vacuum
states) and leaving the important charge-creating operators and the physical
particle-like states they create from the vacuum outside the physical range of
the quantum gauge setting.

This makes it desirable to turn to another description which the previously
mentioned alternative suggests: \textit{abandon pointlike localization and
keep instead the Hilbert space}. Since this is inconsistent with the
quantization of pointlike classical gauge theory, it is not surprising that
such an alternative requires a radical change of the Epstein-Glaser causal
perturbative setting \cite{E-G}. Although this formalism does not depend on
quantization of a classical field structure\footnote{In particular it does not
depend on whether the quantum fields are solutions of Euler-Lagrange
equations.}, it still uses pointlike generating fields in an essential way.
The safest procedure is to try to extract an information from the foundational
localization principles of LQP by asking the following structural questions:
\textit{what is the tightest localization which can be derived solely from the
mass gap property}?

The type of models for which such a question could be relevant are interacting
massive vectormesons. As mentioned before pointlike interactions of such
fields are nonrenormalizable, and since the new Hilbert space setting shows
that the concept of renormalizability is intimately related to the short
distance aspects of localization, The weakening from point- to string-
localization is the result of the restrictive Hilbert space positivity which
is absent in the Krein space setting of gauge theory.

Interestingly it is not necessary to use weaker than string-localized fields
in order to describe a QFT; this is part of a theorem by Buchholz-Fredenhagen
\cite{Haag}: all LQP with a mass-gap (which are known to admit scattering
theory) can be generated by spacelike semi-infinite stringlocal fields.
Covariant generating stringlocal fields $\Psi(x,e),$ $e^{2}=-1$ are localized
on $x+\mathbb{R}_{+}e$ and commute for spacelike separated strings
(appropriately modified for Fermions). In section 6 the string-extended E-G
perturbation theory will be exemplified in massive gauge theories. Whereas the
local observables (field strengths, currents) remain pointlocal and the
interacting physical matter fields are stringlocal, the S-matrix turns out to
be $e$-independent. Massive vectormesons also permit a coupling to neutral
matter (scalar Hermitian fields $H$).

These couplings reveal what was known to some researchers for a long time: the
Higgs mechanism about a mass-creating symmetry breaking is not supported by
QFT; the intrinsic property of all couplings of massive vectormesons to matter
(independent of whether the latter is charged or neutral) is the
"Schwinger-Higgs screening" of the Maxwell charge which is directly related to
to the field strength of the massive vectormeson.. Although this is consistent
with the BRST gauge setting, the new Hilbert space setting using
renormalizable couplings of stringlocal massive vectormesons lead to these
results without having to rely on unphysical Krein space methods (section 6).
Computations need not any more be based on sucessful but (from the quantum
viewpoint) somewhat miraculous descriptions. A surprising new structure which
results from the Hilbert space positivity for renormalizable interactions of
massive $s\geq1$ stringlocal fields is the appearance of lower spin "escort"
fields. In the case of massive vectormesons this is a stringlocal neutral
scalar $\phi$ field which share many properties with the Hermitian $H$ fields
of the Higgs model apart from the fact that they have no relation to any
mass-generating symmetry breaking (section 6).\ 

The fundamental idea which is behind the ongoing radical changes for
interaction$~$involving stringlocal fields is a much deeper understanding of
\textit{quantum} causal locality in the algebraic operator setting of modular
localization. Individual quantum fields never played a similar distinguished
physical role as they do in classical field theory. As mentioned before they
are not directly measured (measuring a hadronic field ?) and the particles
which are identified with counter events are always associated with an
infinite class of (composite) fields which carry the same superselected charge
and are relatively local with respect to each other. Whereas in QM it makes
sense to distinguish in terms of elementary particles and their bound states,
such a hierarchy is rather meaningless in QFT; the omnipresence of vacuum
fluctuations only respects the superselected charges but couples all states
which have the same such charge. The fields within one superselected class are
distinguished by their short distance scale dimensions, and the renormalizable
Lagrangian couplings highlight fields with low $d_{s.d.}.$ but the particle
field relation is based on infinite timelike separations (time-dependent
scattering theory) for which low $d_{s.d.}$ values are irrelevant. QFT is a
quantum theory in which \textit{everything which according to the
superselection rules can be coupled is actually coupled} (there is always a
process in which this coupling is activated). This explains why methods of
quantum mechanics are rather useless in QFT but at the same time this is the
prize to be paid for a fundamental theory. Modular localization theory brings
all these foundational properties (which still remain somewhat hidden in the
perturbation theory in terms of individual fields) into the forefront.

The central issue in LQP refers to two physically motivated requirements on
the local net of operator algebras%
\begin{align}
\left[  \mathcal{A(O}_{1}),\mathcal{A(O}_{2})\right]   &  =0,~\mathcal{O}%
_{1}><\mathcal{O}_{2},~Einstein~causality\label{cau}\\
\mathcal{A(O})  &  =\mathcal{A(O}^{\prime\prime}),~causal\text{ }%
completeness\nonumber\\
\mathcal{A(O}^{\prime})  &  =\mathcal{A(O})^{\prime},~Haag\text{
}duality\nonumber
\end{align}
The first line is a condensed notation for the commutativity of operators from
spacelike separated regions; it is only required for observable fields. The
commutation property for \textit{non-observable} operators, as those coming
from spinor fields or fields carrying superselected charges, are determined by
the local representation properties of the observables (the superselection
theory to their associated observable subalgebras \cite{Haag}).

The \textit{causal completeness property} (\ref{cau}) is a local adaptation of
the old time-slice property \cite{H-S}. In classical relativistic field theory
the field values in the relativistic "causal shadow" (or causal completion)
$V^{\prime\prime}~$(two-fold causal complement of $V$) is the region in which
the classical field values are uniquely determined in terms of the (properly
defined) initial values in a finite volume $V$ at fixed time. Its quantum
adaptation in the LQP setting is the algebraic causal completeness property.
Often particle theoreticians only consider the simpler Einstein causality
property and ignore causal completeness. But there are situations which are
consistent with Einstein causality but violate causal completness\footnote{In
quantum physical terms a completeness violating situation exhibits a
"poltergeist" behavior: new degrees of freedom (which were not present in
$\mathcal{A(O})$) enter $\mathcal{A(O}^{\prime\prime}$)~from "nowhere".}. In
fact in \cite{H-S} the simplest model, a so-called generalized free field with
a suitable continuous mass distribution was used as an illustrative example
for a physically unacceptable Einstein-causal field. Whereas in the setting of
Lagrangian quantization causal completeness is the formal consequence of the
quantization of causally propagating relativistic fields, this property needs
special attention in situations in which classical analogs are not available
as e.g. ideas coming from string theory.

This affects in particular relations between QFTs in different spacetime
dimensions. The fact that in some cases they are backed up by a mathematical
isomorphism does not imply that they are physically acceptable. One such
trend-setting case is the Maldacene conjecture which originally arose in
string theory. Its mathematical basis is an algebraic isomorphism \cite{Reh}
which extends the well-known equality of the spacetime conformal symmetry of a
conformal field theory (CFT$_{n}$) in n spacetime dimensions with the
spacetime symmetry of an anti de-Sitter space in n+1 dimensions (AdS$_{n+1}$)
to a mathematical isomorphism which between suitably chosen local subalgebras
on both sides. But it turns out that this relation only preserves Einstein
causality but violates the causal completeness requirement; if one starts from
an AdS theory which fulfills both, the resulting conformal field theory
fulfills Einstein causality but violates causal completeness and a similar
problem exists if one uses the isomorphism in the opposite direction; a
physical correspondence requires more than a mathematical isomorphism between
certain localized subalgebras.

Unfortunately the knowledge about these important properties (the relevance of
causal completeness) which was attained in the early 60s \cite{H-S} has been
lost within the string-theory community, otherwise Maldacena would not have
been able to convince a world-wide community that the mathematical
$AdS_{n+1}\longleftrightarrow CFT_{n}$ isomorphism can be lifted to a physical
correspondence. Only holographic projections onto a $n$-$1$
\textit{null-surface} lead to a right "thinning out" of degrees of freedom
(loss of information). As a consequence of a loss of some informations one
cannot return to the original theory; nevertheless most informations are in
the holographic projection.

There exist however situations in certain quantum field theories, which
contain massless $s\geq1$ in which for multiply connected spacetime regions
the Haag duality is violated in a specific way; the prototype is the quantum
Aharonov-Bohm effect for the net of algebras generated by the quantum
electromagnetic field strength \cite{charge}. In the case of zero mass field
strengths for $s\geq1$ this is directly related to the clash between pointlike
localization of potentials and the positivity of Hilbert space and its
resolution in terms of stringlocal potentials \cite{Hilbert}.

Mathematically it is very easy to construct Einstein-causal theories which
violate causal completeness and as a consequence (apart from the
aforementioned topological exceptions) lead to pathological physical
properties with respect to their "degrees of freedom" behavior. Well known
cases in addition to the mentioned Maldacena conjecture arise from embedding
lower dimensional quantum field theories and its reverse: Kaluza-Klein
dimensional reductions.

As a result of a subtle relation between the cardinality of phase-space
degrees of freedom with localization (split property, causal completeness,..),
the nuclearity property (introduced by Buchholz and Wichmann \cite{Haag}) in
conjunction with modular theory ("modular nuclearity") became an important
concept for the classification and nonperturbative construction of models of
QFT \cite{Lech} \cite{integrable}.

After having presented some of the physical requirements of the LQP
formulation of QFT, we now pass to a brief description of its main
mathematical support: the Tomita-Takesaki modular operator theory. This theory
has its origin in the operator-algebraic aspects of group representation
algebras from which Tomita took the terminology "modular" (originally
referring to properties of Haar measures). A conference in the US (Baton
Rouge, 1967), which was attended by mathematicians (Tomita, Takesaki,
Kadison,..) and mathematical physicists (Haag, Hugenholz, Winnink,
Borchers,..), is marks the beginning of the Tomita-Takesaki modular operator
theory as a joint project \cite{Bo}. The participating physicists had already
obtained important partial results of that theory through their project of
formulating quantum statistical mechanics directly in the thermodynamic limit
(statistical mechanics of \textit{open systems}) \cite{Haag}. In their new way
of thinking, the Kubo-Martin-Schwinger property (originally an analytic
shortcut for computing Gibbs traces) assumed a conceptual role in the new
formulation of thermal equilibrium states for \textit{open quantum systems}.
Although these ideas originated independently, this conference united them;
there is hardly any area in which the contribution of mathematicians and
physicists have been that much on par as in modular operator theory/modular localization.

One reason for this perfect match was that the area of physical application of
modular theory widened the scope of statistical mechanics and, combined with
\textit{causal localization,} became the most important
mathematical/conceptual tool of LQP. The basic fact which led to this new
connection was the Reeh-Schlieder theorem \cite{Haag} which secures the
validity of the "standardness" requirement for the applicability of the
Tomita-Takesaki theory. Standardness of a pair $(\mathcal{A},\Omega)$ (algebra
and state) means that the action of the operator algebra $\mathcal{A}$ on the
state vector $\Omega~$generates the Hilbert space (cyclicity of $\Omega$) and
that there are no annihilators of$~\Omega$ in $\mathcal{A}$ ($\Omega$ is
separating)%
\[
cycl.:~\overline{\mathcal{A}\Omega}=H,~~~separ.:~A\Omega=0\curvearrowright
A=0,~A\in\mathcal{A}~
\]
The Reeh-Schlieder theorem guaranties the validity of this property for any
pair $(\mathcal{A(O)},\Omega),~\mathcal{O}^{\prime\prime}\subset\mathbb{R}%
^{4};~$in fact this even holds if the vacuum is replaced by any finite energy
state. The importance of the relation between localization and the T-T theory
was noted a decade after then Baton Rouge conference by Bisognano and Wichmann
\cite{Haag}; these authors found that in the context of localization in a
wedge region $\mathcal{O}=W$ the Tomita-Takesaki theory makes contact with
known geometrical/physical objects.

The general T-T theory is based on the existence of an unbounded antilinear
closable involution $S~$with a dense domain $domS$ in $H$ which contains all
states of the form $A\Omega,~$in case of a standard pair $~$\cite{Do-Lo}
\cite{Sum}$.~$Whereas the cyclicity secures the existence of its dense domain,
the absence of annihilators of $\Omega$ in $\mathcal{A}$ guaranties its uniqueness.%

\begin{align}
&  S_{\mathcal{O}}A\Omega=A^{\ast}\Omega,~A\in\mathcal{A\subset}%
B\mathcal{(}H\mathcal{)},~S=J\Delta^{\frac{1}{2}}=\Delta^{-\frac{1}{2}}J~\\
&  J~antiunit.,\text{ }\Delta^{it}~mod.\text{ }unitary,~\sigma_{t}%
(A)=Ad\Delta^{it}A\nonumber
\end{align}
The existence of a polar decomposition in terms of a antiunitary $J$ and a
positive generally unbounded operator $\Delta$ follows from the closability of
$S$ (in the following $S$ stands for the closure)$.$ The modular unitary gives
rise to a \textit{modular automorphism} group of the localized algebra
$\mathcal{A}$.

The physical interpretation in massive theories is only known only for
$\mathcal{O}=W=$ wedge regions, which are Poincar\'{e} transforms of the
standard $t$-$z$ wedge $W_{0}=\left\{  z>\left\vert t\right\vert
;\mathbf{x\in}\mathbb{R}^{2}\right\}  .$ In that case the modular objects are
the unitary transformation representing the W-preserving Lorentz ("boost")
subgroup $\Delta_{W}^{it}=U(\Lambda_{W}(-\pi t))$ and the reflection on the
edge of the wedge $J$ which is, up to a $\pi$-rotation within the edge, equal
to the TCP operator. Since in a theory with a complete particle interpretation
(to which the considerations of this paper are restricted, unless stated
otherwise) the interacting TCP operator and its incoming (free) counterpart
are known to be related by the scattering operator $S_{scat}~$\cite{Jost}$,$
we obtain for all $J~$for arbitrary $W$ \cite{AOP}%
\[
J_{W}=S_{scat}J_{W,in}\text{ ~}for~all~W
\]
This expresses a property of $S_{scat}$ which turns out to be indispensable
for the constructive use of modular localization in QFT, namely $S_{scat}%
$\textit{ is a relative modular invariant between the interacting and the
associated free (particle) wedge algebra.} This property was recently used
within a more physical derivation \cite{Mund} of the Bisognano-Wichmann
theorem which reduces the interacting case in theories with mass gaps and a
complete particle interpretation to that of free fields (see below).

The relative modular invariance of $S_{scat}$ is the crucial property which
accounts for the analyticity of on-shell objects as $S_{scat}~$and the related
formfactors. These on-shell analytic properties find their important
manifestation in the \textit{particle crossing property}. It is also the
starting point of the algebraic construction of integrable QFT \cite{AOP}. The
connection between algebraic and analytic properties is much more subtle for
on-shell objects as the S-matrix and formfactors than for off-shell
correlation function. Since most of these properties were not understood in
the 60s, it is not surprising that Mandelstam's project of formulating
particle physics as a quantization-avoiding on-shell project failed on the
lack of understanding of the relevant on-shell analytic properties.

The misunderstandings about the particle crossing property in the construction
of the \textit{dual model}, which later entered string theory, have their
origin in confusions about the meaning of localization in QFT as opposed to
QM. In section 7 these misunderstandings will be analyzed in the light of
recent progress.

Since it is not possible to present a self-consistent complete account of the
mathematical aspects of modular localization and its physical consequences in
a history-motivated setting as the present one, the aim in the rest of this
section will be to raise awareness about their physical origin.

It has been known for a long time that the algebraic structure associated to
free fields allows a functorial interpretation in which operator subalgebras
of the global algebra $B(H)$ are the \textit{functorial images of certain real
subspaces} of the Wigner space of one-particle wave functions (the famous
so-called "second quantization"\footnote{Not to be confused with quantization;
to quote a famous saying by Ed Nelson: "quantization is an art, but second
quantization is a functor".}), in particular the spacetime localized algebras
are the images of localized real subspaces. This means that the issue of
localization to some extend can be studied in the simpler form of localized
subspaces of the Wigner particle representation space (unitary positive energy
representations of the $\mathcal{P}$-group).

These localized subspaces can be defined in a intrinsic way \cite{BGL} i.e.
without quantization, only using operators from the positive energy
representation $U$ of the proper Poincar\'{e} group $\mathcal{P}%
_{+}\mathcal{~}(det=+1)\mathcal{~}$on the direct sum of two copies of the
Wigner representation $u~$of the connected component (proper orthochronous
$\mathcal{P}_{+}^{\uparrow}$) on the one-particle space $H_{1}.$ For
simplicity of notation the transformation formulas are limited to the case of
a spinless charged particle:%

\begin{align}
H_{1}  &  :\left(  \varphi_{1},\varphi_{2}\right)  =\int\bar{\varphi}%
_{1}(p)\varphi_{2}(p)\frac{d^{3}p}{2p_{0}},~~\hat{\varphi}(x)=\frac{1}{\left(
2\pi\right)  ^{\frac{3}{2}}}\int e^{ipx}\varphi(p)\frac{d^{3}p}{2p_{0}}\\
&  U(g)(\varphi_{1}\oplus\varphi_{2})=u(g)\varphi_{1}\oplus u(g)\varphi
_{2},~u(a,\Lambda)\varphi(p)=e^{ipa}u(\Lambda^{-1}p)\nonumber\\
\Theta &  \equiv TCP,~\Theta(\varphi_{1}\oplus\varphi_{2})=C\varphi_{2}\oplus
C\varphi_{1},~C\varphi(p)=\overline{\varphi(p)}%
\end{align}
Any $\mathcal{P}_{+}$ transformation can be generated from $U(g)$ and
$\Theta.$ For representations with $s>0$ the Lorentz group acts through Wigner
rotations (Wigner's "little group") on the "little Hilbert space" which in the
massive case is the 2s+1 component representation space of rotations. The
massless case leads to a 2-dimensional Euclidean "little space" whose
degenerate representation (with trivially represented "little translations")
form a two-component little helicity space, whereas faithful representation
acts in an infinite dimensional Hilbert space ("infinite spin") \cite{MSY}.
The Lorentz transformations as well as $\Theta$ act also (through
representations of the little group) on the little Hilbert space.

It is precisely through the appearance of this little Hilbert space that the
problem of causal localization of states (wave functions) cannot be simply
solved by Fourier transformation and adding positive frequency contributions
of particles with those of negative frequency from antiparticles. Whereas in
the case of the two classes of finite little spaces (the massive and zero mass
finite helicity class) of positive energy Wigner representation, their
"covariantization" was easily achieved in terms of group theoretic methods
\cite{Weinbook} and led to local pointlike generating wave functions and
fields, this third infinite spin class posed a series obstacle. Attempts to
convert its members into covariant pointlike wave functions and corresponding
fields remained unsuccessful and there was no understanding of the origin of
this failure\footnote{Reference \cite{Y} is an exception in that certain
aspects of the localization problem were already noted.}. Weinberg dismissed
this large positive energy representation class by stating that nature does
not make use out of it \cite{Weinbook}. Since all important physical
properties are connected to aspects of localization which are precisely those
properties which at that time remained poorly understood, this dismissal could
be premature, in particular in times of dark matter.

The localization problems of the infinite spin class were finally solved
\cite{BGL} with the help of modular localization, a concept which for
different problems was already used in \cite{AOP}. In fact the main theorem in
that paper states \cite{BGL} that all positive energy wave functions are
localizable in noncompact spacelike cones and only the first two classes
permit the sharper localization in double cones (the causal shadow of a 3-dim.
sphere). Since the (topological) core of arbitrarily small double cones is a
point and that of arbitrary narrow spacelike cones a semiinfinite spacelike
string, the remaining problem consisted in the actual construction of the
generating fields of these representation; this was achieved in \cite{MSY}.
The result can be described in terms of operator-valued distributions
$\Psi(x,e)$ which depend in addition to the start $x$ of the semiinfinite
string also on the the spacelike direction $e,~e^{2}=-1.$ They are covariant
under simultaneous transformations of $x$ and $e$ and fulfill Einstein
causality for spacelike separated strings
\begin{equation}
\left[  \Psi_{1}(x_{1},e_{1}),\Psi_{2}(x_{2},e_{2})\right]  _{gr}%
=0,~~x_{1}+\mathbb{R}_{+}e\ \rangle\langle\ x_{2}+\mathbb{R}_{+}e_{2}
\label{loc}%
\end{equation}
where $gr$ stands for graded (fermionic strings anticommute).

The modular localization of states uses the following construction. With a
wedge $W=\left(  x~|~x_{3}>\left\vert x_{0}\right\vert \right)  ~$there comes
a wedge-preserving one-parametric group of Lorentz-transformation $\Lambda
_{W}(\chi=-2\pi\tau)$ where $\chi~$is the hyperbolic boost parameter and
$\Theta_{W}$ denotes the $x_{0}$-$x_{3}~$reflection. The latter differs from
the total reflection $\Theta$ by a $\pi$-rotation $r_{W}$ around the $x_{3}%
~$axis (in the $x_{1}$-$x_{2}$ plane ) and therefore acts on the wave
functions as $J_{W}=U(r_{W})\Theta$. \ Both transformations $\Lambda_{W}$ and
$J_{W}$ commute. Since the generators of one-parametric strongly continuous
unitary groups are selfadjoint operators, there exists an "analytic
continuation" in terms of positive unbounded operators with dense domains
which decrease with the increase of distance from the real axis. This forces
the W-localized wave functions to have certain analyticity properties in the
momentum space rapidity $\theta,$ $\left(  p_{0},p_{3}\right)  =\sqrt
{m^{2}+p_{\bot}^{2}}(ch\theta,sh\theta)$ which relate the analytic
continuation of particle wave function to the complex conjugate of the
antiparticle wave function\footnote{If there exists an operator creating a
particle, the negative frequency part associated with the antiparrticle
annihilation must be related to the positive frequency part of the
antiparticle creation of its hermitian adjoint.}. Using the notation
$\Delta_{W}^{i\tau}\equiv U(\Lambda_{W}(-2\pi\tau)),$ the commutation with the
antiunitary $J_{W}$ leads to
\begin{align}
&  S_{W}=J_{W}\mathfrak{\Delta}_{W}^{\frac{1}{2}}=\mathfrak{\Delta}%
_{W}^{-\frac{1}{2}}J_{W},~S_{W}^{2}\subset1,~acts~on~H_{1}\oplus
H_{1}\label{mod}\\
&  S_{W}\psi=\bar{\psi},~K_{W}\equiv\{\varphi\in domS_{W};S_{W}\varphi
=+\varphi\},~S_{W}i\varphi=-i\varphi\nonumber\\
&  K_{W}~"is~standard":K_{W}\cap iK_{W}=0,~K_{W}+iK_{W}~dense~in~H_{1}\oplus
H_{1}\nonumber
\end{align}
where $\bar{\psi}$ denotes the localization-independent $S$-conjugate wave
function (the complex conjugate for the case at hand)\footnote{Although the
action of $S_{W}$ is diagonal, the definition of the $J_{W}$ needs the
antiparticle doubling of the Wigner space.$~$}$.$ The properties are
straightforward consequences of the commutation between the boost and the
associated reflection \cite{BGL}. The important point here is that $S$ relates
wave functions to their conjugates in a way which involves analytic
continuation where the analyticity came from spacetime $W$-localization.

The properties in (\ref{mod}) result simply from the commutativity of
$\Lambda_{W}(\chi)$ with the reflection $J$ on the edge of the wedge; since
$J~$is anti-unitary it commutes with the unitary boost, there will be a change
of sign in its action on the analytic continuation of $u.$ Hence it has all
the properties of a modular Tomita operator. The K-spaces $K(\mathcal{O})$ for
causally closed subspaces localized in $\mathcal{O}$ can be obtained by
intersections i.e. $\cap_{W\supset\mathcal{O}}K(W)$; this intersection may
however turn out to be trivial (see below) if the region is "too small".

The surprise resides in the fact that the transformation of wave functions to
their $S$-conjugate (\ref{mod}, second line) does not only encode the
information about two geometric objects: a one-parametric modular group
leaving a wedge invariant and a reflection on that wedge into its opposite,
but (and at this point the positive energy property of the Wigner
representation becomes relevant \cite{BGL}) it also contains the information
about the spacetime localization of the wave function. This is certainly
something which has no counterpart in QM; it points to an incomplete
understanding of the foundations of QFT which becomes fully revealed in the
relation between localized subalgebras and modular operator theory in the
presence of interactions.

The connection with causal localization is of course a property which only
appears in the physical context. The general setting of modular real subspaces
is a Hilbert space which contains a real subspace $K\subset H$ which is
standard in the sense of (\ref{mod}). The abstract S-operator is then defined
in terms of $K$ and $iK.$

The above application to the Wigner representation theory of positive energy
representations\footnote{The positive energy condition is absolutely crucial
for obtaining the prerequisites (\ref{mod}) of modular localization.} also
includes the \textit{infinite spin representations} which lead to semiinfinite
string-localized wave functions. In this case there are no pointlike covariant
wave function-valued distributions which generate these representations; they
are genuinely string-localized (which the superstring representation of the
Poincar\'{e} group is not; so beware of misleading terminology!). The
application of the above mentioned second quantized functor converts the
modular localized subspaces into a net of $\mathcal{O}$-indexed
interaction-free subalgebras $\mathcal{A(O}).$ Interacting field theories can
clearly not be obtained in this way. The relation between particles and fields
becomes much more subtle in the presence of interactions and this applies even
to models which have a complete particle interpretation i.e. in which the
particles related to fields via the LSZ large time behavior of fields (the LSZ
scattering formalism \cite{Haag}) lead to the identification of the Hilbert
space as a WignerFock particle space (section 7).

The algebraic setting in terms of modular localization also gives rise to a
physically extremely informative type of inclusion of two algebras which share
the vacuum state, the so-called \textit{modular inclusions} $(\mathcal{A}%
\subset\mathcal{B},\Omega_{vac}\mathcal{)}$ where modular means that the
modular group of the bigger $\Delta_{\mathcal{B}}^{it}$ compresses (or
extends) the smaller algebra \cite{K-W}. A modular inclusion automatically
forces the two algebras to be of the monad type. The above mentioned "GPS
construction of a QFT" from a finite number of monads positioned in a common
Hilbert space uses this concept in an essential way. It is perhaps the most
forceful illustration of the holistic nature of QFT.

There are two properties which always accompany modular localization and which
are interesting in their own right. Both are related to the statistical
mechanics nature of impure $\mathcal{A(O})$-restricted vacuum

\begin{itemize}
\item \textit{KMS property}. By ignoring the world outside $\mathcal{O}$ one
gains infinitely many KMS modified commutation properties with modular
Hamiltonians $\hat{K}$ associated to the $\widehat{\mathcal{O}}$ restricted
vacuum.%
\[
\left\langle AB\right\rangle =\left\langle Be^{-K}A\right\rangle
,~\ \Delta=e^{-K},~A,B\in\mathcal{A}\emph{(}\mathcal{O}%
),infinitly~many~\widehat{K}~~for~\widehat{\mathcal{O}}\supset\mathcal{O}%
\]

\end{itemize}

In contrast to the inert factorizing vacuum of QM in the Fock space ("2nd
quantization") description, the spatially restricted QFT vacuum fulfills
infinitely many KMS relations associated with modular Hamiltonians of larger
spacetime regions.

\begin{itemize}
\item Area law for localization-entropy, see (\ref{area})%
\[
Entr=f(\frac{area}{\varepsilon^{2}}),~\varepsilon=split\ size
\]

\end{itemize}

As mentioned in the previous section, one needs to invoke the so-called split
property in order to approximate the singular KMS state by a sequence of
density matrix states; this is similar to the construction of the
thermodynamic limit state in statistical mechanics. In contrast to the
approximation of the latter in terms of box-quantized finite volume Gibbs
states, the split formalism for open subsystems is a part of the (presently
computational rather inaccessible) modular localization theory. It is in
particular not clear whether the density matrix from the split property leads
to a plain dimensionless area law $f\simeq area/\varepsilon^{2}$
\footnote{This is suggested by the vacuum polarization clouds of smeared
fields in the limit of a aharp cut-off smearing function (see previous
section).} as in (\ref{area}) or to a logarithmically modified area law
\cite{BMS}. For chiral conformal theories on the lightray there is a rigorous
derivation of the localization entropy for an interval with vacuum attenuation
length $\varepsilon$ (surface fuzziness) from the well-known linear length
$l\rightarrow\infty$ behavior (the "one-dimensional volume factor" $l$). They
are related as $ln\varepsilon^{-1}\sim l\times kT.$ This \textit{inverse Unruh
effect} plays an important role in the full understanding of the E-J conundrum
and will be presented in the next section.

Great care needs to be taken in identifying the modular localization
"temperature" with that measured with a thermometer. This is because the
notion of thermometer temperature is based on the zeroth thermodynamic law
(the \textit{local temperature} in \cite{B-S}), whereas the KMS temperature
refers to the second law according to which it is impossible to gain energy
from equilibrium states by running a Carnot cycle (the absolute temperature).
In inertial systems those two definitions coalesce (after proper
normalization), whereas in a accelerated systems (used e.g. in the Unruh
Gedankenexperiment to achieve the Rindler-wedge localization) this is not the case.

A closer examination shows \cite{B-S} that the conclusion about "egg-boiling"
and particle radiation claimed to be observed by an accelerated observer are
incorrect, a fact which has been ignored in the literature on the Unruh
effect. The correct local temperature, different from the Carnot temperature,
does not depend on the acceleration and since it vanishes at spacelike
infinity, it vanishes everywhere. Although the black hole situation is
different, the application of Einstein's equivalence principle suggests
caution about the relation of a rescaled modular temperature with that
measured by a thermometer. This includes also the presently very popular ideas
about \textit{firewalls} which are allegedly created by restricting
generically locally normal states to a causal/event hotizon.

In order to facilitate the reader's accessibility to philosophical and
historical aspects, but also to maintain a lighthearted touch in dealing with
issues which by some are considered to be controversial, the following will be
presented in the form of Galileio's famous dialogs between Sagredo and
Simplicio. Since fundamental properties of nature are expected to be based on
simple physical principles, the role of the presenter of foundational
viewpoints in this dialog is Simplicio.

\textbf{Sagredo}: Dear friend Simplicio, I noticed that you have some critical
opinions about the topic of extra dimensions and dimensional reductions. Can
you explain your arguments against these extremely popular ideas?

\textbf{Simplicio}: Kaluza and Klein observed that in classical field theories
and quasiclassical approximations one may relate models in different spacetime
dimensions by appropriately reinterpreting the field content. In this way the
combined gravitation+electromagnetism may be obtained by dimensional reduction
from a five dimensional pure gravity theory. However the recent more
foundational understanding of the issue of causal localization in its precise
form of \textit{modular localization of quantum matter} reveals that the
localization aspects are a characteristic part of quantum matter and one
confronts grave problems if one tries to reduce spacetime dimensions. A first
indication comes from Wigner's theory of positive energy representations of
the Poincar\'{e} group which has a functorial relation to quantum matter in
the absence of interactions. The latter depend in an essential way on the
representation theory of Wigner's "little group" which changes with spacetime
dimension. The fact that dimensional regularization can be used as a technical
trick in renormalization theory and that in case of spinless matter Wilson's
dimensional $\varepsilon$-expansion led to reasonable approximate results for
critical indices should not be taken as an indication that causal quantum
matter can be "transplanted" by an imagined dimensional reduction.

\textbf{Sagedo}: But there \textit{are} rigorous relations between theories in
different spacetime dimensions, as the famous $AdS-CFT$ correspondence.

\textbf{Simplicio}: The $AdS_{n+1}-CFT_{n}~$correspondence is indeed a
mathematical isomorphism between the algebraic structure of QFT on two
different spacetimes which extends the prior known equality of their symmetry
groups; in fact it is the only known case in which two spacetime manifolds in
different dimensions share the same symmetry group. What prevents this
mathematical isomorphism from defining a physical correspondence is that it
does not preserve an important aspect of causality. Starting from a causal AdS
theory the corresponding CFT maintains the spacelike Einstein causality but
violates the causal completeness property. There are more degrees of freedom
in the algebra of the causal closure $\mathcal{A(O}^{\prime\prime})$ than
there are in $\mathcal{A(O}).$ For an observer living in such a world there
are degrees of freedom in $\mathcal{O}^{\prime\prime}$ which according to the
causal completeness property should have been already present in $\mathcal{O}%
$. A QFT in which new degrees of freedom come apparently from nowhere is
physically not acceptable. In the opposite direction, i.e. started from a
causal CFT, it was shown by Rehren \cite{Reh} that the resulting AdS theory
does not have enough degrees of freedom in order to support the existence of
nontrivial algebras of observables for compact localization regions; in such
situations nontrivial algebras only exist for noncompact spacetime
localization regions in the AdS spacetime.

The intuitive picture behind this violation of causal completeness is that the
cardinality of degrees of freedom of causal quantum matter depends on the
spacetime dimensionality and hence the concept of causal quantum matter cannot
be separated from spacetime. The algebraic stuff which the above isomorphism
generates from physical matter is not the expected causal quantum matter after
having applied the isomorphism. This shows that Maldacena's conjecture,
claiming that the isomorphism connects two physical theories, is not correct.
This failure of causal completeness is symptomatic for all attempts of
relating QFTs via dimensional reduction.

The AdS-CFT isomorphism shown that even under optimal mathematical conditions
such ideas run into serious problems with causal localization. It is
worthwhile to mention that there is only one relation between QFTs to which
the present critique does not apply; this is the holographic projection onto
null-surfaces \cite{BMS}. The important point here is that in an projection
(instead of an isomorphism) the cardinality of degrees of freedom is reduced
to that which is appropriate for the lower dimensional null-surface.

\textbf{Sagredo}: The classical Kaluza-Klein dimensional reduction idea
entered particle theory when it became clear that the high dimensional
solutions of string theory remain academic unless one finds a way to extract
properties which are relevant for the real world. The attempts to adjust the
dimensional reduction in classical field theory to the requirements of QFT led
to the idea to compactify a spacetime coordinate and "curl it up" to a tiny
circle so that the resulting QFT appears as one which lives on a reduced
spacetime. Therefore my question: is a such a dimensional curling up also flawed?

\textbf{Simplicio}: It is correct that for QFTs which permit a Euclidean
description one can formally compactify a coordinate. Physically this means
that one passes from the vacuum expectation values to expectation values in a
thermal state whose inverse temperature is proportional to the radius of the
circular compactified coordinate. What is however not correct is to relate
this thermal QFT with increasing temperature to a Klein-Kaluza reduction.
There is simply no classical analog of increasing thermal fluctuations;
passing to a thermal state with a high temperature has little to do with a
dimensional reduction a la Kaluza-Klein.

The continued uncritical use of the dimensional reduction idea is more of a
sociological problem; as long as the protagonists and leading defenders of
string theory accept dimensional reduction as a way which allows to obtain
properties of real particle theory from theories with extra dimensions, the
members of the string theory community will continue to use it with the result
that the understanding of local quantum physics will becomes increasingly metaphoric.

Of course particle physics at its foundational frontiers was always
speculative and errors are sometimes unavoidable, but the old "Streitkultur"
between equals at the time of Pauli, Landau, Feynman, Schwinger Jost,
K\"{a}llen and many others prevented a long term solidification of incorrect ideas.

\textbf{Sagredo}: Thank you my dear friend for your enlightening comments.//

\section{The E-J conundrum, Jordan's model}

With the \textit{locally restricted vacuum} representing a highly impure state
with respect to \textit{all} modular Hamiltonians $H_{mod}(\mathcal{\check{O}%
}),$ $\mathcal{\check{O}\supseteq O~}$on local observables $A\in
\mathcal{A(O)=A(O}^{\prime\prime}\mathcal{)},$ a fundamental conceptual
difference between QFT and QM has been identified. QM (type I$_{\infty}$
factors) is the conceptual home of \textit{quantum information theory}%
\footnote{Another subject which would have taken different turn with a better
appreciation of the problems in transfering notions of quantum information
theory to QFT is the decades lasting conflict about the problem of "black hole
information loss".}, whereas in case of localized subalgebras of QFT a direct
assignment of entropy and information content to a monad, if possible at all,
can only be done in a limiting sense. The present work shows that although QFT
started with this conceptual antagonism in the E-J conundrum, its foundational
understanding only began more than half a century later and is still far from
its closure.

For this reason it is more than a historical retrospection to re-analyze the
E-J conundrum from a contemporary viewpoint. In a modern setting Jordan's
two-dimensional photon\footnote{This terminology was quite common in the early
days of field quantization before it was understood that that in contrast to
QM the physical properties depend in an essential way on the spacetime
dimension. Jordan's two-dimensional photons and his later neutrinos (in his
"neutrino theory of light" \cite{Jor}) bear no relation to objects in the real
world.} model is a chiral current model. As a two-dimensional zero mass field
which solves the wave equation it can be decomposed into its two u,v lightray components%

\begin{align}
&  \partial_{\mu}\partial^{\mu}\Phi(t,x)=0,~\Phi
(t,x)=V(u)+V(v),~u=t+x,~v=t-x~\label{chiral}\\
&  j(u)=\partial_{u}V(u),~j(v)=\partial_{v}V(v),~\left\langle j(u),j(u^{\prime
})\right\rangle \sim\frac{1}{(u-u^{\prime}+i\varepsilon)^{2}}\nonumber\\
&  ~~\ T(u)=:j^{2}(u):,~T(v)=:j^{2}(v):,~~\left[  j(u),j(v)\right]
=0~~~~~~\nonumber
\end{align}

The scale dimension of the chiral current is $d(j)=1$, whereas the
energy-momentum tensor (the Wick-square of $j$) has $d(T)=2$; the u and v
worlds are completely independent and it suffices to consider the fluctuation
problem for one chiral component. The logarithmic infrared divergence problems
of zero dimensional chiral $d(V)=0$ fields arise from the fact that the zero
mass field $V$, different from what happens in higher dimensions\footnote{The
$V$ are semiinfinite integrals over the pointlike $j^{\prime}s,$ just as the
stringlike vectorpotentials in QED are semi-infinite integrals over pointlike
field strength \cite{charge}.}, are really stringlike instead of pointlike
localized. In fact the V is best pictured as a semiinfinite line integral (a
string) over the current \cite{Jor}; this underlines that the connection
between infrared behavior and string-localized quantum matter also holds for
chiral models on the lightray. It contrasts with QM where the infrared aspects
are not related to the infinite extension of quantum matter but rather with
the \textit{range of forces} between particles. Exponentials of
string-localized quantum fields involving integration over zero mass string
localized d=1+3 vectorpotentials share with the exponentials of integrals over
d=1+1 currents $expi\alpha V$ the property that their infrared behavior
requires a representation which is inequivalent to the vacuum representation
of the field strength or currents; the emergence of superselection rules
("Maxwell charges") is one of the more radical consequences of string-localization.

The E-J fluctuation problem can be formulated in terms of $j$ (charge
fluctuations) or $T$~(energy fluctuations). It is useful to recall that vacuum
expectations of chiral operators are invariant under the fractionally acting
3-parametric acting M\"{o}bius group (x stands for u,v)%
\begin{align}
U(a)j(x)U(a)^{\ast}  &  =j(x+a),~U(\lambda)j(x)U(\lambda)^{\ast}=\lambda
j(\lambda x)~\ ~dilation\\
U(\alpha)j(x)U(\alpha)^{\ast}  &  =\frac{1}{(-sin\pi\alpha+cos\pi\alpha)^{2}%
}j(\frac{cos\pi\alpha x+sin\alpha}{-sin\pi\alpha x+cos\pi\alpha}%
)~rotation\nonumber
\end{align}

The next step consists in identifying the KMS property of the locally
restricted vacuum with that of a global system in a thermodynamic limit state.
For evident reasons it is referred to as the \textit{inverse Unruh effect,}
i.e. finding a localization-caused thermal system which corresponds (after
adjusting parameters) to a heat bath thermal system. In the strong form of an
isomorphism this is only possible under special circumstances which are met in
the Einstein-Jordan conundrum, but not in the actual Unruh Gedankenexperiment
for which the localization region is the Rindler wedge.

\begin{theorem}
(\cite{E-J}) The global chiral operator algebra $\mathcal{A}(\mathbb{R})$
associated with the heat bath representation at temperature $\beta=2\pi$ is
isomorphic to the vacuum representation restricted to the half-line chiral
algebra such that
\begin{align}
(\mathcal{A}(\mathbb{R}),\Omega_{2\pi})  &  \cong(\mathcal{A}(\mathbb{R}%
_{+}),\Omega_{vac})\label{map}\\
(\mathcal{A}(\mathbb{R})^{\prime},\Omega_{2\pi})  &  \cong(\mathcal{A}%
(\mathbb{R}_{-}),\Omega_{vac})\nonumber
\end{align}
The isomorphism intertwines the translations of $\mathbb{R}$ with the
dilations of $\mathbb{R}_{+}$, such that the isomorphism extends to the local
algebras:
\begin{equation}
(\mathcal{A}((a,b)),\Omega_{2\pi})\cong(\mathcal{A}((e^{a},e^{b}%
)),\Omega_{vac}) \label{exp}%
\end{equation}

\end{theorem}

This can be shown by modular theory. The following proof extends prior work by
Borchers and Yngvason \cite{Bo-Yng}. $.$ Let $\mathcal{A}$ denote the
$C^{\ast}$ algebra associated to the chiral current $j$\footnote{One can
either obtain the bounded operator algebras from the spectral decomposition of
the smeared free fields $j(f)$ or from a Weyl algebra construction.}. Consider
a thermal state $\omega$ at the (for convenience) modular temperature $2\pi$
associated with the translation on the line. Let $\mathcal{M}~$be the operator
algebra obtained by the GNS representation and $\Omega_{2\pi}~$the state
vector associated to $\omega.$ We denote by $N$ the half-space algebra of
$\mathcal{M}$ and by $\mathcal{N}^{\prime}\mathcal{\cap M}~$the relative
commutant of $\mathcal{N}~in~\mathcal{M}.~$The main point is now that one can
show that the modular groups $\mathcal{M},~\mathcal{N}$ and $\mathcal{N}%
^{\prime}\mathcal{\cap M~g~}$generate a "hidden" positive energy
representation of the M\"{o}bius group $SL(2,R)/Z_{2}$ where hidden means that
the actions have no geometric interpretation on the thermal net. The positive
energy representation acts on a hidden vacuum representation for which the
thermal state is now the vacuum state $\Omega.$The relation of the previous 3
thermal algebras to their vacuum counterpart is as follows:
\begin{align}
&  \mathcal{N}=\mathcal{A}(1,\infty),~\mathcal{N}^{\prime}\mathcal{\cap
M=A}(0,1),~\mathcal{M}=\mathcal{A}(0,\infty)\\
&  \mathcal{M}^{\prime}\mathcal{=A}(-\infty,0),~\mathcal{A}(-\infty
,\infty)=\mathcal{M\vee M}^{\prime}\nonumber\\
&  \mathcal{M}(a,b)=\mathcal{A}(e^{2\pi a},e^{2\pi b})
\end{align}
Here $\mathcal{M}^{\prime}$ is the "thermal shadow world" which is hidden in
the standard Gibbs state formalism but makes its explicit appearance in the so
called \textit{thermo-field} setting i.e. the result of the GNS description in
which Gibbs states described by density matrices or the KMS stated resulting
from their thermodynamic Llimits are described in a vector formalism. The last
line expresses that the interval algebras are exponentially related.

In the theorem we used the more explicit notation
\[
\mathcal{M}(a,b)=(\mathcal{A}(a,b),\Omega_{th})=(\mathcal{A}(e^{2\pi
a},e^{2\pi b}),\Omega_{vac})
\]

Moreover we see, that there is a natural space-time structure also on the
shadow world i.e. on the thermal commutant to the quasilocal algebra on which
this hidden symmetry naturally acts. Expressing this observation a more
vernacular way: the thermal shadow world is converted into virgin living
space\footnote{In \cite{Haag} it is shown how to extract the shadow world
description from the density matrix (Gibbs states) formalism with the help of
the canonical GNS construction.}. In conclusion, we have encountered a rich
hidden symmetry lying behind the tip of an iceberg, of which the tip was first
seen by Borchers and Yngvason.

Although we have assumed the temperature to have the Hawking value $\beta
=2\pi,$ the reader convinces himself that the derivation may easily be
generalized to arbitrary positive $\beta$ as in the cited Borchers-Yngvason
work. A more detailed exposition of these arguments is contained in a paper
\textit{Looking beyond the Thermal Horizon: Hidden Symmetries in Chiral Models
}\cite{S-W}.

In this way the semi-infinite interval ($-\infty,-L)$ of a thermal system in a
one dimensional interval $(-L,L)$ of length $2L$ (one-dimensional "box")
passes to the split interval $\varepsilon$ (the size of the Heisenberg's
vacuum polarization cloud) $\varepsilon\sim e^{-2\pi L}.$ As a result the
thermodynamic $L\rightarrow\infty$ corresponds to the limit of sharp
localization $(e^{-2\pi L},~e^{2\pi L})\overset{L\rightarrow\infty
}{\rightarrow}(0,\infty)$ on the vacuum side. From this one draws the
conclusion that the thermal heat bath entropy for large $L~$passes to the
localization-entropy in the vacuum state for small split distance
$\varepsilon$%
\begin{equation}
En_{kT=2\pi}\simeq L=-\frac{1}{2\pi}\ln\varepsilon\simeq En_{loc}%
\end{equation}

Though it is unlikely that a localization-caused thermal system is isomorphic
to a heat bath thermal situation in higher dimensions, there may exist a
"weak" inverse Unruh situation in which the volume factor corresponds to a
logarithmically modified dimensionless area law i.e. $(\frac{R}{\Delta
R})^{n-2}ln(\frac{R}{\Delta R})~$where $R$ is the radius of a double cone and
$\frac{R}{\Delta R}=\varepsilon$ its dimensionless surface roughness; the
volume in this case is that of a box with two transverse- and one lightlike-
directions is the counterpart of the spatial box so that the volume factor $V$
corresponds to a box where one direction, the one responsible for the
logarithmic factor, is lightlike. But the analogy with the area
proportionality of vacuum fluctuations in Heisenberg's partial charges
$Q(R,\Delta R)$ favours the area law which also agrees with the result from 't
Hooft's proposal of a brickwall picture \cite{brick}.

Although the thermal aspect of a restricted vacuum in QFT is a structural
consequence of causal localization, the general identification of the
dimensionless modular temperature with an actual temperature of a heat bath
system, or, which is equivalent, the modular "time" with the physical time is
not correct; the modular Hamiltonian does not describe the inertial time for
which the local temperature defined in terms of the zeroth thermodynamic law
agrees with the "Carnot temperature" of the second law \cite{B-S}.

Properties of states in QFT depend on the nature of the algebra: a monad does
not have pure states nor density matrices, but only admits rather singular
impure states as singular (non Gibbs) KMS states. The identification of states
with vectors in a Hilbert space up to phase factors becomes highly ambiguous
and physically impractical outside of QM. The state in form of a linear
expectation functional on an algebra and the unique vector (always modulo a
phase factor) obtained by the intrinsic GNS construction \cite{Haag} leads
always to a vector representation, but this depends on the particular state
used for the GNS construction. In QM the algebras are always of the $B(H)$
type where this distinction between vector states and state vectors is not necessary.

\section{Particle crossing, on-shell constructions from modular setting}

An important new insight into "particles \& fields" comes from a derivation of
the \textit{crossing property} of particle physics from the modular properties
of wedge-localization. The formfactor crossing states that the
n-particle-to-vacuum matrixelement of a local operator $B$ is analytically
related to to the \textit{connected part} of the formfactors of $B~$between
$k~$incoming and $n$-$k$ outgoing particles in terms of the following
identity
\begin{align}
&  \left\langle 0\left\vert B\right\vert p_{1},..p_{n}\right\rangle
^{in}=~^{out}\left\langle -\bar{p}_{k+1}..,-\bar{p}_{n}\left\vert B\right\vert
p_{1},..p_{k}\right\rangle _{con}^{in}\label{cro}\\
&  B\in\mathcal{A(O}),~\mathcal{O}\subseteq W,~\bar{p}=antiparticle~of\text{
}p\nonumber
\end{align}
Here the momenta $-\bar{p}~$on the backward mass-shell refer to the
anti-particles of the $n$-$k$ crossed particles of the original n-particle
state where the transition to the negative momenta involves an analytic
continuation within the complex mass-shell. The analyticity following from
principle of modular wedge-localization is however not in the $s,t,u$
Mandelstam invariants associated to the momenta, but rather in the rapidity
$\theta$ variables. It turns out that the better-known crossing properties of
the S-matrix do not have to be considered separately, they can be related to
those of formfactors by the use of the LSZ reduction formalism. The nontrivial
aspect is the posssibility to relate a scattering amplitude to that of its
crossed form by an analytic continiation which remains on the complex mass shell.

The physical content of formfactor crossing is that the different $k$ to
$n$-$k$ formfactors are analyticlly related to one master formfactor which may
be taken to be the n-particle to vacuum formfactor. The only known
non-perturbative general derivation of formfactor crossing uses modular
theory\footnote{For a special case (elastic scattering) Bros, Epstein and
Glaser \cite{BEG} derived crossing of the S-matrix within the rather involved
setting of functions of several analytic variables.}, to be more precise the
modular theory of a wedge-localized subalgebra. Before a sketch of its
derivation will be given, some remarks about its conceptual relation to other
consequences of modular localization of wedge regions may be helpful. The
conceptual proximity of the particle crossing propertx to the Unruh
\cite{Unruh} effect through the shared wedge localization is somewhat
unexpected. Whereas the latter together with the Einstein-Jordan subvolume
fluctuations will probably remain a "Gedankenexperiment" which illustrates
consequences of vacuum entanglement, the particle crossing has observational
consequences \cite{Martin} and constitutes an important concept of high energy
particle physics. This changes the conceptual setting of crossing from that
attributed to it in the dual model and ST, a topic which will be taken up in
section 7.

The modern conceptual understanding of crossing came from the recognition that
in models of QFT with a mass gap and a complete particle interpretation the
S-matrix is more than a global operator of scattering theory, it also is
possesses a less conspicuous local property namely it is a \textit{relative
modular invariant} which intertwines between the interacting wedge algebra
$\mathcal{A}(W)$ and its interaction-free incoming counterpart constructed
from the incoming free fields $\mathcal{A}(W)_{in}$. Namely the two modular
reflections are related through \cite{AOP}\cite{integrable}%
\begin{equation}
J_{W}=J_{W,in}S_{scat},~or~S_{W}=S_{W,in}S_{scat}\text{ }using~S=J\Delta^{1/2}%
\end{equation}
a relation which can be traced back to Jost's proof \cite{Jost} of the TCP
theorem and the fact that $J_{W}$ is only different from Jost's TCP by a $\pi
$-rotation within the edge of the wedge (which commutes with the Poincar\'{e}
invariant $S_{scat}$).

Another idea from modular wedge-localization which is used in the derivation
of formfactor crossing is \textit{emulation} of interacting wedge-localized
states (state vectors obtained by applying interacting smeared fields
$B(f)~$with $suppf\subset W$ to the vacuum $\Omega$) in terms of
interaction-free wedge-localized states obtained by applying operators
$A_{in}(f)$ to the vacuum \cite{integrable} \cite{cau}. Emulation involves
different algebras acting in the same Hilbert space and sharing the same
$\mathcal{P}$-goup representation.

To get some technicalities out of the way, let us first formulate the
\textit{free field KMS} relation in the way we need it for later purpose. With
$B$ a W-smeared composite of a free field, the modular KMS relation for
wedge-localized free fields reads
\begin{align}
&  \left\langle BA^{(1)}A^{(2)}\right\rangle =\left\langle A^{(2)}\Delta
BA^{(1)}\right\rangle ,~\Delta^{it}=U(\Lambda(-2\pi t))\label{fr}\\
&  A^{(1)}=:A(f_{1})..A(f_{k}):,~~A^{(2)}=:A(f_{k+1})..A(f_{n}):\nonumber\\
&  \Delta A^{(2)}{}^{\ast}\left\vert 0\right\rangle =\Delta SA^{(2)}\left\vert
0\right\rangle =\Delta^{1/2}JA^{(2)}\left\vert 0\right\rangle \nonumber
\end{align}
A smeared free field can be written in terms of creation/annihilation
operators integrated with wavefunctions which are the mass-shell restriction
of the Fourier transforms of W-supported test functions (for economy of
notation $f~$will also be used for the Fourier transform)%
\begin{align}
&  A(f)=\int(f(p)a^{\ast}(p)+\bar{f}_{a}(p)b(p))\frac{d^{3}p}{2p_{0}},~p\in
H_{m}\\
&  A(f)^{\ast}=\int(f_{a}(p)b^{\ast}(p)+\bar{f}(p)a(p))\frac{d^{3}p}{2p_{0}%
}\nonumber
\end{align}
where $f_{a}$ is the wavefunction of the antiparticle. We take the wedge
$W~$in the $0$-$3$~directions, so that it is left invariant by $\Lambda
_{0\text{-}3}$ Lorentz boosts, and parametrize the mass-shell momenta in terms
of W-affiliated rapidities. It is well-known that the Fourier transforms of
$W$-supported testfunctions lead to wavefunctions $f(p)$ which are boundary
values of functions $f(p(z)),$ holomorphic in the rapidity strip in such a way
that the analytic continuation of the particle wave function to the other side
of the strip is equal to the complex conjugate of the antiparticle
wavefunction.
\begin{align}
p(z)  &  =(mshz,mchz;p_{\perp}),~~0<\operatorname{Im}z<\pi\\
&  f(p(\theta+i\pi))=\bar{f}_{a}(p(\theta))\nonumber
\end{align}

Rewriting the KMS relation (\ref{fr}) in terms of particle states we obtain
\begin{align}
&  \int..\int\left\langle 0\left\vert B\right\vert p_{1},..p_{n}\right\rangle
f_{1}(p_{1})..f_{n}(p_{n})\frac{d^{3}p_{1}}{2p_{0,1}}..\frac{d^{3}p_{n}%
}{2p_{0,n}}+contr.=\\
&  \int..\int(\Delta^{1/2}J\left\vert \bar{p}_{k+1}..\bar{p}_{n}\right\rangle
,B\left\vert p_{1},..p_{k}\right\rangle )f(p_{1})..f(p_{n})\frac{d^{3}p_{1}%
}{2p_{0,1}}..\frac{d^{3}p_{n}}{2p_{0,n}}+contr.\nonumber
\end{align}
where the round bracket in the second line denotes the scalar product between
the bra and ket vectors and $contr.$ stands for the contraction terms between
two Wick-products. They contain a lower number of particles and hence do not
contribute to the n-particle terms. The third line in (\ref{fr}) was used
inside the inner product in order to rewrite the right hand side of the KMS
relation as a matrix element of $B$ between particle states.

To pass to the crossing relation (\ref{cro}) we must show that one can omit
the integration with the dense set of strip-analytic wavefunction. Since
formfactors in interacting models are generally distributions, this is not
possible without knowing that the formfactors are locally square integrable;
in this case the relation on a dense set of wave functions implies its
validity on all locally L$^{2}~$integrable$~$functions and hence (\ref{cro})
follows. Here $B$ is any composite of a free field.

In the presence of interactions the extraction of the particle crossing from
the KMS relation is more demanding. Particles are related to
(incoming/outgoing) free fields, whereas the fields in the KMS relation are
interacting. The crossing relation (\ref{cro}) which we want to derive
contains in and outgoing particles which are associated with in/out free
fields. We need to know a relation between incoming and interacting wedge
localized states. Using the notation: $\mathcal{A}(W),~\mathcal{A}_{in}%
(W)~$for the interacting and incoming free field wedge-local algebra and
recalling that both algebras share the same representation of the Poincar\'{e}
group, one obtains from the equality of the W-preserving Lorentz boosts the
equality of the domains of their Tomita operators $domS_{\mathcal{A}%
(W)}=domS_{\mathcal{A}_{in}(W)}.$ This means that for a vector state created
by applying a wedge-local operator from $\mathcal{A}_{in}(W)$ to the vacuum
there will be a corresponding uniquely defined operator in $\mathcal{A}(W)$
operator which, applied to the vacuum creates the same vector. Existence and
uniqueness is secured by modular theory applied to the wedge region
\cite{BBS}. We refer to this bijection between wedge local operators as
\textit{emulation of wedge localized free fields within the interacting wedge
algebra \cite{cau}\cite{integrable} }and denote the emulated image by a
subscript $\mathcal{A}(W)$
\begin{equation}
\left\vert f_{1},..,f_{k}\right\rangle =A_{in}(f_{1})...A_{in}(f_{k}%
):\left\vert 0\right\rangle =(:A_{in}(f_{1})...A_{in}(f_{k}):)_{\mathcal{A}%
(W)}\left\vert 0\right\rangle ,~suppf\subset W
\end{equation}
where, as before, the $f$\ inside the bracket state vectors are the wave
functions associated with the W-supported testfunctions.

The KMS relation for interacting fields, from which the particle crossing is
to be derived, reads now \cite{S1}%
\begin{align}
&  \left\langle B(A_{in}^{(1)})_{\mathcal{A}(W)}(A_{in}^{(2)})_{\mathcal{A}%
(W)}\right\rangle =\left\langle (A_{in}^{2})_{\mathcal{A}(W)}\Delta
B(A_{in}^{1})_{\mathcal{A}(W)}\right\rangle \label{rel}\\
&  \Delta(A_{in}^{(2)})_{\mathcal{A}(W)}^{\ast}\left\vert 0\right\rangle
=\Delta^{\frac{1}{2}}JA_{out}^{(2)}\left\vert 0\right\rangle ,~J=S_{scat}%
J_{in}\nonumber
\end{align}
The identification of the right hand side with a (analytically continued)
particle formfactors is similar to the free case; the difference is the
presence of the scattering matrix which converts an incoming bra-state into an
outgoing state
\begin{equation}
\left\langle BA_{in}^{(1)}(p_{1},..p_{k})_{\mathcal{A(}W)}|p_{k+1}%
,..p_{n}\right\rangle ^{in}\simeq~^{out}\left\langle -\bar{p}_{k+1},..-\bar
{p}_{n}\left\vert B\right\vert p_{1},..p_{k}\right\rangle ^{in} \label{int}%
\end{equation}
The equivalence sign expresses the fact that the equality according to
(\ref{rel}) only holds after integration with wavefunctions from a dense set
of $W$-localized wave functions, and the $\Psi$ stands for a state obtained by
applying an emulated $k$-particle operator to an $n$-$k$ incoming state. It
depends on $n$ on-shell particle momenta but is not an incoming n-particle
state (+ contributions from contractions)\footnote{A outgoing free creation
operator applied on a n-1 incoming state is not an n-particle state. Similarly
the action of emulated incoming fields on an incoming state is an infinite
superposition of incoming particle states even though the emulated momenta are
on-shell.}; the product of emulations of free field states is not the
emulation of the product of the latter. In order to relate the action of an
"$k$ emulat" on a $n$-$k$ particle state one needs an additional idea.

There exists a concept which achieves this: \textit{the analytic on-shell
order change.} It arose in the setting of integrable models \cite{Kar} and
consists in an analytic interchange of particle momenta within formfactors
which, in the presence of interactions, is different from the kinematical
interchange in terms of statistics. For simplicity of notation we restrict to
d=1+1 in which case on-shell formfactors are fully described by rapidities
$\theta.~$We define a new object (denoted by a superscript $an$) in a special
configuration%
\begin{equation}
\left\langle B|\theta_{1}..\theta_{n}\right\rangle ^{an}\equiv\left\langle
B|\theta_{1},..,\theta_{n}\right\rangle _{in}~for\text{ }\theta_{1}%
>....>\theta_{n}%
\end{equation}
Using (bosonic) particle statistics, formfactors can always be written in this
naturally ordered form. An analytic ordering change along a certain path leads
from the natural order to a different formfactor function which depends not
only on the new order but also on the path of the analytic continuation which
was used to achieve it. The resulting object is still on-shell, but one
generally does not know its interpretations (or representation) in terms of
particle states.

Fortunately for the derivation of the momentum space crossing one does not
have to know the particle content after the analytic changes. If the
formfactors are locally square integrable one can, by using wave functions
with ordered $\theta$-supports, always "filter out" the natural order. This is
achieved by passing from wedge-local wave functions, which are spread
(\ref{rel}) over $\theta$-line, to wave functions supported in naturally
ordered $\theta$-intervals. In other words the on-shell analytic ordering
property permits to reduce the derivation of the crossing property in the
presence of interactions to that of the interaction-free case; the presence of
interactions would only show up in the unknown contributions from different
orders. Before we attempt to algebraize the analytic ordering idea it is
helpful to take a look at the simpler case of integrable models.

Integrable models permit an explicit illustration of the previous arguments,
including an operator-encoding of analytic ordering changes into a
representation of the permutation group (with the analytic transpositions
being defined in terms of the 2-particle elastic scattering matrix). In fact
the emulated free fields\footnote{In earlier publications the special case of
an emulated incoming field was referred to as a
vacuum-polarization-free-generators (PFG) \cite{BBS}.} turn out to be
identical to the Fourier transforms of the Zamolodchikov operators which obey
the Zamolodchikov-Faddeev algebra (see \ref{emul} below).

This simplicity has its mathematical origin in restrictive domain properties
of emulats which characterize integrability \cite{BBS}. Emulats in general QFT
only inherit the invariance property of their domains under the
wedge-preserving subgroup. The requirement that the domain is also invariant
under translations turns out to be extremely restrictive \cite{BBS}. In
$d>1+1~$the definition of integrability in terms of domain properties of PFG's
forces the S-matrix to be trivial $S_{scat}=1,$ whereas in $d=1+1$ it allows
nontrivial S-matrices which are suitable combinatorial products of elastic
2-particle S-matrices which fulfill the bootstrap properties (matrix-valued
scattering functions)\footnote{In d=1+1 the cluster factorization does not
distinguish a nontrivial elestic scattering amplitude from $S_{scat}=1.$}. In
other words the \textit{connected} higher particle scattering contributions
vanish, which is the standard definition of integrability in terms of
properties of the S-matrix (the infinite number of conservation laws is a
consequence). The elastic S-matrices are given in terms of (possibly
matrix-valued) scattering functions which have to obey certain analytic
properties in order to come from a field theory; these scattering functions
permit a classification.

Using these scattering functions as structure functions in a
Zamolodchikov-Faddeev algebra \cite{Zam} one obtains the creation/annihilation
components of wedge-localized temperate PFGs. At this point one realizes that
the above abstract definition in terms of domain properties of PFGs coalesces
with the standard definition of d=1+1 integrability. Such models are
susceptible to solutions in closed form and are therefore called "integrable".
Compared with the classical integrability which requires to find a complete
set of "conservation laws in involution" (and where integrable systems exist
in every dimension), integrability in QFT is limited to d=1+1.

The so-called "bootstrap-formfactor construction program" relates the
scattering functions to explicitly computed formfactors \cite{Kar}. The last
step consists in showing that these formfactors really belong to an existing
model of LQP. In order to achieve this on has to establish the nontriviality
of double cone localized intersections of wedge-local algebras. This is a very
nontrivial step which has been accomplished in terms of the use of modular
nuclearity in the work of Lechner \cite{Lech}. The same author also showed how
(in the absence of bound-states ) one can construct the wedge-algebra
generating PFG's in terms of deformations of free fields \cite{Le}. The
existence proof for some integrable models is considered to represent a
progress compared to the old existence proofs which were limited to
unrealistic short distance restrictions in the form of superrenormalizability
\cite{Gl-Ja}.

This simplicity of integrable S- matrices (the absence of connected parts for
$n>2$) keep integrable models in the proximity of interaction-free models.
Hence it is not so surprising that their wedge-generators (the
Zamolodchikov-Faddeev algebra generators) can be obtained by
\textit{deformations} of free fields instead of the more complicated emulation
\cite{Le}.

For the convenience of the reader and for later use we add some details on the
algebraic structure of emulated free fields for integrable models.%

\begin{align}
&  \left(  A_{in}(f)\right)  _{\mathcal{A}(W)}=\int_{C}f(\theta)Z^{\ast
}(\theta)d\theta,~C=\partial strip,~p=m(ch\theta,sh\theta)\label{emul}\\
&  strip=\left\{  z~|~0<Imz<\pi\right\}  ,~\ Z(\theta)\equiv Z^{\ast}%
(\theta+\acute{\imath}\pi)\nonumber\\
&  Z^{\ast}(z_{1})Z^{\ast}(z_{2})=S(z_{1}-z_{2})Z^{\ast}(z_{2})Z^{\ast}%
(z_{1}),~z\in C\nonumber
\end{align}

Since integrable models preserve the particle number in scattering processes,
the n-fold application of the creation parts $Z^{\ast}(\theta)$ to the vacuum
are n-particle states. Identifying the velocity-ordered particle state with
the incoming states
\begin{align}
&  Z^{\ast}(\theta_{1})Z^{\ast}(\theta_{2})..Z^{\ast}(\theta_{n})\left\vert
0\right\rangle =\left\vert \theta_{1},\theta_{2},.,\theta_{n}\right\rangle
_{in},~\theta_{1}>\theta_{2}>..>\theta_{n}\label{ana}\\
&  anal.~transpos.~\left\langle 0\left\vert B\right\vert \theta_{2},\theta
_{1},..\theta_{n}\right\rangle _{in}=S(\theta_{1}-\theta_{2})\left\langle
0\left\vert B\right\vert \theta_{1},\theta_{2},..\theta_{n}\right\rangle
_{in}\nonumber
\end{align}
the old degenerate representation related to (bosonic) statistics has been
encoded into the natural order while the other orders describe analytic
changes inside formfactors. For integrable models the transposition of two
adjacent $\theta$ uses the two-particle S-matrix.

It follows from repeated application of (\ref{ana}) that the analytic change
of a $\theta$ through a k-cluster of $\theta$ on its right hand side will be a
\textit{product} of of scattering functions which rewritten \textit{in terms
of the full k+1 S-matrix} corresponds to a \textit{grazing shot}
\textit{S-matrix~}defined as \cite{on-shell}
\begin{equation}
S_{g.s.}(\theta;\theta_{1},..\theta_{k})=S^{k}(\theta_{1},..\theta_{k}%
)^{-1}S^{k+1}(\theta,\theta_{1},..\theta_{k}) \label{g.r.}%
\end{equation}
This grazing shot concept has been used to generalize the properties of
integrable emulations to the generic situation \cite{cau}\cite{integrable} by
converting the idea of analytic changes of ordering into an algebraic
structure; in this sense this is an attempt to generalize the structure of the
Zamolodchikov-Faddeev algebra.

The first attempt of an on-shell construction of particle theory after the
failure of the S-matrix bootstrap was that by Mandelstam. It ignored the
subtlety of analytic on-shell properties by trying to guess their structure
based on a postulated double spectral representation for elastic scattering
amplitudes (the Mandelstam representation) instead of a derivation from the
foundational causal locality principles of QFT. It finally came to an end when
Mandelstam supported the incorrect idea of identifying the meromorphic
function of Veneziano's dual model with the particle crossing of scattering
amplitudes (more in section 7).

The idea of the present work is suggested from properties of the modular wedge
localization and consists in relating on-shell analytic order changes to the
action of emulats. For two relatively naturally ordered clusters, the analytic
ordering idea for the left hand side in (\ref{int}) reads%

\begin{align}
&  \left\langle BA_{in}^{(1)}(\theta_{1},..\theta_{k})_{\mathcal{A(}W)}%
|\theta_{k+1},..\theta_{n}\right\rangle ^{in}~=\left\langle B\theta_{1}%
,\theta_{2},..\theta_{n}\right\rangle ^{in}+contr.\label{form}\\
&  ~~~~~~~\ \ ~\ \ \ \ ~~for~(\theta_{1},..\theta_{k})>(\theta_{k+1}%
,..\theta_{n})\nonumber
\end{align}
i.e all$~\theta~$in$~$left$~$hand$~$cluster$~$are$~$larger$~$than$~$the those
on the right hand side. The contractions result from the incoming Wick product
$A_{in}^{(1)}(\theta_{1},..\theta_{k})$ acting on the n-k particle state; they
are delta function contact terms in rapidity space and hence do not contribute
if all $\theta$ are different. Fortunately other orders are not needed for the
crossing relation, but they contain the dynamic information and enter which is
important for the full understanding of crossing and enter any constructive
approach which tries to generalize what has been learned from integrable models.

That the ordering prescription is crucial for the derivation of the standard
form of the LSZ property is corroborated by the derivation of the
time-dependent LSZ reduction formula from the foundational properties of QFT
\cite{Bu-Su}. In that derivation overlapping wave functions have to be avoided
because through such overlaps threshold singularities enter into the problem.
This result supports the picture of analytic changes of moving through new
threshold singularities at points of coalescence of two $\theta.~$It is an
indication that ordering changes of two $\theta~$lead to nontrivial changes
which affect the derivation of the LSZ formula and the crossing relation. The
present arguments suggest that both these changes should have their
explanation in a better understanding of the consequences of modular
localization for wedge-local algebras in QFT. From a modern viewpoint it is
clear that the conceptual tools for its solution were simply not available at
the time of its proposal.

The ideas about PFGs and of wedge-localized particle states in terms of
emulated fields can (and in my opinion should) be viewed as an extension of
Wigner's representation-theoretical approach for noninteracting particles and
its functorial relation ("second" quantization)~with quantum fields but now in
the presence of interactions. The conceptual distance between the functorial
particle-free field relation and emulation in the presence of interactions is
immense. Modular localization, as a mathematical precise formulation of the
causal locality principle of LQP is the only intrinsic property which has the
necessary conceptual pugnancy to eventually solve this field-particle problem
in the presence of interactions.

\section{Impact of modular localization on gauge theories}

It is well-known that the \textit{Hilbert space formulation} for
renormalizable couplings of \textit{pointlike} fields is limited to spin
$s<1$. For $s=1$ vectorpotentials, one is forced to use a Krein space
formulation, either in the form of the Gupta-Bleuler formalism, or for massive
gauge theories in terms of the ghost field formalism of the
Becchi-Rouet-Stora-Tyutin (BRST) operator gauge setting. Usually textbooks on
QED do not explain that this deviation from the Hilbert space setting of
quantum theory comes from an incompatibility of pointlike zero mass
vectorpotentials with the positivity of Hilbert space (closely related to
quantum probability) which leads to limitations of viewing models of QFT as
obtained by quantizing classical gauge theories. In fact this problem arises
for all massless $s\geq1$ tensor potentials; only their associated pointlike
field strengths are fields acting in a Hilbert space. This problem of loss of
the Hilbert space setting for interactions which use pointlike
vectorpotentials is the origin of gauge theory.

For our aim to present a new formulation which permits to describe the full
theory in Hilbert space, it is helpful to recall first some facts about the
BRST gauge setting. We will use the so-called BRST operator formalism as it
can be found in Scharf's book "Gauge theory, a true ghost story"
\cite{Scharf}, but present it in a form which highlights analogies between the
nilpotent cohomological BRST $s$-formalism in Krein space and the
$d$-operation on differential forms of the new Hilbert space setting which
requires the use of stringlocal field (the SLF formalism).

The BRST description of massive vectormesons relates the physical Proca field
$A_{\mu}^{P}~$with $\partial^{\mu}A_{\mu}^{P}=0$ with short distance dimension
$d_{P}=2~$to an unphysical field in Krein space $A_{\mu}^{K}$ with $d_{K}=1$
and a negative metric scalar St\"{u}ckelberg field $\phi^{K}$~with
$d_{sd}=1.\ $The idea is to compensate the highest short distance singularity
in terms of the $d=2~$derivative $\partial_{\mu}\phi^{K};~$for this
compensation one needs the opposite sign of the $\phi^{K}$ two-point-function;
this just uses the well-known (already before gauge theory) short-distance
softening effect of indefinite metric which is of cause inconsistent with
quantum physics but helpful for renormalization. The idea of gauge-invariance
is to formulate a restriction which permits at least to return to a "smaller"
physical setting. The relation which intuitively achieves this is of the form%

\begin{equation}
A_{\mu}^{K}(x)\simeq A_{\mu}^{P}(x)+\partial_{\mu}\phi^{K},~\curvearrowright
~\partial^{\mu}A_{\mu}^{K}(x)+m^{2}\phi^{K}\simeq0 \label{pro1}%
\end{equation}
The equivalence sign is meant to indicate that relation between the Krein
space vectorpotential and its physical Proca counterpart is not yet an
operator equality but rather a relation which requires a cohomological
interpretation. The reader recognizes in the second relation the Lorentz
condition; by relating physical states with suitably defined equivalence
classes of Krein states these relations represent cohomological equivalences.
In the following we restrict the formalism to massive vectormesons; in this
case the space is a Fock-Krein particle space and the BRST formalism can be
formulated in terms of indefinite metric free fields.

The BRST formalism extends the Krein space setting by additional indefinite
metric free fields: the ghost and anti-ghost fields $u,\tilde{u}~$fields; this
permits the reformulation of the content of (\ref{pro1}) as operator
equalities in terms of a nilpotent ghost charge $Q$ which in turn allows the
definition of a nilpotent $s$-operation%

\begin{align}
sA_{\mu}^{K}  &  =\partial_{\mu}u,~s\phi^{K}=u,~s\tilde{u}=-(\partial
A^{K}+m^{2}\phi^{K})\label{free}\\
sB^{K}  &  =[Q,B^{K}]_{\operatorname{grad}},~Q\text{ }ghost~charge,~s^{2}%
=0\nonumber
\end{align}
where the graded commutator is an anti-commutator if $B$ contains an odd
number of ghost fields $u,\tilde{u}.$ The first line leads to $s(A_{\mu}%
^{K}-\partial_{\mu}\phi^{K})=0~$which according to (\ref{pro1})~is consistent
with the physical nature of the Proca field. As shown in \cite{Scharf}
\cite{Aste} \cite{Garcia} this leads to an operator formulation of
\textit{renormalizable gauge theories for massive}\footnote{The free field
transformation rules (\ref{free}) refer to the incoming free fields of
scattering theory. In massless gauge theories as QED the ghost charges depend
on the coupling \cite{D-F}.}\textit{ vectormesons} coupled to charge-carrying-
or neutral matter fields. The S-matrix in such a setting is characterized by
$sS=0;$ for details we refer to the cited papers.

The action of the $s~$via the graded commutator with the ghost charge defines
the quantum gauge symmetry transformation so that gauge-invariant operators
are annihilated by the action of $s$. As mentioned the limitation of the
operator gauge formalism shows up in the attempt to construct
\textit{physical} matter fields\footnote{The gauge.variant matter fields have
no physical content and it is also not possible to extract physical matter
fields in a perturbative setting.} which couple to vectormesons. The
well-known nonrenormalizability of pointlike massive vectormeson interactions
in a \textit{Hilbert space }indicates that pointlike physical fields are more
singular than Wightman fields (operator-valued tempered distributions). The
literature on the BRST formalism contains no informations about their
construction. An illustration can be found in \cite{FMS} where its was shown
that the use of unphysical matter fields leads to the wrong result that the
Maxwell charge (associated to the identically conserved Maxwell current
$j_{\mu}=\partial^{\nu}F_{\mu\nu})~$of the gauge-variant matter field vanishes
which contradicts rigorous results about states created by physical matter
fields acting on the vacuum. Buchholz \cite{B80} has used an appropriated
formulation of the quantum Gauss law in order to prove that physical
charge-carrying operators cannot be better localized than in arbitrarily
narrow spacelike cones whose cores are semiinfinite spacelike strings.

Some more comments on the BRST operator gauge formalism and its relation to
classical gauge theory may be helpful. The terminology gauge "principle" is
sometimes misunderstood as a special \textit{physical} property of $s=1$
fields. Its role is however of a pure \textit{technical} kind; working with a
formulation in a Krein space, one needs to extract from such an unphysical
description physical data referring to objects which act in a Hilbert space;
in the past there has been simply no renormalizable formalism in terms of
pointlike fields in a Hilbert space setting.

The BRST gauge formalism in Krein space achieves its limited validity in the
vacuum sector (generated by the local gauge-invariant fields) by constructing
a "symmetry" which involves in addition to the Krein space counterparts of
matter fields also "ghost" and anti-gost operators (\ref{free}). This formal
symmetry (sometimes referred to as a local gauge symmetry) \textit{is by
itself not a physical symmetry} in the usual sense; even though its formal
invariants are the physical \textit{local observables }whose application to
the vacuum state generate the Hilbert space of the vacuum sector. Important
physical fields, as those which transfer electric charge, remain outside the
quantum gauge formalism. Neither does one know a physically useful
generalization of gauge symmetry to higher spin. Indefinite metric spaces
entered QFT through quantization of QED (the Gupta-Bleuler formalism), and the
BRST setting resulted from generalizing the gauge formalism to interactions
involving \textit{massive} vectormesons.

Before describing some of the conceptual-mathematical details of the new
setting it is helpful to recall how physical stringlocal charge-carrying
matter fields have been formally envisaged in the BRST gauge setting. The
formal expressions in the Krein space setting are well known%

\begin{equation}
\varphi(x,e)=\varphi^{K}(x)expig\int_{x}^{\infty}A_{\mu}^{K}(x+\lambda
e)e^{\mu}d\lambda,~e^{\mu}e_{\mu}=-1, \label{Jo}%
\end{equation}
they already appeared in publications of Jordan and Dirac during the 30s. But
anybody who (besides playing formal games) tried to obtain a perturbative
computational control on the basis of such nonlocal composite formal
expressions within renormalized perturbation theory knows that this is an
impossible task.

The new SLF formalism solves this problem by converting it from its head to
its feet; instead of trying to represent physical charge-carrying fields in
terms of pointlike gauge-variant fields in a Krein space setting, it bases
renormalized perturbation theory direct on stringlocal fields in Hilbert
space. In this way the stringlocal physical fields become the basic fields in
terms of which renormalized perturbation theory is formulated \cite{charge}
\cite{Hilbert}. For free massive pointlike potentials (Proca potentials) the
short distance dimension $d_{\Pr oca}=2$ poses no problems. The problems start
if such fields interact, since it is impossible to define an interaction
density which stays within the power-counting limit $d_{int}=4;$ all
interactions of Proca fields are nonrenormalizable.

In classical field theories Hilbert space positivity plays no role; the
vectorpotential is a perfectly legitimate and useful classical object; the
fact that many different vectorpotentials correspond to the same field
strength and the formalization of this observation in terms of the
introduction of a classical gauge group does not change this. However the
quantum Hilbert space structure and in particular its \textit{positivity
property} (related to the quantum probability) has no classic analog from
which it could arise via the quantization parallelism, This changes the whole
game; for zero mass quantum vectorpotentials there is a \textit{clash between
covariant pointlike zero mass vectorpotentials and the Hilbert space
positivity;} this clash extends to $s>1$ tensorpotentials\footnote{Example:
for s=2 the tensorpotential is the $g_{\mu\nu}$ and the associated field
strength is a tensor field with 4 indices (the linearized Riemann tensor).},
whereas the associated observable pointlike quantum field strengths remain
pointlike fields in Hilbert space. In fact stringlike potentials $A_{\mu
}(x,e)$ arise by integrating pointlike field strengths over semi-infinite
spacelike lines in the direction $e$ (see below) starting at the spacetime
point $x$. This process can be extended to higher spin field strengths (for
s=2 the linearized Riemann tensor); each time the short distance dimension
improves by one unit until the process ends at a covariant ($e$ is a Lorentz
vector) stringlocal sibling of dimension $d_{string}=1~$of a pointlocal tensor
potential which has $d_{pont}=s+1;~$some details will be presented later on.

Whereas the clash in the zero mass case already occurs for \textit{free}
tensorpotentials\footnote{The counterpart of vectorpotentials to higher spin.
E.g. for $s=2\,\ $the field.strength is a 4-index tensor (the linearized
Riemann tensor) and the associated tensorpotential is the 2-index $g_{\mu\nu
}~$tensor.} and hence is of a kinematic nature, the weakening of pointlike
localization in the \textit{massive} case is a dynamic phenomenon which
manifests itself in a subtle \textit{connection between renormalization and
locality} in the sense that nonrenormalizability of certain fields implies
that they do not exist as pointlike Wightman fields but that their
perturbative interactions becomes renormalizable if formulated in terms of
\textit{stringlike} Wightman fields.

The indefinite metric gauge formalism for pointlike massless tensorpotentials
can be related via quantization to the classical pointlike formalism, but it
is not immediately clear what is the tightest localization which is consistent
with the Hilbert space positivity. The before mentioned structural theorem for
localization in QED suggests that it should be semi-infinite stringlike. There
is a powerful general theorem which states that in theories with mass gaps and
pointlike observable algebras the generating fields (fields act always act in
Hilbert space unless otherwise stated) which carry superselection charges, are
generically stringlocal \cite{BF}; in other words, in order to generate the
operator algebras of QFT, one does not need generating fields which live e.g.
on spacelike hypersurfaces. Stringlike $\Psi(x,e)$ fields are covariantly
localized on a semiinfinite spacelike string: $x+\mathbb{R}_{+}e,~~e\cdot
e=-1.~$In this new setting pointlike fields $\Psi(x)$ are considered as
special $e$-independent cases. By definition local observables are always
pointlike generated (currents, field strength,..).

It is the main point of this section to abandon the gauge description in
favour of a Hilbert space formulation; for $s\geq1$ this requires to replace
pointlocal vectormesons by their stringlike counterpart. The Krein space gauge
setting and the SLF Hilbert space formulation meet on the level of local
observables where the property of gauge invariance corresponds to
$e$-independence, whereas the Hilbert space setting provides the missing
higher sectors beyond the vacuum sector which cannot be generated by local
observables but need stringlike generating operators (e.g. the physical
electron fields). The gauge setting arises naturally from the Langrangian
quantization of the classical electromagnetism, whereas stringlocal
vectorpotential have no Euler-Lagrange description.

Fortunately perturbative QFT does not depend on an Euler-Lagrange description.
The Epstein-Glaser (E-G) formulation \cite{E-G} of perturbation theory
("causal perturbation theory") accepts Lorentz-invariant interaction densities
in terms of covariant fields independent of whether these fields are of
Lagrangian origin or results of representation-theoretic (Wigner) local
quantum physical constructions. However the use of covariant stringlocal
fields requires a nontrivial extension of the E-G inductive construction from
pointlike to stringlike crossings; such an extension has been achieved in
recent but yet unpublished work by Mund \cite{J}.

The Hilbert space positivity restricts the existing pointlike formulation to
$s<1$. According to the aforementioned theorem \cite{BF} generating fields of
LQP in theories with a mass gap are at most string-localized. In the following
it will be shown that nonrenormalizable couplings involving massive pointlike
$s\geq1~$fields can be rewritten in terms of stringlocal Wightman fields. We
believe that perturbative nonrenormalizable pointlike couplings which cannot
be converted into stringlike renormalizable couplings (see next section) do
not define models consistent with principles of QFT.

Our SLF setting requires to describe interactions of zero mass vectormesons
(QED, QCD) as limiting cases of massive interactions; in the limit only the
stringlocal Wightman fields survive; their presence is essential for the
understanding of physical consequences of infrared divergencies and they are
the only physical fields which carry a nontrivial Maxwell charge. As already
mentioned a nonperturbative proof of string-localization as the tightest
possible localization of charged fields which carry a nontrivial Maxwell
charge is based on the quantum Gauss law \cite{B80}. In contradistinction to
massive strings in different directions which are unitarily equivalent,
charged QED strings are "rigid"; in particular they lead to a spontaneous
breaking of Lorentz covariance \cite{Froe}.

The new SLF setting bases renormalized perturbation theory direct on
stringlocal physical fields \cite{charge} \cite{Hilbert}. For free massive
pointlike potentials (Proca potentials) the short distance dimension
$d_{Proca}=2$ poses no problems. They start if such fields interact since it
is impossible to define an interaction density which stays within the
power-counting limit $d_{int}=4,$ i.e. all interactions of Proca fields are nonrenormalizable.

The first hint into which direction to look comes from the observation that
there are two other fields which belong to the localization class of the Proca
field and have the short distance dimension $d=1$ instead of $2)$. They are
constructed from the Proca potential in terms of the following definitions%

\begin{align}
F_{\mu\nu}(x)  &  :=\partial_{\mu}A_{\nu}^{P}(x)-\partial_{\nu}A_{\mu}%
^{P}(x),~~A_{\mu}(x,e):=\int_{0}^{\infty}F_{\mu\nu}(x+\lambda e)e^{\mu
}d\lambda\label{semi}\\
\phi(x,e)  &  :=\int_{0}^{\infty}A_{\mu}^{P}(x+\lambda e)e^{\mu}%
d\lambda,~e^{2}=-1\nonumber
\end{align}
All three covariant free fields are written in terms of the same basic Wigner
$s=1$ creation/annihilation operators $a^{\#}(p,s_{3}),$ $s_{3}=-1,0,1;~$%
unlike in the BRST setting no additional St\"{u}ckelberg degrees of freedom
are introduced, so that the Hilbert space remains identical to that which the
Proca field generates from the vacuum\footnote{This renders the SLF setting
more similar to the Ginsberg-Landau phenomenological theory of
superconductivity tnan the relation of the latter to the Higgs mechanism for
which the "fattened" vectormeson need the presence of the Higgs particle.}.~In
the presence of interactions the stringlocal scalar $\phi$ may potentially
interpolate particles of any integer spin \cite{MSY}, including $s=0$%
\begin{equation}
\left\langle p,s_{3}\left\vert \phi\right\vert 0\right\rangle \neq0,~-s\leq
s_{3}\leq s
\end{equation}
Which boundstate particles actually appear in addition to the "elementary"
$s=1$ vectormeson and the matter field with which it interacts depend on the
interactions of massive vectormesons with other matter or among themselves.

The important point here is that the covariant string-local nature of $\phi
~$permits a \textit{linear} interpolation, whereas covariant pointlike fields
achieve this only by forming (nonlinear) composite fields \cite{MSY}. Solely
\textit{zero mass stringlocal} fields maintain the standard connection between
spinorial indices and physical spin\footnote{For massless pointlike fields the
relation is more restricted} $s=\left\vert h\right\vert $ which is the same as
that for pointlike \textit{massive} fields; in particular they have no
$\phi~"$escorts". The semiinfinite line integrals in (\ref{semi}) lowers the
dimension by one unit, so that the stringlocal potential and its stringlocal
escort field permit to define formal interaction polynomials within the
power-counting restriction. The string-localization shows up in the
commutation relation; bosonic strings commute if and only if the entire
strings $x+\mathbb{R}_{+}e$ are spacelike relative to each other.

Between the pointlocal Proca field and its stringlike relatives there exists a
(easy verified) linear relation
\begin{equation}
A_{\mu}(x,e)=A_{\mu}^{P}(x)+\partial_{\mu}\phi(x,e),~\text{ }d_{sd}(A_{\mu
})=1,~d_{sd}(\phi)=1,~d_{sd}(A_{\mu}^{P})=2 \label{class1}%
\end{equation}
In contrast to the equivalence relations (\ref{pro1}) in Krein space, these
relations are bona fide operator equations in Hilbert space which (in case of
free fields) are direct consequences of the above definitions. The $\phi$ are
similar to the St\"{u}ckelberg fields in the BRST gauge setting (\ref{pro1})
but in contrast to the letter they are physical i.e. they interpolate physical
states. In the free field case these escort fields (of stringlocal
vectorpotentials) $\phi~$generates the same $s=1$ Wigner particle as $A_{\mu
},$ but in the presence of interactions they may potentially interpolate any
integer spin particle, including a scalar bound state. The compensation of the
most singular part (in the present case of the Proca field) by the derivative
of a lower dimensional field (\ref{class1}) is the mechanism by which later on
the singular nonrenormalizable pointlike interaction density will be converted
into its less singular renormalizable stringlocal counterpart.

In contrast to the role of the scalar Higgs field, which must be added to the
zero order field content, the Hermitian stringlocal scalar $\phi^{\prime}s$
are inexorable companions ("intrinsic escorts") of renormalizable massive
vectormesons. Together with the Proca field they disappear in the massless
limit in which the relation (\ref{class1}) breaks down and only stringlocal
vectorpotentials remain.

Before presenting illustrative second order perturbative model calculations in
the new SLF Hilbert space formulation, it is helpful to know how the local
equivalence class relation between point- and string-local fields can be
extended to the matter fields. Looking at the "gauge theoretic
appearance"\footnote{This is not a gauge transformations between fields of the
same kind, but rather an equation which connects string-and point-like fields
which are members of the same localization class.} of (\ref{class1}) it is not
surprising that this relation takes the form of a gauge transformation%

\begin{equation}
\psi(x)=~e^{-ig\phi(x,e)}\psi(x,e) \label{class2}%
\end{equation}
The $g$-dependent exponential dependence on the physical $\phi$ field changes
the renormalizable stringlocal matter field; the result is a very
\textit{singular pointlike field} with unbounded short distance dimensions
(non-polynomial increase in momentum space). Such fields have been introduced
in \cite{Jaffe}; they are more singular\footnote{In fact they only allow
smearing with a dense class of localized testfunctions.} than operator-valued
Schwartz distributions ("Wightman fields") and indicate their presence in
terms of a breakdown of renormalizability. Any attempt to calculate them
directly (i.e. without using the relation to their stringlocal renormalizable
siblings) will lead to the well-known problems of nonrenormalizable
perturbation theory with infinitely many counterterm parameters, whereas their
calculations as objects \textit{within} the renormalizable stringlocal
perturbation theory will maintain the same number of parameter as those
appearing in the stringlike formulation. In fact they provide a very singular
"coordinatization" of the same physical situation. In particular they do not
allow the construction of localized operator algebras by smearing with
arbitrary compact spacetime supported smooth testfunctions.

The intrinsic nature of the stringlocal physical $\phi$ fields strengthens the
analogy with the massive gauge fields in the Ginsberg-Landau theory of
superconductivity. In contradistinction to the Higgs mechanism, which adds
additional degrees of freedom (namely the extrinsic Higgs fields) in the
belief that vectormesons need them in order to be massive, the SLF setting
describes massive vectorpotentials coupled to charged matter \textit{without}
adding degrees of freedom, just as the quantum mechanical theory of
superconductivity describes short range vectorpotential without requiring
additional degrees of freedom. What is not clear at this point but will become
evident in the following subsections, is that these scalar stringlocal fields,
which together with the other two fields (\ref{class1}) are members of the
same relative localization class, play a crucial role in the interaction of
massive vectormesons with matter.

It is interesting to note that the local equivalence class picture permits a
generalization in which the linear relation between $s=1$ free fields is a
special case a more general relation for integer spin $s>1$ fields
\[
A_{\mu_{1}..\mu_{n}}(x,e)=A_{\mu_{1}..\mu_{n}}^{P}(x)+\partial_{\mu_{1}}%
\phi_{\mu_{2}..\mu_{n}}+\partial_{\mu_{1}}\partial_{\mu_{2}}\phi_{\mu_{3}%
..\mu_{n}}+...+\partial_{\mu_{1}}...\partial_{\mu_{n_{n}}}\phi
\]
The left hand side represents a stringlocal spin $s=n$ tensor potential
associated to a pointlike tensor potential with the same spin. The
$\phi^{\prime}s$ $s=n-i,~i=1,..,n$ are tensorial stringlocal fields of
dimension $d=n-i+1$. \ Each $\phi$ "peels off" a unit of dimension so that at
the end one is left with the desired spin $s$ stringlocal $d=1$ counterpart of
the tensor analog of the Proca field. The main problem of using such
generalizations is to identify those couplings which guaranty the existence of
sufficiently many (generally composite) local observables generated by
pointlike Wightman fields (operator-valued Schwartz distributions). This may
be important in attempts to generalize the idea of gauge theories in terms of
SLF couplings involving massive $s>1~$fields.

The two-point functions of the above $s=1$ stringlocal fields are
$e$-dependent
\begin{align}
&  \left\langle \Phi_{1}(x,e)\Phi_{2}(x^{\prime},e^{\prime})\right\rangle
=\frac{1}{(2\pi)^{3/2}}\int e^{-ip(x-x^{\prime})}M_{\Phi_{1},\Phi_{2}%
}(p;e,e^{\prime})\frac{d^{3}p}{2p_{0}}\label{mixed}\\
&  M_{A_{\mu}^{P},A_{\nu}^{P}}=-g_{\mu\nu}+\frac{p_{\mu}p_{\nu}}{m^{2}%
},~M_{\phi,\phi}=\frac{1}{m^{2}}-\frac{ee^{\prime}}{(pe-i\varepsilon
)(pe^{\prime}+i\varepsilon)}\nonumber\\
&  M_{A_{\mu},A_{\nu}}=-g_{\mu\nu}-\frac{p_{\mu}p_{\nu}}{(pe-i\varepsilon
)(pe^{\prime}+i\varepsilon)}+\frac{p_{\mu}e_{\nu}}{pe-i\varepsilon}%
+\frac{p_{\nu}e_{\mu}^{\prime}}{pe^{\prime}+i\varepsilon}\nonumber
\end{align}
Besides these three diagonal expectations there are also mixed $e$-dependent
two-point functions of which only%
\begin{equation}
M_{A_{\mu},\phi}=-i(\frac{e_{\mu}^{\prime}}{pe^{\prime}+i\varepsilon}%
-\frac{p_{\mu}ee^{\prime}}{(pe-i\varepsilon)(pe^{\prime}+i\varepsilon)})
\label{off}%
\end{equation}
will be needed later on. The $\varepsilon$-prescription defines the
distributions as boundary values of analytic functions. A systematic
derivation of such relations in the context of the intertwiner formalism for
stringlike fields \cite{MSY} will appear in \cite{M-S1}.The appearance of
$e$-dependent time-ordered correlations complicates analytic perturbative
calculations as compared to the BRST setting.

But the extra computational effort is unavoidable, because it is the only
possibility to construct correlation function involving \textit{physical} zero
mass matter fields since the latter \textit{exclusively exist as stringlocal
objects}\footnote{Even the singular pointlike fields of the massive case
disappear in the massless limit.}\textit{ }and the massive vectormeson
theories offer a natural covariant way (without ad hoc cutoffs) to analyze the
infrared behavior. Such constructions are necessary if one wants to show that
confinement is a property of zero mass gluon-matter interactions. In fact the
expected result is that $m\rightarrow0~$limiting$~$correlations vanish if
besides pointlike observable (composite) fields they also contain stringlocal
gluons/quarks; the only expected exception are quark-antiquark pairs with an
$e$ which matches the direction of the spacelike separation between the pair
(a stringlike bridge). One knows from infrared problems in QED that the
leading logarithmically divergent contributions must be re-summed before one
takes zero mass limits \cite{YFS}.

One should also note that the apparent simplicity of the pointlike BRST
perturbation theory as compared to the Hilbert space setting is deceiving; the
difficult part in the gauge setting is not the perturbation theory itself, but
rather the extraction of the physical results. Physical operators, as the
S-matrix, inevitably contain unphysical fields, and to compute their
matrixelements between physical particle states is a nontrivial and even
ill-defined task since the physical space is not simply a subspace but rather
results from a cohomological construction. The $s$-invariant BRST S-operator
in the Bogoliubov formulation depends not only on the physical matter
operators but also on the unphysical $A_{\mu}^{K}$ and $\phi^{K}$ free fields
and even if one finds a way to compute scattering amplitudes by "sandwiching"
S~between physical Wigner particle states it is not clear whether it would
agree with the scattering amplitudes which are calculated by doing the same
with $S_{phys}$ obtained from the Hilbert space formulation where such
problems do not occur. The only secure result of the gauge approach is the
physical nature of the gauge-invariant local observables, but from those alone
it is not possible to derive the S-matrix.$~$

\subsection{SLF perturbation theory involving massive vectormesons}

For the perturbative study of interactions of massive vectorpotentials with
charged matter, one needs to establish the validity of relations as
(\ref{class1}\ref{class2}) in every order of perturbation theory. The zero
order matter fields are pointlike but, as a result of their interaction with
the stringlocal vectorpotential, they become stringlike in higher orders, in
fact they turn out to be even "more stringy" than the vectorpotentials which
mediate the interactions. The important idea which permits to establish these
relation in every order within the St\"{u}ckelberg-Bogoliubov-Epstein-Glaser
(SBEG) setting of renormalized perturbation theory will be referred to as
"adiabatic equivalence" (AE) since it involves the adiabatic limit in which
the spacetime-dependent compact supported coupling $g(x)~$of the SBEG
functional formalism approaches the spacetime-independent everywhere constant
physical coupling strength $g$; this will be explained in the sequel.

Before we turn to concrete model illustrations of perturbation theory in terms
of stringlike fields, a historical remark about the origin of these ideas may
be appropriate. It had been known for a long time that Wigner's infinite spin
representations of the Poincar\'{e} group cannot be generated by pointlike
wave functions \cite{Y}. Further progress had to await the concept of modular
localization, which first appeared in the context of integrable models
\cite{AOP}. Of significant importance was the systematic application of
modular localization to positive energy Wigner representations in \cite{BGL}.
In that paper it was shown that all such representations permit a causal
localization in (arbitrary narrow) spacelike cones. Since the core of such a
conic region is a semi-infinite spacelike string, the only remaining
computational problem was the construction of covariant fields $\Psi(x,e)$
which are causally localized on $x+\mathbb{R}_{+}e,~e^{2}=-1~$and generate
operator localized in (arbitrary narrow) spacelike cones \cite{MSY}$.$ This
finally led to a solution of the age-old problem concerning the field content
of Wigner's "infinite spin" representation class.

It then turned out that the construction of stringlocal fields is also useful
for the pointlike localizable representations since it resolves the
\textit{clash between pointlike localization and the Hilbert space positivity
for zero mass }$s\geq1$\textit{ fields} which one encounters in passing from
pointlike field strength to their associated potentials\footnote{A
corresponding result holds for massless higher halfinteger integer spin
fields.with $s\geq3/2.$}.

The use of stringlike potentials also lowers the short distance dimension;
instead of $d_{sd}=s+1$ for pointlike spin $s~$fields, one can always
construct a free stringlike field with $d_{sd}=1$ for all $s.$This allows to
convert interactions between massive nonrenormalizable pointlike fields into
renormalizable interaction involving their stringlocal analog. It also shows
that the this conversion can be used in the opposite direction; the stringlike
renormalization theory permits to construct well-defined (but more singular)
higher order pointlike interaction densities via the detour of renormalizable
stringlike Wightman fields; this roundabout way cannot prevent the singular
(non-Wightman) nature known from direct use of pointlike perturbation theory
with first order interaction densities beyond the power-counting limit
$d_{int}\leq4,$ but at least the number of parameters stays the same as in its
stringlike counterpart\footnote{The growth of the number of independent
counterterms parameters wizh the perturbative oder in the direct pointlike
setting renders nonrenormaliable interactions rather useless.}.

Although \textit{modular localization} was important for the understanding of
the role of stringlocal fields and their role in the reformulation of gauge
theory, their renormalization theory can nowadays be carried out without
direct use of modular methods. The latter remain present in the background;
they furnish the conceptual-mathematical fundament for the ongoing changes in
QFT. They shows in particular, that the perturbative use of SLF in Hilbert
space is more than a computational substitute of the BRST gauge formulation;
in fact It is the only perturbative formulation in which the full field
content (and not just the local observables of the gauge-invariant vacuum
representation) complies with the physical principle of causal localization in
a Hilbert space.

After having explained the philosophy behind SLF, we will now illustrate these
ideas in three different models. As a preparatory step the reader is first
reminded of the SBEG setting of perturbation theory. Its central object is
Bogoliubov's perturbative operator-S-functional which generates the
\textit{time-ordered products} associated with the scalar interaction density
$L\mathcal{(}x\mathcal{)}$. The scattering matrix $S_{scat}~$and the quantum
fields are then defined in terms of the adiabatic limit of the following
definitions
\begin{align}
S(gL)  &  \equiv\sum_{n}\frac{i^{n}}{n!}T_{n}(L,\ldots,L)(g,\ldots
,g)=:Te^{i\int L(x)g(x)},~S_{scat}=\lim_{g(x)\rightarrow g}S(gL)\label{Bo}\\
\psi_{g}(f)  &  :=S(gL)^{-1}\,\sum_{n}\frac{i^{n}}{n!}T_{n+1}(L,\ldots
,L,\psi)(g,\ldots,g,f),~\psi(f)=\lim_{g(x)\rightarrow g}\psi_{g}(f)\nonumber
\end{align}
Here $g(x)\rightarrow g$ is the adiabatic limit in which the spacetime
dependent coupling approaches the coupling constant and the S-matrix and the
fields become covariant. A sufficient condition is the existence of mass-gaps,
which is satisfied if all fields in the Lorentz-invariant interaction density
are massive\cite{Haag}. Since quantum fields are not operator-valued functions
but rather operator-valued distributions, the definitions of the S-matrix and
quantum fields must be subjected to renormalization which has to be carried
out order by order.

In the case of massive scalar QED \cite{Jens} \cite{Hilbert} we have two
$L^{\prime}s$, a pointlike interaction $L^{P}$ and its stringlike counterpart
$L$
\begin{align}
&  L^{P}(x)=j^{\mu}(x)A_{\mu}^{P}(x)=L(x,e)-\partial^{\mu}V_{\mu}\label{S}\\
&  L(x,e)=j^{\mu}(x)A_{\mu}^{S}(x,e),~V_{\mu}=j^{\mu}(x)\phi(x,e),~j_{\mu
}(x)=:\varphi^{\ast}(x)i\overleftrightarrow{\partial}_{\mu}\varphi
(x):\nonumber\\
&  S(gL^{P}+f\psi)\simeq S(gL+f\psi^{S})\nonumber\\
&  A_{\mu}^{P}(x)=A_{\mu}^{S}(x,e)-\partial_{\mu}\phi(x,e),\text{ }\psi
^{P}(x)=e^{-ig(x)\phi(x,e)}\psi^{S}(x,e)\nonumber
\end{align}
The $L^{P}$ is the singular pointlike Proca interaction, whereas $L$ is the
new stringlike interaction which, as a result of $d_{sd}(A_{\mu}^{S})=1,$
stays within the power-counting limit of renormalizable couplings; both
$L^{\prime}s$ act in the Hilbert of the free fields which were used in the
definition of $L^{P}$. The vector $V_{\mu}$ contains the previously introduced
intrinsic escort field $\phi~$of $A^{S},$ and $\partial^{\mu}V_{\mu}~$with
$d_{sd}(\partial^{\mu}V_{\mu})=5$ plays a similar role with respect to $L^{P}$
as $\partial_{\mu}\phi$ in (\ref{class1}) with respect to $A_{\mu}^{P}$,
namely it "peels off" the highest short distance dimension from $L^{P}$ and
converts it into the renormalizable $d_{sd}=4$ interaction density
$L$\footnote{For convenience of notation we omit the superscript $S~$for
stringlocal objects.}. The highest divergence is now carried by the derivative
$\partial^{\mu}V_{\mu}~$term which, integrated with $g(x),$ becomes a boundary
term and hence vanishes (in massive theories) in the adiabatic limit
$g(x)\rightarrow g.$ In this way one arrives at the equality (up to problems
of normalization) of the first order pointlike scattering matrix with its
string counterpart
\begin{equation}
\int L^{P}d^{4}x=\int Ld^{4}x~~or~L^{P}\overset{AE}{\simeq}L \label{S1}%
\end{equation}
which defines the concept of "adiabatic equivalence" of the two interactions.

For notational conveniences, and also in order to maintain formal analogy to
the BRST formalism, one views $A_{\mu}(x,e)$ and $\phi(x,e)$ as zero forms in
$e,$ with $d_{e}$ denoting the differential operator which maps $n$-forms into
$n+1$ forms so that $d_{e}^{2}=0.$ Then the basic relation of
string-independence (\ref{class1}) reads
\begin{align}
&  d_{e}(A_{\mu}(x,e)-\partial_{\mu}\phi(x,e))=0,~u:=d_{e}\phi\label{point1}\\
&  d_{e}(L\mathcal{(}x,e\mathcal{)}-\partial_{\mu}V^{\mu}(x,e))=0.~Q_{\mu
}=d_{e}V_{\mu}\nonumber
\end{align}
and the last relation, in which the $d_{e}$ acts on composites, is a
consequence of the $d_{e}~$action on the basic free fields. For all
interactions of massive vectormesons with matter such pairs $L$, $V_{\mu}$
exist. The content of the bracket in the second line is simply the lowest
order nonrenormalizable pointlike interaction; for massive QED see (\ref{S}).

The differential form calculus is \textit{formally} similar to the nilpotent
$s$-operation of the cohomological BRST gauge formalism (see below). Its
conceptual role remains however quite different; in the case at hand the
differential formalism separates pointlocal observables from stringlocal
fields in Hilbert space, whereas the main purpose of the BRST $s$-operation is
to allow the return from an unphysical Krein space to a quantum theoretical
Hilbert space in which (only) gauge invariant observables act. Operators as
(\ref{Jo}), which in the BRST terminology may be called "gauge invariant
nonlocal matter fields", are outside the range of the perturbative gauge
formalism, whereas in the SLF setting they define the basic renormalizable
matter fields of perturbation theory. In contrast to the nilpotent
$s$-operation, which is needed for the construction of a Hilbert space, the
$d_{e}$ acting on classical differential zero forms is directly related to the
physical localization properties.

If the $T$-products would not involve distributions with singularities at
coinciding points as well at string crossings which impede to pull the
$\partial_{\mu}~$through the $T$, higher order string independence relations
as
\begin{equation}
(d_{e}+d_{e^{\prime}})(TL~L^{\prime}-\partial_{\mu}T\ V^{\mu}L^{^{\prime}%
}-\partial_{\nu}^{\prime}TL\ V^{\nu\prime}+\partial_{\mu}\partial_{v}^{\prime
}TV^{\mu}V^{\nu\prime})=0 \label{bra1}%
\end{equation}
would be an automatic consequence. This relation may be somewhat simplified by
splitting it (using the symmetry in $x,e\leftrightarrow x^{\prime},e^{\prime}%
$) into:
\begin{equation}
d_{e}(TLX^{\prime}-\partial_{\mu}TV^{\mu}X^{\prime})=0,~~~X^{\prime}%
=L^{\prime},V^{\mu\prime} \label{bra2}%
\end{equation}
The ambiguities of time-ordering at point or string-crossings make the
fulfillment of these relations a nontrivial renormalization problem. Their
validity as distributional relations, including coalescent $x^{\prime}s$ and
string crossings$,$ would imply the string-independence of the second order
scattering matrix, since all derivative terms lead to vanishing boundary terms
in the AE limit.

The vanishing of the bracket in (\ref{bra1}) also provides a second order
definition of a T-product of singular "pointlike"\footnote{The $TL^{P}%
L^{P\prime}~$is generally not pointlike as an interaction density, since there
remain $e$-dependent contact terms which only vanish after integration (i.e.
in the AE limit).} interactions $TL^{P}(x)L^{P}(x^{\prime}),$ which in the
standard pointlike setting would be outside the range of renormalization
theory.
\begin{equation}
TL^{P}L^{P\prime}\overset{AE}{\simeq}TL~L^{\prime},~~TL^{P}L^{P\prime}\equiv
TL~L^{\prime}-\partial_{\mu}T\ V^{\mu}L^{^{\prime}}-\partial_{\nu}^{\prime
}TL\ V^{\nu\prime}+\partial_{\mu}\partial_{v}^{\prime}TV^{\mu}V^{\nu\prime}
\label{bra3}%
\end{equation}
The derivative terms, which in massive theories lead to vanishing surface
contributions after integration over spacetime, account for the fact that this
$e,e^{\prime}~$independent definition of a second order pointlike interaction
leads to the same scattering matrix as its stringlike counterpart.
Renormalization means the construction of a time-ordering which fulfills
$e$-independence in the sense of (\ref{bra3}).

This is conveniently done by decomposing the time-ordered products in terms of
Wick-ordered products. The resulting operator contributions can be ordered
according the number of contractions. The term with no contraction obviously
fulfills the above identity. The so-called tree-contribution contains one
contraction; for contractions containing the time-ordering of derivative of
fields this leads to a renormalization problem. The only massive vectormeson
coupling in which this problem is absent is massive spinor QED \cite{Jens}.
Loop contributions are as usual absorbed in mass- and coupling- renormalization.

The interesting new phenomena of SLF in Hilbert space happen in the
"tree"-component. In the following this problem and its solution will be
sketched for three models: scalar massive QED, its chargeless counterpart
(coupling to a Hermitian field $H$) and some comments on the massive
Yang-Mills coupling (interacting massive gluons). In the following three
subsection we will be content with the calculation of the second order
S-matrix. The calculation of off-shell correlation of quantum fields and the
relation between singular pointlike and renormalizable stringlike matter
fields (\ref{class2}) will be left to a separate publication.

For new interesting problems of mathematical physics arising from stringlocal
perturbation theory, in particular problems related to the extension of
Epstein-Glaser causal renormalization theory to string-crossings, we refer to
forthcoming work by Mund \cite{J}.

\subsection{Scalar massive QED}

According to the traditional view, massless scalar QED is a pointlike model
with two coupling parameter\footnote{The electromagnetic coupling and a
parameter related to a quadrilinear scalar field coupling.}; it is known to be
renormalizable in the unphysical pointlike BRST Krein space setting. Unlike
its classical counterpart, this quantum gauge description is severely
restricted; the positivity requirements of the Hilbert space clash with the
pointlike localization and quantum gauge theory is the result of a compromise;
the description is limited to local observables which constitute the gauge
invariant part, physical matter fields remain outside.

As a consequence, quantum gauge theory is not capable to provide a spacetime
description of collisions between electrically charged particles; however
there exist calculational successful infrared regularized momentum space
recipes for photon-inclusive cross sections. There is presently no spacetime
understanding of collision theory analogous to that provided by the LSZ
scattering theory\footnote{The large-time LSZ limits vanish for infraparticle
fields \cite{Hilbert}.} in case of models with mass gaps. The traditional
point of view is that zero mass interactions are simpler than their massive
counterparts; but this refers to purely formal aspects of renormalization
theory and ignores the physical-conceptual problems. The latter point into the
opposite direction.

The problems of infraparticles in QED \cite{Bu} and confinement in QCD still
belong to the conceptual demanding unsolved problems of particle theory,
whereas the incorporation of renormalization problems of its massive
counterparts can be achieved by extension of the renormalization theory to the
new SLF setting in Hilbert space. Apart from some remarks at the end of next
section, the construction of massless limits and new ideas to tackle infrared
problems will be left to a separate publication.

The defining first order stringlocal interaction density of massive scalar
QED
\begin{align}
L(x,e)  &  =gA_{\mu}(x,e)j^{\mu}(x)=L^{P}+\partial^{\mu}V_{\mu}\label{L}\\
j^{\mu}  &  =\varphi^{\ast}\overleftrightarrow{\partial^{\mu}}\varphi,~V_{\mu
}=\phi j_{\mu}\nonumber
\end{align}
is according to (\ref{bra1}) $d_{e}$-equivalent to its pointlocal counterpart
$L^{P}$. This secures the $e$-independence of the first order S-matrix in the
AE limit. In these equivalences the stringlocal intrinsic escort fields $\phi$
which appears explicitly in $V_{\mu}$ play an essential role. Whereas the
first order relation is a result of the definition of a "stringlocal"
interaction, the second order relation (\ref{bra1}) is a nontrivial
restriction on the renormalization.

One defines a reference time-ordering $T_{0}$ of two-pointfunctions of
derivatives of the complex scalar field $\varphi$ by taking the derivatives
outside the two-point function e.g.%
\[
\left\langle T_{0}\partial_{\mu}\varphi^{\ast}(x)\partial_{\nu}^{\prime
}\varphi(x^{\prime})\right\rangle =i\frac{\partial_{\mu}\partial_{\nu}%
^{\prime}}{\left(  2\pi\right)  ^{4}}\int d^{4}pe^{-ipx}\frac{1}{p^{2}%
-m^{2}+i\varepsilon}%
\]
On the other hand the time ordering in Epstein and Glaser's renormalization
approach permits delta function counterterms of the same scaling degree as the
integrand, for the present case
\begin{equation}
\left\langle T\partial_{\mu}\varphi^{\ast}(x)\partial_{\nu}^{\prime}%
\varphi(x^{\prime})\right\rangle =\left\langle T_{0}\partial_{\mu}%
\varphi^{\ast}(x)\partial_{\nu}^{\prime}\varphi(x^{\prime})\right\rangle
-aig_{\mu\nu}\delta(x-x^{\prime}) \label{c1}%
\end{equation}
where $a$ is a free parameter.

If we were to treat the defining first order interaction $A_{\mu}j^{\mu}$ as
involving a pointlike $A_{\mu}$ field in the Krein space of pointlike massless
vectorpotentials, the interaction is renormalizable in the perturbative
inductive Epstein-Glaser renormalization setting where it leads to two
counterterms. The first counterterm (\ref{c1}) appears in the second order
tree approximation and amounts to a modification of the interaction through a
second order contact term (all operator products are meant to be Wick-ordered)%
\begin{equation}
aA_{\mu}(x)A^{\mu}(x)\varphi^{\ast}(x)\varphi(x) \label{c2}%
\end{equation}
with an \textit{independent} coupling parameter $a.$ There is an additional
quadrilinear counterterm with a coupling parameter of the form%
\begin{equation}
b\left(  \varphi^{\ast}(x)\varphi(x)\right)  ^{2} \label{c3}%
\end{equation}
which appears for the first time in $4^{th}~$order; these two counterterm
exhaust the possibilities of counterterm structures (primitively divergent
contributions in the Feynman graph setting), which means that the renormalized
theory is 3-parametric.

To recuperate local oberservables acting in a Hilbert space (at the expense of
charge-carrying matter fields which remain unphysical fields in Krein space)
one has to \textit{extend the Krein space formulation by ghost operators} as
explained in the previous section; in this way one arrives at the \textit{BRST
gauge formulation which fixes the parameter }$a$\textit{ in (\ref{c2}) to a
numerical value~}$a=1$\textit{ }according to the rules of a formal "gauge
symmetry". By itself this term has no direct physical interpretation apart
from its role in the extraction of local observables from an unphysical
description. For the formal description and the perturbative calculations of
the two-parametric massive scalar QED one needs the full BRST "ghost program",
even though the physics is only contained in the small subalgebra generated by
"gauge invariant" local observables. The gauge symmetry is a technical trick
and not a physical symmetry; in particular its spontaneous breaking is
physically meaningless.

In the SLF Hilbert space setting on the other hand, the second order with the
correct value of $a$ is "induced" from the model-defining first order $A\cdot
j~$interaction; it is simply the result of the implementation of locality in
Hilbert space setting. No additional principle as gauge symmetry has to be
invoked in order to fix $a$ to its correct numerical value; models QFT are
realizations of the foundational causal localization principle. The difficult
task is to trace the richness of models back to different physical
manifestations of this principle. The \textit{induction mechanism}~exists only
for higher spins $s\geq1,$ for lower spins the renormalization theory is the
well-known counterterm formalism of pointlike interactions.

For the case at hand this is done as follows. From the results in the previous
section we know that the second order locality requirement for the S-matrix in
the presence of stringlike fields amounts to the vanishing of the $d_{e}%
~$operation on the renormalized tree component%
\begin{align}
&  d_{e}(TA\cdot jA^{\prime}\cdot j^{\prime}-\partial^{\mu}T\phi j_{\mu
}A^{\prime}\cdot j^{\prime})_{1-con}=0\label{c4}\\
-A_{e}  &  :=d_{e}(T_{0}A\cdot jA^{\prime}\cdot j^{\prime}-\partial^{\mu}%
T_{0}\phi j_{\mu}A^{\prime}\cdot j^{\prime})_{1-con}=N_{e}+\partial^{\mu
}N_{e\mu},~A=A_{e}+A_{e^{\prime}}\nonumber
\end{align}
and a similar expression in which the unprimed and primed $x,e$ are
interchanged; the total anomaly $A$ from the one-contraction terms is simply
the sum of the two contributions. They originate from the divergence of propagators%

\begin{align}
\partial^{\mu}\left\langle T_{0}\partial_{\mu}\varphi^{\ast}(x)\partial_{\nu
}^{\prime}\varphi(x^{\prime})\right\rangle  &  =-i\partial_{\nu}^{\prime
}\delta(x-x^{\prime})+reg\\
\partial^{\mu}\left\langle T_{0}\partial_{\mu}\varphi^{\ast}(x)\varphi
(x^{\prime})\right\rangle  &  =-i\delta(x-x^{\prime})+reg\nonumber
\end{align}
where $reg$ stands for the regular contributions which results from applying
the wave operator to the free field $\varphi^{\ast}(x)~$inside the time
ordering. The anomaly contribution is not the only delta contribution, the
$T_{0}LL^{\prime}$ also contributes since according to the rules of minimal
scaling we are required to introduce a counterterm (\ref{c1}) with an
undetermined parameter $a.$ According to the minimal short distance scaling
rules of renormalization for two derivatives the $T_{0}~$passes to a $T~$which
contains a free renormalization parameter $a~$(\ref{c1}) ~whereas we keep
$T=T_{0}~$for $\varphi~$propagators with a lower number of derivatives.
$T_{0}$ propagators also appear in the $\varphi$-contractions of the tree
contribution $TLL^{\prime}|_{1-con}.~$Instead of presenting the lengthy but
straightforward calculation of the $N^{\prime}s$ we only write the result%
\begin{equation}
N=2\delta\varphi^{\ast}\varphi A\cdot A^{\prime},~N_{\mu}=\delta\varphi^{\ast
}\varphi(\phi A_{\mu}^{\prime}+\phi^{\prime}A_{\mu})
\end{equation}
where the $\delta$ stands for $\delta(x-x^{\prime}).$

By inspection one now realizes that the choice $a=1~$in (\ref{c1}) leads to a
compensation of the $N$-anomaly with the normalization term from $TLL^{\prime
}.$ The $N_{\mu}\ $contributes to the renormalization of the $TV_{\mu
}L^{\prime}~$operator but does not contribute to the renormalization of the
second order S-matrix. As a consequence of the identity $d_{e}\partial^{\mu
}\phi=d_{e}A^{\mu}$ there are no delta anomaly- contributions from $\phi
$-$A_{\nu}$ contractions. One obtains the expected second order quadratic
in$~A_{\nu}~$contributions which in the gauge formalism results from imposing
gauge invariance%

\begin{align}
&  TLL^{\prime}=T_{0}LL^{\prime}+2i\delta(x-x^{\prime})L_{2},~L_{2}%
=2\varphi^{\ast}(x)\varphi(x)A\cdot A^{\prime}\label{2nd}\\
&  S=\int(igL-\frac{1}{2}g^{2}L_{2})-g^{2}\frac{1}{2}\int\int T_{0}LL^{\prime
}+higher=ig\int L-\int\int\frac{g^{2}}{2}TLL^{\prime}+higher\nonumber
\end{align}
The separate $e,e^{\prime}$ dependence is a consequence of the independent
directional fluctuations i.e. reminder that $e$ is not the gauge parameter of
the noncovariant axial gauge but rather the fluctuating string variable of a
covariant stringlocal potential. Since the anomaly contributions are
Wick-ordered quadrilinear Terms there is no problem with setting $e=e^{\prime
};$ the only problematic aspect is to identify the $e^{\prime}s$ in
propagators.$~$In momentum space scattering amplitudes one can always avoid
the dangerous $e$-directions by choosing the $e=e^{\prime}$ such that the
denominator in the propagator does not vanish. Then the formalism guaranties
that each contribution to a 2-particle scattering amplitude is well-defined
and their sum (the scattering amplitude) is the $e$-independent sum of these
contributions is guarantied by the formalism (the string-independence of the S-matrix).

In (\ref{2nd}) the last equation has absorbed the $L_{2}~$contribution into a
redefinition of the $T$-product. This is a notational simplification for tree
contributions of arbitrary high order which the gauge description does not
suggest. As mentioned in (\ref{bra3}) the SLF setting permits a "backdoor"
construction of pointlike interaction densities; their momentum space behavior
corresponds to \ what one expects from the pointlike counterterm formalism
but, different from the latter it introduces no new undetermined coupling
parameters. For such computations it is necessary to use the $N_{\mu}~$for the
renormalization of the $TV_{\mu}L^{\prime}~$derivative terms. This observation
is restricted to pointlike interaction densities of arbitrary order but does
not yet extend to field correlations; for the latter one has to extend the
Bogoliubov S-matrix formalism, a step which is well-known for pointlike
fields, but still needs to be elaborated for stringlocal fields.

The structure of the definition (\ref{bra3}) shares with the naive expression
obtained from second order pointlike perturbation theory the large momentum
increase, but the mass-shell restriction of the former leads to the
cancellation of leading high momentum contributions which is the momentum
space counterpart of the on-shell "peeling property" in x-space. This
difference between off-shell correlations and the on-shell lowering of the
p-increase has no counterpart in the pointlike Feynman formalism. As
everything which is different from the standard pointlike formalism, its
origin is the powerful Hilbert space positivity which starts to assert itself
in massive $s\geq1$ interaction; as such it is a completely new phenomenon
with no counterpart in the Krein space gauge setting.

\subsection{Couplings to Hermitian fields and the Higgs model}

Although having no counterpart in classical theory, one may ask whether it is
possible to couple a massive vectormesons to a Hermitian scalar fields $H$ as
kind of "charge-neutral" counterpart of massive scalar QED. A second order
BRST operator gauge treatment of such a situation which is suitable for a
comparison with our SLF setting has been given by the University of Z\"{u}rich
group (\cite{Scharf} and references therein) and more recently in
\cite{Garcia}, It is appropriate to briefly recall their results before
presenting the solution in the SLF Hilbert space setting. For comparison it is
helpful to reformulate their derivation in analogy to our Hilbert space
formulation \cite{M-S1}.

The first order pair $L$, $V_{\mu}$ which corresponds to the lowest pointlike
interaction with a Hermitian field $H$ is\footnote{A term $A^{P}\partial
H^{2}~$turns out to be a total derivative since $\partial A^{P}=0.$}
($\phi_{Scharf}\sim m\phi$)
\begin{align}
L^{P}  &  =m(A^{P}\cdot A^{P}H+cH^{3})=L-\partial_{\mu}V^{\mu}%
~with:\label{first}\\
L  &  =m(A\cdot AH+\frac{1}{2}A\cdot(\phi\overleftrightarrow{\partial}%
H)-\frac{m_{H}^{2}}{2}\phi^{2}H+cH^{3}+u\tilde{u}H)\nonumber\\
V_{\mu}  &  =m(A_{\mu}\phi H+\frac{1}{2}\phi^{2}\overleftrightarrow
{\partial_{\mu}}H)\nonumber
\end{align}
where the superscripts $K$ on $A_{\mu},$ $\phi,~L$\ and $V_{\mu}$ have been
omitted for notational convenience (for the notation see (\ref{free})). The
mass factor $m$ (the vectormeson mass) has been introduced in order to keep
track of the overall "engineering dimension" $d_{en}=4$. Even though the
conceptual content of the Hilbert space approach is quite different from the
gauge theoretical approach, there are close formal correspondences between the
differential form $d_{e}~$formalism with the BRST\ nilpotent $s$-operation.
Instead of starting with pointlike trilinear interaction $L^{P}$ and
converting it into an $L$ and a divergence of a $V_{\mu},~$there is the more
general looking possibility to start with a trilinear Ansatz for a stringlike
$\hat{L}$ within the power-counting restriction and find the correct $L$ and
$V~$which fulfills the string-independence string-independence $d_{e}%
(L-\partial V)=0$ \ in a unique way (up to a contribution in $V$ whose
divergence vanishes). We may call this the first order "induction".

We now pass to the second order implementation of the BRST $s$-symmetry (the
operator form of gauge invariance in Krein space). The appearance of a
$u\tilde{u}H~$term, which only vanishes on $Kers/Ims,$ has no counterpart in
the SLF setting; it simply does not occur in the Hilbert space setting. Again
one computes the anomalies of the one-contraction ($1$-$c$) contributions of
the $s~$operation according to the rules (\ref{free}) and compensates them
with corresponding normalization terms by choosing the free normalization
parameter in $TLL^{\prime}~$in such a way that, in analogy to massive QED they
match the well-defined anomaly $A$. The induced counterterms which together
with the $T_{0}$-product define the renormalized $T$-product ($Q_{\mu
}:=sV_{\mu}$)%

\begin{align*}
&  A=sT_{0}LL^{\prime}-(\partial^{\mu}T_{0}Q_{\mu}L^{\prime}%
+{\small (x\longleftrightarrow x}^{\prime}{\small ))}=sN+N_{\mu}\\
&  with~TLL^{\prime}=T_{0}LL^{\prime}+N,~~TQ_{\mu}L^{\prime}=T_{0}V_{\mu
}L^{\prime}+N_{\mu}%
\end{align*}
For the calculation of the renormalized S-matrix we only need the
one-contraction ("tree") component of the anomaly. As has been shown in
(\cite{Scharf} page 147) this leads to 4 induced delta function anomaly terms
$N=\delta(x-x^{\prime})L_{2}~$with
\begin{equation}
L_{2}=AAH^{2}+AA\phi^{2}-\frac{1}{4}m^{2}m_{H}^{2}\phi^{4}-\frac{1}{2}%
m_{H}^{2}\phi^{2}H^{2}+cH^{3}+c^{\prime}H^{4} \label{s1}%
\end{equation}
Here the $c^{\prime}$ is an additional coupling which, although still free in
second order, is needed for the compensation of anomalies in 3rd order which
leads to the value $c^{\prime}=-\frac{1}{4}\frac{m_{H}^{2}}{m^{2}}.$ In other
words the Mexican hat potential is fully induced by the gauge $s$-invariance
of the S-matrix. There is no place for symmetry breaking shifts in field space.

Again the sum of the local second order term $\frac{g^{2}}{2}L_{2}$ is not
physical by itself, but the sum
\begin{equation}
\frac{g^{2}}{2}(\int L_{2}(x)d^{4}x+\frac{1}{2}\int\int T_{0}L(x)L(x^{\prime
})d^{4}xd^{4}x^{\prime}) \label{s2}%
\end{equation}
is the second order contribution to the gauge-invariant S-matrix. As in
(\ref{2nd}) the form of the induced interaction $L_{2}$ depends again on the
definition of the $T_{0}$ with which the anomalies were computed; and as in
the previous case of scalar massive QED one can absorb the quadratic terms in
$A$ in (\ref{s1}) into a change $T_{0}\rightarrow T.$ What remains is the
quadrilinear $H$-$\phi~$potential which together with the $A$-independent
terms from $L_{1}$ can brought into the form of a Mexican hat potential as
shown in \cite{Scharf}. But here this is a result of a \textit{second order
gauge induction and not of a symmetry-breaking interaction}; the numerical
coefficients of the induced potential are ratios of the masses and do not
depend on a symmetry-breaking field shift. Together with the first order
$H$-$\phi~$contributions they can be written in in the form of a Mexican hat potential.

\textit{Contrary to the destruction of gauge invariance by a numerical field
shift in the gauge-dependent field of scalar QED} and subsequent adjustments
in terms of special gauges, \textit{the induced Mexican hat potential results
from the preservation of gauge invariance for the coupling of a Hermitian
field to a massive vectormeson} \textit{using the BRST gauge invariance of the
S-matrix through the relation }$sS=0.$ Since all operators are massive there
are no infrared problems. As a result the second order inductions of a Mexican
hat potential from the implementation of the BRST $s$ gauge invariance is
totally different from the introduction by hand of a Mexican potential whose
purpose is the (impossible task) to break the gauge symmetry in order to
generate a mass. The work of the university of Z\"{u}rich group \cite{Scharf}
\cite{Aste} should have caused the ringing of bells with respect to the Higgs
issue, but it was ignored.

In the SLF setting the calculation proceeds in a similar fashion. The
$e$-independence first order argument results in%

\begin{align}
L  &  =m(A\cdot(A^{P}H+\phi\partial H)-\frac{m_{H}^{2}}{2}\phi^{2}H+cH^{3})\\
V_{\mu}  &  =m(A_{\mu}^{P}\phi H+\frac{1}{2}\phi^{2}\overleftrightarrow
{\partial}_{n}H)\nonumber
\end{align}
Again the $m$ factors keep track of the engineering dimension. Different from
the previous case there are off-diagonal propagators between $A,A^{P}$ and
$\phi.$ It turns out that the best way to handle this problem is to use one
$A^{P}$ instead of only $A^{\prime}s$; this can be done as long as the
power-counting restriction $d_{int}\leq4$ is obeyed.$~$Apart from the absence
of the $u\tilde{u}H$ term and the difference in normalization between the
negative metric $\phi^{K}$ and the physical escort field $\phi,$ the algebraic
steps of the implementation of second order string independence in the
spacetime $d_{e}~$differential form calculus are following the same steps as
those of the nilpotent $s$ calculus. Therefore it is not surprising that also
the results are accordant. One expected difference is the appearance of both
$e$ and $e^{\prime}$ in the Mexican hat potential; this is similar to the
second order expression of the previous massive scalar QED model. The
appearance of in $e,e^{\prime}~$asymmetric term in addition to the symmetric
Mexican hat contribution is however unexpected. This term vanishes on the
diagonal $e=e^{\prime}.$ There is no problem to let the directions coalesce in
the Wick-ordered Mexican hat potential, the problem is the propagator of the
tree-component. For each momentum space region this is possible by choosing
$e=e^{\prime}$ such that the denominator does not vanish. This suffices to
insure that each contribution to the second order tree approximation is
well-defined and by construction the result of adding up the various
contributions is independent of any $e.$ The full second order contribution
will be presented in a joint paper with J. Mund \cite{M-S1}.

The calculation in the stringlocal Hilbert space setting confirms the results
of the BRST gauge setting. This confirmation is important because the physical
content of the gauge formalism is restricted to the gauge-invariant local
observables but the the S-matrix is a global object.

As mentioned before, it is not necessary to go through detailed calculation if
one only wants to see the inconsistency of the Higgs-mechanism with the
principles of QFT. From a conceptual viewpoint the fastes way is to argue that
couplings of massive vectormesons to any matter cannot produce conserved
currents with diverging charges (the spontaneous broken symmetry condition).
Their "Maxwell charge" is always screened and in case of only Hermitian
matter, \textit{the identically conserved Maxwell current is the only
current}. In zero order i.e. for a free massive vectormeson one has%
\begin{equation}
\partial^{\mu}F_{\mu\nu}=j_{\nu}^{Maxwell}\sim m^{2}A_{\mu}^{P}%
\end{equation}
and higher order corrections can be computet by using the SBEG renormalization
theory for fields. It is somewhat strange that the followers of the Higgs
mechanism did not at least check the zero order Maxwell current of a massive
vectormeson and verify that its charge is screened and not spontaneously
broken. In the following table all possible situations related to conserved
currents have been collected%

\begin{align}
&  screening:~Q=\int j_{0}(x)d^{3}x=0,~\partial^{\mu}j_{\mu}=0\\
&  spont.symm.-breaking:\int j_{0}(x)d^{3}x=\infty~\nonumber\\
&  symmetry:\int j_{0}(x)d^{3}x=finite~\neq0\nonumber
\end{align}

In order to avoid any misunderstanding, the present critique is not directed
against discoveries made by metaphoric arguments; many discoveries, including
Dirac's important idea of antiparticles, were based on incorrect models or
theories (the hole theory). Metaphoric observations are valuable placeholders
but start to be harmful if in due time they are not replaced by arguments
compatible with the foundational principles of QFT. The idea that QFT can say
something about the masses of elementary particles (masses of the
interaction-defining fields) is incorrect; its causality principles are
expected to determine masses of bound states which are interpolated by
composite fields but for there description one still has to rely on methods of
lattice approximations.

The case of Goldstone's spontaneous symmetry breaking is no exception. The
definition of a spontaneous symmetry breaking is the existence of a conserved
current whose charge (the would be generator of a symmetry ) can not generate
a symmetry because the integral over the zero component of the current
diverges. The Goldstone theorem says that this can only happen if the energy
momentum spectrum starts at zero; for selfinteracting bosons its content is
more specific in that there must be a zero mass Goldstone boson which couples
to the conserved current and prevents the convergence of the integral over the
zero component of the current for large distances. The shift in field space is
not the definition but only a mnemonic trick (a "pons asini") to find a model
of selfinteracting scalar fields whose first order interaction leads to such a
current; the intrinsic observable properties of such a situation (the
correlation functions) do not contain a field shift but only scalar fields and
their physical masses. Nevertheless it is within the range of metaphorical
tolerance to say that "the shift breaks the symmetry". The problem with the
Higgs mechanism is that this metaphor created the idea if a spontaneous mass
creation which is not compatible with the structure of QFT.

The situation of coupling of massive vectormesons to matter is totally
different. In that case there is the Schwinger-Swieca theorem which says that
the charge of the Maxwell current is screened and in the $H$-model this is the
only current. This case involves $s\geq1$ for which is known that the coupling
of massive vectormesons in Hilbert space leads to nonrenormalizable
interactions as a result of violation of the power-counting bound for
interacting $d=2$ Proca potentials. The trick for $s=1~$is to use the BRST
Krein space gauge setting but this severely limits the confiable range of
validity to the small subset of gauge invariant obervables (the vacuum
sector). The new Hilbert space formulation requires to replace the pointlike
by stringlocal vectorpotentials and in this way secures the physicality of all
operators; the important field operators are stringlike. But it makes good
sense to use stringlike operators to generate particle states because the
difference between point-and stringlocal disappears on the level of particle
states; there simply are no point- and stringlike particles. Different from
the the Goldstone situation where the issue is the construction of a first
order interaction (in terms of free fields which already have the masses of
the physical particles) which leads to a lowest order current with the
Goldstone properties, the lowest renormalizable order in the s=1 situation
must be constructed according to the above $L-\partial V~$requirement (the $s$
respectively the $d_{e}~$invariance) whose validity must be insured in higher
orders. In both cases there appears a second order induced Mexican hat
potential, but in contrast to the Goldstone case this has no relation with a
symmetry-breaking field shift.

To discover something important \textit{together with the correct and final
theoretical explanation }is an unreasonable requirement on the discoverer (in
this case Peter Higgs), this is rather the responsibility of the particle
theoreticians who use the observation. Critique is the live-blood of any
highly speculative theoretical research especially if it takes place on the
frontiers of particle physics. Interestingly a critical view was already
around at the time of Higgs' discovery: namely Schwinger's suggestion that
currents of massive vectormesons lead to "screened charges" and Swieca's
subsequent proof which led him to the terminology "Schwinger-Higgs screening"
\cite{Sw}, see also \cite{BF1}. Unfortunately these early attempts were
ignored and vanished in the maelstrom of time. Schwinger did not mention that
in massive gauge theories there are two currents: the Maxwell current and the
particle-antiparticle counting current which only coalesce in the massless
limit. For $H$-couplings there is no counting current and the coupling
disappears (decomposes into free fields) in the massless limit.

The case of Y-M interactions is significantly different from abelian gauge
theories. In this case the BRST gauge formalism requires the presence of a
coupling to a $H$-field and this can be shown without recourse to the
conceptually incorrect Higgs mechanism of symmetry breaking \cite{Scharf}. It
is of foundational importance for nonabelian gauge theoretical consistency.
The appearance of a Mexican hat like $H$-selfinteraction is a consequence of
gauge invariance and has no relation to symmetry breaking. Hopefully the
necessitiy of its presence can be shown to be a direct consequence of the
causal localization principles in a Hilbert space setting \cite{Mund} because
to base physical arguments on the use of unphysical fields outside the
physical vacuum sector of gauge invariant local observables is physically not
trustworthy (although the perturbative BRST formalism is mathematically consistent).

The appearance of an interacting escort $H$ of nonabelian massive
vectorpotentials breaking prescription starting from the coupling of zero mass
Y-M fields to comples scalar matter is a phenomenon which needs better
foundational understanding. The broken symmetry picture is inconsistent with
the principles of QFT and fortunately the breaking prescription starting from
the coupling of zero mass Y-M fields is not needed. Interestingly the presence
of $H~$produces a pointlike observable Maxwell-field $F_{\mu\nu}^{a}H^{a}$
which again leads to an identically conserved current with a screened charge
(the Schwinger-Swieca-Higgs screening). This suggests that the old screening
idea may have a direct relation with the consistency of massive
selfinterinteracting vectormesons. Hence the Higgs-Englert issue may be a
metaphor for a not yet understood manifestation of particle theory involving
selfinteracting massive higher spin fields (presence of lower spin escorts?).

The arguments against spontaneous mass creation through spontaneous symmetry
breaking (the Higgs mechanism") do in no way rule out the possibility that the
H-fields may be needed for other foundational \ reasons. Whereas the
calculations on massive QED in this section show that the BRST gauge setting
does not need their presence for interacting abelian massive vectormesons with
matter, the implementation of gauge invariance of the second order S-matrix
for selfinteracting nonabelian massive vectormesons needs their presence.
Again no symmetry breaking is needed, the quadrilinear selfinteracting
H-contributions follow from second order gauge invariance. In other words

\subsection{Selfinteracting massive gluons and remarks on confinement}

For abelian massive gauge theories in the SLF Hilbert space formulation there
are no structural theoretical reasons for enlarging the field content beyond
the matter fields to which one wants to couple the massive vectormesons since
its escort fields do not create any additional degrees of freedom. This is
less clear in case of selfinteracting massive gluons. Although the arguments
against the consistency of the Higgs mechanism are generic (independent of the
kind of vectormeson interactions), there could be other consistency
requirements coming from the foundational modular localization properties in
Hilbert space which make it necessary to introduce additional degrees of
freedom. Present calculational attempts indicate that this is not the case; to
excludes such situations higher order calculations are necessary \cite{M-S2}.

The escort fields $\phi$ fields do not count in this balance since they are
part of the Hilbert space formalism for all higher spin interactions; they are
already present in the interaction-free case where they enter the relation
between the pointlike Proca potential and its stringlike sibling. The masses
of self-coupled massive vectormesons are totally independent and the mass of
each escort is equal to that of the stringlocal vectormeson which it escorts.
On the other hand the masses of coupled $H$-fields are independent and (as all
masses, except for that of the $A$-dependent escorts which must be identical
to the $A$ masses) are part of the interaction-defining free field content.

In the remainder of this subsection we will present the first order
stringlocal Y-M interactions which are obtaind from the $d_{e}(L-\partial
V)=0~$argument which also determines the first order pointlike interaction
density $L^{P}$ with its $d_{sd}>4$ scaling degree. For simplicity we take the
equal mass $O(3)$ Y-M model. The starting point is the reduction of the
power-counting violating $d=5~$dimension pointlike interaction $L^{P}$ by
peeling off the highests dimension 5 and in this way obtaining a $d=4$
stringlike interaction density $L$%
\begin{align}
L^{P}  &  =\sum\varepsilon_{abc}F_{a}^{\mu\nu}A_{b,\mu}^{P}A_{c,\nu}%
^{P}=L-\partial^{\mu}V_{\mu},~or~d_{\varepsilon}(L-\partial^{\mu}V_{\mu
})=0\label{gauge}\\
L  &  =\sum_{1}^{3}\varepsilon_{abc}\left\{  F_{a}^{\mu\nu}A_{b,\mu}A_{c,\nu
}+m^{2}A_{a}^{P\mu}A_{\mu}^{b}\phi_{c}\right\}  ,~V_{\mu}=\sum\varepsilon
_{abc}F_{a}^{\mu\nu}(A_{b,\nu}+A_{b,v}^{P})\phi^{c}%
\end{align}
Actually we could have started with the most general trilinear Ansatz for
$\hat{L}$ in terms of $A$ and $\phi.~$Since there are 4 such terms, this
Ansatz would contain 4 different types of yet undetermined $f_{abc}%
^{i},~i=1,..,4$. Then asked the question would be whether within this general
Ansatz for $\hat{L}$ there exists a $\hat{V}_{\mu}~$such that%
\begin{equation}
~d_{e}(\hat{L}-\partial^{\mu}\hat{V}_{\mu})=0~\text{ }%
\end{equation}
The only solution (up to additional divergence free contributions to $V_{\mu}%
$) of this requirement in the case of equal masses turns out to be the first
line of (\ref{gauge}). Defining the content of the bracket as $L^{P}$ we
realize that the first order stringlocal S-matrix is equal to the first order
pointlike counterpart since the two different first order interaction
densities are adiabatically equivalent (the boundary term from the divergence
of $V_{\mu}$ vanishes in the adiabatic limit)
\begin{equation}
\int L^{P}=\int L,~~~L^{P}\overset{AE}{\simeq}L
\end{equation}
This is the beginning of an extremely restrictive \textit{induction mechanism}
which has no counterpart in the nonrenormalizable pointlike $s\geq1$ setting.
For the full Lie-algebra structure (\ref{gauge}) one has to proceed to the
induced second order which will be done in \cite{M-S2}.\ 

These observations generalize those which were already made in the abelian
case in subsection 6.2; the locality principle \ together with Hilbert space
positivity leads to restrictions between couplings which are analogous to
those of classical gauge theory (the geometry of fibre bundles). Here they are
simply the result of the Hilbert space positivity which for interactions which
couple $s\geq1$ fields requires the use of string-localization. There is
absolutely no need for any support from the fibre-bundle setting of classical
gauge theory; QFT does not need any "crutches" from classical field theory
such as those wich are provided by the classical-quantal parallellism of
quantization. Any quantum fields obtained from covariantizing Wigner's
classification of positive energy representation of $\mathcal{P}$ can be
coupled to a scalar density which defines the first order interaction density
of a QFT and in case its short distance dimension falls within the
power-counting range $d_{sd}\leq4~$the interaction density is on the best way
to define a renormalizable model of QFT. The above "self-induction" mechanism
also works for unequal masses; in this case the $f^{\prime}s$ depend also on mass-ratios.

The potentially most important consequence of the Hilbert space SLF
formulation is a profound insight into hitherto incompletely or not understood
infrared phenomena as "infraparticles" and confinement. Concerning the latter,
the remarks one finds in the literature do not go beyond the statement that
the perturbative expressions for the massless gauge-variant correlations of
gluon- or quark- fields are infrared divergent and that this indicates the
breakdown of perturbation theory. But the behavor of pointlike gauge-variant
matter fields in a BRST gauge setting is physically irrelevant; what one needs
is an understanding of \textit{the infrared property of massless limits of
massive stringlocal gluon correlations} and the only way to do this is offered
by the SLF formalism in Hilbert space, a task which is outside the physical
range of gauge theory. One expects that all correlations vanish which contain
besides pointlocal composites also gluon/quark fields; in fact this seems to
be the only way in which the localization principles of QFT can realize
confinement. Pointlike physical fields never lead to confinement.

The infraparticle situation is slightly more accessible . The
Yennie-Frautschi-Suura (YSF) proposal (generalizing previous model
calculations by Bloch and Nordsiek) introduces an ad hoc infrared
regularization $\varepsilon$ in terms of which the scattering amplitudes
involving charged particles are logarithmically divergent for $\varepsilon
\rightarrow0.$ The leading logarithmic divergencies are then summed up to a
coupling-dependent power behavior containg factors $\lambda^{f(g)}$ which
vanishes for $\lambda\rightarrow0.$ The vanishing of the scattering amplitude
shows that the LSZ scattering theory is not the correct concept for obtaining
nontrivial scattering information for "infraparticles"; in fact the presence
of milder cut-type singularities which replace the one-particle mass shell
poles confirm that such milder singularities cannot counteract the large-time
dissipation of wave packets in the LSZ time-dependent scattering theory so
that one obtains zero for $t\rightarrow\infty.~$Low order perturbative
calculations also show that the vanishing can be prevented by passing from
scattering amplitudes to photon inclusive cross sections before letting
$\lambda\rightarrow0.~$This limit constructions should be viewed as a
perturbative analog of recent more abstract represenrarional proposals to
describe charged states \cite{B} \cite{B-R}.

Although both QED and Y-M gluons couplings lead to stringlocal fields without
singular pointlike counterparts, their mathematical structure and physical
manifestations are very different. \textit{Interacting vectorpotentials in QED
are integrals over pointlike observable zero mass field strength whereas this
property is lost in massless Y-M interactions}. This implies in particular
that massless gluon strings cannot be approximated by local observables. Such
objects are inherently nonlocal in an irreducible sense. This severe
nonlocality cannot occur in $s<1~$models, even global objects as charges
(integrals over pointlike currents) can always be approximated by compact
localized matter. The emergence of Inherently noncompact fields from
collisions of ordinary matter would create havoc with causality; this only can
be avoided if they remain virtual objects whose use is necessary in order to
formulate the interaction density but disappear in the correlation functions.
Confinement in the sense of vanishing correlation functions which contain in
addition to pointlike composits also irreducible stringlike gluons solves this
problem in a radical way\footnote{Only quark-antiquark pairs separated by a
finite string can avoid confinement since their compact nature avoids the
problem cause by noncompact matter.};

The definition of interacting zero mass vectormesons as limits of their much
simpler massive counterparts in terms of their correlation functions (from
which one may reconstruct the operator formulation) accounts for the fact that
the limit represents an inequivalent representation in which the Wigner-Fock
structure of the Hilbert space is lost. Structures which are expected to be
independent of the mass, as the Callen-Symanzik beta-function $\beta(g)$,
should be computed in the massive case since a direct perturbative derivation
of the Callen-Symanzik is not possible due to the presence of infrared
divergencies. A derivation of the C-S relations for stringlocal and hence
renormalizable massive vectormesons should be possible and provide a proof
(and not only a consistency argument based on additional assumptions outside
mathematical control\footnote{The assumption that in deriving the
callan-Symanzik equation one can separate low from high momenta.}) of
asymptotic freedom and in this way close that old but unfinnished subject.

The above confinement scenario presents an interesting contrast to another
kind of stringlocal matter: the QFT of Wigner's zero mass "infinite spin"
positive energy representation class. Actually the understanding of the
importance of string-localization for the conceptual progress of QFT started
with a paper \cite{MSY} in wwhich the main point was the presentation of the
QFT behind this mysterious 1939 Wigner representation. As a positive energy
representation it shares properties as the stability of matter and coupling to
the gravitational field with the massive and massless finite helicity
representations. It turns out that the infinite spin Wigner representations
contains no pointlike covariant wave functions at all and there are convincing
arguments that the associated net of local algebras admits no compact
localized subalgebras generated by composite pointlike fields; such
representations describe noncompact matter par excellence.

Whereas gluon or quark matter cannot emerge from collisions of normal matter
(which interacts in a compact region), Wigner's noncompact free infinite spin
matter, once it got inside our universe, cannot be registered in earthly
particle counters. In fact it is totally inert apart from gravitational
manifestations \cite{dark}. This means that the presence of such inherent
noncompact matter would change the gravitational balance of normal matter in a
galaxy. When Weinberg wrote his book on QFT he rejected the infinite spin
matter because "nature does not make use of it"; at that time its strange
noncompact localization properties were not yet known, apart from the fact
that all attempts to describe this matter in terms of pointlike covariant
fields had failed. Although its property of eluding registration in particle
counters would still cause stomachaches with high energy physicists, it seem
that astrophysicists should like such inert matter whose only arena of action
are galaxies.

It may be helpful for the reader to use again Galileo's method of codification
in terms of a dialog between Sagredo and Simplicio.

\textbf{Sagredo}: Dear friend Simplicio, are you still claiming that the Higgs
mechanism is only a metaphor for the coupling of Hermitian (chargeless) scalar
fields to massive vectorpotentials i.e. the neutral analog of the massive
scalar QED? And would this mean that the mass of the massive vectormeson and
the Hermitian Higgs field in the simplest (abelian) coupling does not
originate from a spontaneous symmetry breaking of the scalar two-parametric
QED\footnote{Different from spinor QED which only has one coupling parameter,
the application of the standard pointlike renormalization formalism to a
scalar gauge coupling leads to an additional quadrilinear selfcoupling of the
matter field.} in terms of a "field shift" (the "gauge-breaking" defining
Mexican hat potential) ? Is the picture of a distinguished particle whose
interaction does not only create the mass of the vectormeson but also its own
mass (often referred to as the self-creating "God particle") inconsistent with
the principles of QFT?

\textbf{Simplicio}: This is more or less my point of view, but I would suggest
to look at the present situation in a historical context. History is more
lenient, in particular it explains how the protagonists of the "Higgs
mechanism" were led to their ideas within the prevalent Zeitgeist which
dominated the post QED particle theory. For a long time (and to a certain
extend even nowadays) its was believed that interactions of zero mass
vectormesons are simpler than those involving their massive counterpart
(spinor or scalar QED) and that therefore one should try to understand the
massive interaction by starting from massless models and think about ideas of
how to generate masses.

\textbf{Sagredo}: But isn't this true, are massless propagators and their use
in Feynman loop integrations not much simpler than integration involving
massive propagators; and above all isn't "massive QED" nonrenormalizable
because any coupling of a massive vectorpotential (a Proca field) would lead
to power-counting violating interactions of short distance scale dimensions
$d_{int}>4$?

\textbf{Simplicio}: Not quite, at least if you recall what QFT is about,
namely to understand particle theory in terms of the foundational localization
principle of QFT. The most basic structure of quantum physics is the Hilbert
space positivity and this is violated in both cases. In QED the violation
enters explicitly through the use of pointlike massless vectorpotentials which
only exist in indefinite metric Krein spaces, and its massive counterpart is
nonrenormalizable as a result of the $d=2~$Proca potential and only becomes
formally renormalizable by the use of a $d=1~$potential (together with a
negative metric scalar St\"{u}ckelberg field) in Krein space. The required
formalism is the BRST gauge setting which is somewhat more elaborate than the
QED Gupta Bleuler formalism. In both cases the physics is reduced to the
vacuum sector which the gauge-invariant observables generate by acting on the
vacuum state. The restrictive nature of the quantum gauge formalism (i.e. its
limitation to the vacuum sector) is shared between the massive and massless
case, but the infrared problems from massless vectormesons come on top of
these limitations.

\textbf{Sagredo}: Yet people compute scattering amplitudes, which certainly
cannot be obtained within the vacuum sector of gauge theories. How can one
understand this?

\textbf{Simplicio}; This is indeed a sore point of quantum gauge theory which
has no analog in classical gauge theory. Strictly speaking the S-matrix should
be computed from the LSZ limit of fields, but there is a formalism which goes
back to Bogoliubov which represents the $n^{th~}$order S-Matrix in terms of
formal spacetime integrals over time-ordered products of the (first order)
interaction density. The Krein space gauge setting uses this formalism within
the BRST operator formulation and claims that the BRST condition $sS=0$ in
terms of the nilpotent BRST s-operation insures that the resulting S lives in
the Wigner-Fock space of physical particles. But this requirement cannot be
formulated within the vacuum sector of the local observables, so the
conceptual clarity remains less than perfect.

\textbf{Sagredo}: All these problems arise because one tried to resolve the
conceptual clash between $s\geq1$ pointlike interactions and Hilbert space
positivity at the expense of the Hilbert space in favor of keeping the
pointlike field formalism for tensor potentials. Can one not take the other
direction by letting the Hilbert space positivity decide which is the tightest
covariant localization consistent with it?

\textbf{Simplicio}: Yes one can, provided one is prepared to make a new
conceptual investment of the same caliber as that which led from the old
(Wentzel, Heitler) noncovariant perturbation theory to that of post wwII
covariant QED which included vacuum polarization (loop contributions). It
turns out that the clear answer to your question is to use covariant
stringlocal fields localized on spacelike lines $x+\mathbb{R}_{+}e.$ But this
is much easier said than done, it amounts to a nearly revolutionary change of
QFT of almost all of its perturbative aspects except of its causal
localization principle which becomes strengthened. This is not surprising
because people did not opt for gauge theory because they were unaware of the
physical importance of Hilbert space positivity but rather as a result of lack
of apparent alternatives. There were hints in what direction to look at by
Mandelstam and DeWitt but they consisted in the restriction of the formalism
to field and missed the short distance improving stringlocal potentials.
Others observed that the axial gauge is, together with the noncovariant
Coulomb gauge, consistent with Hilbert space positivity, but failed to treat
the $e$-variable as a fluctuating spacetime variable by assigning to it the
role of a gauge parameter which is the same in all fields. This
misunderstanding caused serious renormalization problems of short- mixed with
long- distances which finally led to the abandonment of this gauge.

The correct understanding came in a roundabout way from the solution of the
localization problem related to the infinite spin Wigner representation by
methods of modular localization \cite{BGL} \cite{MSY}. The related free field
theory turned out to describe noncompact localized "stuff": not only
potentials but all covariant fields are stringlocal. From here arose the idea
that all massless $s\geq1$ free potentials are covariant relatives in the
Hilbert space of the noncovariant Coulomb (radiation) representations. Though
the price to pay in terms of localization is surprisingly little since the
smallest causally closed noncompact localization region is an arbitrarily
narrow spacelike cone whose core is a semi-infinite string, to deal with the
renormalization theory in Hilbert space of stringlocal fields with independent
directional fluctuations is a quite unaccustomed new problem.

Physicists of the older generation as the principle protagonist of the BRST
formulation Raymond Stora knew perfectly that gauge theory is only a
placeholder for a still (up to recently) unknown Hilbert space formulation.
They certainly would have been surprised if the implementation of the
extremely restrictive Hilbert space positivity does not lead to new insights
outside the range of gauge theory (a different view of the "Higgs mechanism",
the ability of Y-M interactions to exist without the classical fibre-bundle
"crutches", a deeper and more specificic understanding of what hides behind
infrared divergencies as the confinement problem).

\textbf{Sagredo}: Are their any new physical concepts which have no
counterpart in the pointlike setting?

\textbf{Simplicio}: Yes there are several. One remarkable new aspect is the
appearance of "escorts" of stringlocal massive vectorpotentials; these are
stringlocal scalar Hermitian fields $\phi$ (one for each massive vectormeson).
Its name refers to the fact that (unlike a $H$-field) it has the same mass and
the same coupling strength as the vectormeson, but (also unlike the $H$) it
does not add new degrees of freedom to those which are already contained in
the stringlocal vectormeson i.e. it is a kind of "massive gluonium field". In
the abelian Higgs model, whose physical content in the Hilbert space SLF
description is just the unique renormalizable $A\cdot AH~$coupling, this first
order interaction density induces a second order Mexican hat like potential in
$H$ and $\phi.~$Naturally the numerical coefficients depend only on the ratio
of the two masses of the massive $A$ and the $H.\ $The new Hilbert space SLF
setting turns the Higgs mechanism from its hat to its feet: instead of
spontaneously creating masses with the help of a symmetry-breaking Mexican hat
potential the renormalizable interaction between massive $A$ and $H~$fields
associated with a string-independent S-matrix \emph{induces} a second order
Mexican hat potential. Its form depends on the masses of the defining free
fields and the requirement that the Hilbert space S-matrix should be
independent of the $e^{\prime}s~$( i.e. it should be a kind of global
counterpart of the local observables).

Whereas the shift in field space in Goldstone's s model is a quasiclassical
device whose physical aim is to prepare a situation in which a conserved
current is prevented from leading to a global charge, its application to the
gauge-dependent scalar field of QED is not supported by any physical idea.
Quantum gauge "symmetry" (misleadingly also called "local symmetry) is,
contrary to classical theory in which Hilbert space positivity is not an
issue, not a physical quantum symmetry but rather a prescription how to
extract a vacuum representation of local observables from an unphysical
setting in Krein space. Renormalized perturbative QFT is also not able to
incorporate metaphors about the origin of masses. The hope that by formally
breaking a symmetry one may save parameters (as compared to using the physical
masses in the definition of the model-defining first order interction) only
lead to frustration upon realizing that such formal devices do not reduce the
number of parameters of electroweak interactions; all ways of breaking
correspond to all possibilities of choosing masses. One avoids all physically
meaningless manipulations by studying directly renormalizable couplings of
Hermitian $H$ fields to massive vectormesons in the BRST gauge setting instead
of breaking the latter for couplings of complex scalar to massless
vectormesons. In this way one realizes that its characteristic intrinsic
physical property is the Schwinger-Swieca-Higgs charge screening and not the
ssb Higgs mechanism.

\textbf{Sagredo}: If it is that simple as you presented it, namely a kind of
neutral counterpart of the Maxwell theory of charged matter, why was such a
coupling not studied before Higgs?

\textbf{Simplicio}: Thinking in terms of quantizing Maxwell fields coupled to
charge-carrying quantum-matter the generalization to massive vectormesons
appears natural; but the idea of coupling neutral (Hermitian) $H$-fields is
quite removed from Lagrangian quantization of classical fields\footnote{More
precisely of classical fields obtained by reading quantum fields (Dirac
spinor,..) and their quantum symmetries back into the classical realm.}, in
particular since such "chargeless" interactions are only possible with massive
vectormesons and disappear for $m\rightarrow0$ i.e. have no counterparts in
classical electromagnetism. The first indications of what may be different
with massive vectormesons came from Schwinger \cite{Schw} who suggested that
in such a case the charge is "screened" (vanishes); as a model which only
exists in the screening phase he proposed the d=1+1 rigorously solvable
"Schwinger model" \cite{Swieca}. In a subsequent structural (nonperturbative)
proof of charge-screening by Swieca \cite{Sw} it became clear that in
couplings of massive vectormesons to complex matter there are two conserved
currents namely the identically conserved Maxwell current from the divergence
of the massive field strength $j_{\mu}=\partial^{\nu}F_{\mu\nu}$ and the
particle-antiparticle counting current of complex fields; they only coalesce
in the massless limit. Swieca emphasized that in case of a selfconjugate
$H$-field the Maxwell charge (the only conserved charged in the abelian Higgs
model) is screened and not spontaneously broken; for this reason he used the
terminology Schwinger-Higgs mechanism in his publications.

Unfortunately the Higgs mechanism of spontaneous mass generation was proposed
by several authors at the same time with identical computational recipes
involving field shifts in the gauge-variant complex field of scalar QED so
that the shared conceptual error was protected by the "many people cannot err"
dictum. Swieca's scientifically successful but sociologically futile attempts
may serve as an illustration that there was a well-founded early scientific
criticism of these ideas, but the beginnings of a correct understanding were
finally lost in the maelstrom of time. After Glashow, Weinberg and Salam
supported the spontaneous symmetry breaking the Higgs issue became sealed and
the chance for a correct understanding within QFT evaporated.

\textbf{Sagredo}: Even if the "Higgs symmetry breaking" is only a metaphor for
a coupling of a Hermitian field to a massive vectormeson as you claim,
couldn't it survive as a mnemonic trick which at the end more or less decribes
what you want ? If a metaphoric idea leads to results which later on finds a
derivation for which every step is consistent with the principles of QFT,
\textit{isn't it justified to credit the discoveres}? After all we attribute
the discovery of antiparticles to Dirac even though his hole theory was later
recognized as being incorrect.

\textbf{Simplicio}: This is an important point, and yes they do deserve
recognition as in many other cases besides Dirac. In a science about
foundational properties of matter, the frontiers are often in a very
speculative state and discoveries via metaphors are helpful as placeholders
for a later understanding. But at the times of Pauli, Feynman, Lehmann,
Landau, Kallen, Schwinger, 't Hooft, Veltman, Jost,...none of the many less
than correct proposals which resulted from conceptual misunderstandings had
the chance to survive for more than a decade (SU(6), peratization,...). The
valuable discoveries, as those of renormalized perturbation of nonabelian
gauge theories, went through many refinements; starting with 't Hooft and
Veltman, passing through Faddeev-Poppov, Slavnov and reaching the level of
technical maturity in the BRST formalism before the recent proposal of the use
of stringlocal potentials in a Hilbert space setting (which is still very much
in its infancy) again returned to it.

There are only two exceptions to the continuous unfolding of a discovery: the
5 decades old discovery of String Theory (which has neither observational nor
theoretical credentials) and the Higgs mechanism which is the only
experimentally successful discovery which managed to survive for more than 4
decades without any theoretical modification . The (often well-founded) early
critique was unable to penetrate the thick protecive sociological layer of
approval and finally vanished in the maelstrom of time. Present Big Science
and a Nobel prize guaranty that the issue will remain protected against
scientific critique.

To be more concrete, QFT is a foundational theory based on the quantum
adaptation of causal localization. Its perturbative implementation in the most
accepted (Bogoiubov) formulation is based on interaction-defining free fields
and a first order interaction density. Hence the definition of a model
includes the masses of these fields; renormalization theory insures that these
masses are identical to the masses of the observed particles which are
considered elementary within that model. QFT is \textit{not a theory which can
say anything about masses of the defining fields}. renormalized perturbation
theory in its present stage is not able to say something about bound states,
for this task one presently uses lattice approximations. In case of the
massive vectormeson-$H$ coupling one only needs to write down the first order
$AAH~$coupling, the rest (which include the second order "Mexican hat" shaped
$H$-selfinteractions) are \textit{induced} by the powerful BRST gauge
conditions (the $s$-invariance of the S-matrix) or by the even more powerful
Hilbert space positivity condition in the new SLF setting (the differential
calculus implementing string independence).

All the numerical aspects of the induced second order potential are fixed in
terms of the masses of the model-defining free fields there is no place for
gauge symmetry breaking field shifts. By ignoring the gauge aspects of scalar
QED one can of course envisage a quasiclassical picture of how to attain such
a $H$-coupling, but when setting up the renormalized perturbation theory
involving a massive vectormeson one has to liberate oneself from such pictures
and follow the BRST rules of gauge theory starting with the $gAAH$ coupling of
massive vectormesons to Hermitian scalar fields and let the BRST operator
formalism do its job (the implementation of $sS=0$) \cite{Scharf}. Otherwise
one may overlook the fact that the Mexican hat potential is induced in order
$g^{2}$ of the gauge-invariant S-matrix and depends on the two masses. $~$Of
course one may re-construct from these two data the strength of a quadrilinear
selfcoupling of an imagined scalar QED and the gauge breaking shift paramter
in field space (also 2 parameters), but why does anybody want to construct
something, which is it best a quasiclassical image, if the first order is
already prepard for renormalized perturbation theory in the BRST setting.

This is totally different from setting up perturbation theory of a Goldstone
spontaneous symmetry breaking. Here the starting point is the intrinsic
definition of spontaneous symmetry breaking. Contrary to popular opinion it is
not the shift in field space but the physical (observable) attribute of a a
conserved current whose charge (the would-be symmetry generator) diverges. A
structural theorem \cite{E-S} says that this can only occur in the presence of
a zero mass Goldstone particle which destroys the convergence of the charge
for large distances\footnote{His Cargese lecture notes \cite{Car} on this
topic are highly recommended since they reveal the clarity and depth on which
these issues were once understood. His profound knowledge about spontaneous
symmetry-breaking led him some years later to the "Schwinger-Higgs charge
screening mechanism" \cite{Sw}.}.

The task is therefore to find renormalizable couplings in which zero mass and
massive fields interact in a way which leads to a conserved "Goldstone
current". A convenient way to achieve this is to use the quasiclassical device
of a field shift which itself is not a physical parameter. This current is
very different from that associated to a massive vectormesons which is the
only current of the abelian $H$-model. The Maxwell current of a massive
vectormeson (the only current of the $H$-model) leads to "screened charge"
$Q=0,$ whereas a spontaneously broken symmetry manifests itself in terms of a
diverging charge $Q=\infty$. Nothing could be more different than that! The
unification of both phenomena under the roof of spontaneous symmetry breaking
is a conceptual misunderstanding.

Interacting massive gauge theories exist with arbitrary masses and in the
Hilbert space setting each massive vectorpotential is accompanied by its
scalar stringlocal escort $\phi$ which carries the same mass. For equal
vectormeson masses (in particular for zero masses) the induction mechanism
imposes a Lie-structure on the self-couplings of vectormesons; solutions with
completely independent couplings mut equal masses would violate the Hilbert
space positivity. $H$-fields may be coupled in addition, but their presence is
not necessary for "fattening" massless gluons and $H$-fields. Presently no
\textit{theoretical }reason for their existence is known; the LHC results may
possibly also be consistent with the existence of a foundational gluonim bound
state of massive vectormesons. Further studies are necessary.

Interacting zero mass vectormesons (QED, QCD) are outside the range of the
standard field -particle setting in a Wigner-Fock Hilbert space. In that case
one need to go the round-about way of computing appropriately
infrared-renormalized correlation functions of stringlocal vectormesons in the
massless limit and then reconstruct the operators using Wightman's
reconstruction theorem. One expects that the characterizing property of
confinement will be the vanishing of correlation functions containing in
addition to local observables also self-interacting massless stringlocal gluon
and quark fields.

\textbf{Sagredo}: In your new setting the stringlocal fields $\Psi(x,e)$ are
renormalizable in the standard sense of the power-counting criterion, which in
particular means that they are localizable in the sense of Wightman's
testfunction smearing with Schwartz $\mathfrak{D}~$functions in the
($x,e$)~variables. On the other hand you claim that your new stringlocal
renormalization theory also allows to construct associated singular pointlike
fields whose short distance scaling degree is unbounded (increasing with
perturbative order) which explains their pointlike nonrenormalizability in
terms of their worsened localizations. Do these singular pointlike fields play
any useful role?

\textbf{Simplicio}: Most of the intuition which comes with the definition of
the model in terms of a pointlike massive field interaction is preserved; in
fact the associated renormalizable stringlike interaction is obtained by
"peeling-off surface terms" which carry the leading short distance
singularities and hence do not contribute in the adiabatic S-matrix limit.
This has the interesting consequence that the high-energy behavior of
scattering amplitudes is better than that which one naively reads off in
momentum space by going in a simple-minded way to the mass shell from the
Fourier transformed pointlike correlation functions. Hence phenomenological
arguments in favor of the presence of Higgs particles based on the use of
Feynman diagrams are not supported; the perturbative content of $s\geq1$
stringlocal interactions cannot simply be encoded into Feynman diagrams
(including contributions from counterterms). The intuitive appeal of pointlike
couplings is however completely lost in the massless limit; in that limit the
singular pointlike fields disappear and the stringlike localization in QED
becomes more stiff since different $e$-directions of the localization lines
along which infrared photons "hover" cease to be unitarily equivalent
(spontaneous breakdown of Lorentz covariance in electrically charged sectors);
this is the regime in which the standard field-particle relation is lost.

QFT is presently undergoing significant changes. There are several forthcoming
papers which promise to clarify the mathematical problems coming from causal
string crossings in addition to the already existing Epstein-Glaser
renormalization theory for point-crossings. The development of these new ideas
will be slow because a lot of foundational knowledge about QFT has been lost.
$~$

\textbf{Sagredo}: I thank you dear friend for sharing your thoughts, and I
hope that your pessimistic assessment about particle theory in the shadow of
Big Science remains a warning and does not become a prediction about its
future. It will take some time to fully comprehend what you told me; lots of
important issues to think about lie before me before I will meet you again.//

\section{The dual model, misunderstandings about particle crossing}

The\textit{ }idea to avoid the use of singular fields, which led to the
problem of ultraviolet divergencies, and instead formulate particle physics in
terms of the S-matrix goes back to Heisenberg. It was abandoned soon
afterwards when the success of renormalized perturbation theory in QED left no
doubts that the conclusion of inconsistency of QFT based on those divergencies
was premature. The problem which perturbative methods had with strong
interactions led to adaptation of the Kramers-Kronig dispersion relations to
particle physics. It was modest in scope\footnote{Its main aim was to make
sure that the causal localitity principle of QFT continues to be valid at the
energies of the newly emerging High Energy Physics.} but after a decade it
came to closure by achieving all its objectives (the only project in particle
theory which came to a successful closure) which included the support of the
validity of the locality principle in the at that time new high energy region.

This success encouraged several theoreticians to formulate a new constructive
S-matrix setting in which the perturbative analytic particle crossing property
for the S-matrix (and later formfactors) played the important role. Together
with unitarity and Poincar\'{e} invariance it became known as the "S-matrix
bootstrap" but it was soon abandoned as a result of the unmanageable nonlinear
problems arising from simultaneously implementing these three properties "by
hand". Without any demonstrable success it nevertheless enjoyed a lot of
support even by people who on different topics had been quite critical as e.g.
Freeman Dyson. A related problem was the insufficient understanding of the
conceptual origin of particle crossing; its derivation from the locality
principle for some very special scattering amplitudes did not lead to
sufficient insights, and the prohibitively difficult method of analytic
functions \cite{BEG} of several complex variables led to an early end of these attempts.

Another attempt to obtain a constructive computational access to particle
theory in terms of an on-shell project based on S-matrix properties was
formulated by Mandelstam \cite{Mandel}. In analogy to the successful use of
the Jost-Lehmann-Dyson spectral representation which led to a rigorous proof
of dispersion relation, Mandelstam postulated the validity of a double
spectral representation for the elastic scattering amplitude as a starting
point for getting access to analytic on-shell properties, including the
crossing property.

The era of genuine misunderstanding of particle crossing started with
Veneziano's \cite{Vec} construction (based on properties Euler's beta
function) of a meromorphic function of two variables which had an infinity of
first order poles in the two variables which were related by an analytic
crossing relation. Although his presentation did not contain any physical
argument why this mathematically constructed function which is meromorphic in
variables which he identified with the invariant s,t,u variables (the
"Mandelstam variables") should be related with the elastic part of a
scattering amplitude, his construction created a lot of excitement within
which a critical attitude had little chance. Apparently the results on
integrable models, which could have revealed that although scattering
amplitudes can be meromorphic in the rapidity variables but not in the
Mandelstam variables, were not known to the dual model community.

Instead of speculating about what went on the mind of peoples who excepted
Veneziano's use of the dual model meromorphic function as an approximation of
an elastic scattering amplitude (to be improved by "unitarization"), it is
much easier to understand what kind of quantum field theoretic idea leads
precisely to such dual model function. This clarification is due to Mack
\cite{Mack}, and his construction is here referred to as the "Mack-machine";
this name is chosen because it cannot only produce Veneziano's dual model and
similar dual models constructed later, but in a certain sense it contains all
dual models (all crossing symmetric \textit{meromorphic}
functions\footnote{Even in the simplest context of integrable models elastic
crossing symmetric scattering amplitud are not meromorphic in the Mandelstam
variables (but rather in the exponential "rapidity variables")} in s,t,u).

The construction uses conformal global operator expansions for pairs of
operators which, in contrast to the Wilson-Zimmermann short distance
expansions, are known to converge%

\begin{align}
&  A(x)B(y)\Omega=\sum_{k}\int d^{4}z\Delta_{A,B_{.},C_{k}}(x,y,z)C_{k}%
(z)\Omega\\
&  \left\langle A_{1}(x_{1})A_{2}(x_{2})A_{3}(x_{3})A_{4}(x_{4})\right\rangle
\rightarrow3\text{ }different\text{ }expansions \label{3}%
\end{align}
and applies them to all pairings inside the 4-point function (second line).
Each pair of operators has a converging expansion on the vacuum in which the
resulting operators $C_{k}$ stand for a list of composites which can be
connected with the given pair through nonvanishing conformal 3-point functions
$\Delta.$ Used inside the 4-point function, this leads to three different ways
of decomposing the 4-point function into a sum over two three-point functions
multiplicatively connected by an integrattion over the $z$-variables. Mack
showed that the Mellin transform of this infinite sum over operators $C$ leads
precisely to the pole representation of the meromorphic functions which define
dual models; the position of the first order poles is given in terms of the
spectrum of scale dimensions of those $C^{\prime}s$ which couple to the $A,B$
pairs. Veneziano's model corresponds to a certain chiral conformal model, but
any conformal 4 point function in any spacetime dimension upon expansion of
its 4-point function and Mellin transforming the resulting series always leads
to a dual model in the sense of defining a meromorphic function with first
order poles which fulfills a crossing relation. The set of contributung poles
is (up to a shared factor) a subset of the anomalous dimension spectrum of the
conformal theory. What initially looked magic and unique\footnote{The
uniqueness, which was already expected to be follow from the bootstrap
principles, was a precursor of the reductionist idea of a theory of everything
(TOE) which originated in connection with ST.} in Veneziano's, is now
"mass-produced" by the Mack-machine; demystified in this way it makes no sense
to identify the dual model with scattering amplitudes.

A scattering functions cannot be meromorphic in the Mandelstam variables but,
under special circumstances (integrability) it is meromorphic in the rapidity
variables. Conformal theories are interesting quantum field theories from
which one can learn a lot about the inner workings of the modular localization
properties, but they certainly contain no information about scattering of
particles; in fact \textit{interacting conformal models contain no particles
at all}, they are rather theories of anomalous scale dimensions which live on
a covering of the compactified Minkowski space. Mellin transforms of their
4-point functions may be called dual models, but this has no bearing on
interactions between particles. It does not make sense to apply ideas of
unitarization to them as if they would define a kind of nonunitary
approximation of an S-matrix.

This could have been the end of a misunderstanding and led to the closure of
this unfortunate chapter of misguided particle physics. In fact it probably
would have been the end if not an even stranger twist would have greatly
increased the mysterious aspects and with it the attractiveness of ST. This
consisted in the observation that the oscillator algebra resulting from the
Fourier decomposition of a certain chiral 10-component conformal current
algebra formally related to supersymmetric version of the Polyakov action%

\begin{align}
&  \int d\sigma d\tau\sum_{\xi=\sigma,\tau}\partial_{\xi}X_{\mu}(\sigma
,\tau)g^{\mu\nu}\partial^{\xi}X_{\mu}(\sigma,\tau),~\sigma,\tau=t\pm
x\label{pol}\\
&  X=potential\ of\ conformaL\ current\text{ }j\nonumber
\end{align}
\textit{permits a positive energy representation of the Poincar\'{e} group}
which decomposes into a discrete infinite sum of irreducible representation
(an infinite $(m,s)$ "tower"). This action is conveniently formulated on the
oscillator variables obtained by Fourier transformations in the standard
circular compactification of conformal theories.

The construction of such a tower (an infinite component field fields) from an
\textit{irreducible algebraic structure} was Majorana's project which he
formulated in 1932 with the idea to achieve something similar to what the
$O(4,2)$ group representation theory does for the hydrogen atom spectrum in
QM. This project was revived in the 60s when it acquired some popularity under
the name "dynamic infinite group representation project" (Fronsdal, Barut,
Kleinert,..\cite{Tod}). Majorana's project as well as its later revival
restricted this search to irreducible representations of extensions of the
Lorentz group. The only known solution up to date is the representation
\textit{on the irreducible oscillator algebra of the supersymmetric 10
component current algebra,} the so-called superstring representation of the
Poincar\'{e} group. This is a group theoretic fact which, although discovered
by string theorists, has no relation to Mandelstam S-matrix based on-shell project.

To understand in a more generic way the prerequisites one need to encounter
the representation of a noncompact group as a kind of internal symmetry group
on the component space of a multicomponent chiral conformal algebra, it is
helpful to be reminded of same basic fact of LQP in which inner symmetries
arise from the local net of observable algebras in the vacuum representation.
The inequivalent local representation classes (superselection sectors) can in
typical cases be combined with the vacuum representation within a larger
\textit{field algebra net~}\cite{Haag}. There are convincing arguments why a
continuous set of superselection sectors (in the presence of zero mass
particles as QED one must pass to charge-classes \cite{B}) and noncompact
internal symmetries of the field algebras cannot occur in higher than two
dimensions. The superselection analysis is very different in d=1+1 dimensions
and such cases can occur; in fact the abelian chiral current models are examples.

As an illustration let us look at a n-component current algebra%

\begin{align}
&  \Phi_{k}(x)=\int_{-\infty}^{x}j_{k}(x),~~~\left\langle j_{k}(x)j_{L}%
(x^{\prime})\right\rangle \sim\delta_{k,L}\left(  x-x^{\prime}-i\varepsilon
\right)  ^{-2}\label{con}\\
&  \Psi(x,\vec{q})=~:e^{i\vec{q}\vec{\Phi}(x)}:~carries~scale\text{
}dimens.~d(q)\sim q^{2}\nonumber\\
&  suggests~analogy~q\sim p,~~~\vec{q}\cdot\vec{q}\sim p_{\mu}p^{\mu}%
,~d_{sd},\sim m^{2}\nonumber
\end{align}
Here we have substituted the somewhat confusing letter $X~$(\ref{pol}) in
favor of $\Phi$ for the multicomponent current potential because we want to
avoid a notation which may suggest the wrong idea of an operator which embeds
a chiral conformal theory on a lightray (or on its compactified circle) into a
n-dimensional Minkowski spacetime so that its development in time it looks
like a 2-dimensional surface (a tube, in case of a chiral theory on a circle).
This picture of a covariant string generating a spacetime tube-like
world-sheet is incorrect inasmuch as it is incorrect to think that the
classical covariant particle Lagrangian $\sqrt{ds^{2}}$ leads to a covariant
quantum embedding described in terms of a covariant operator $x_{\mu}%
^{op}(\tau).~$In$\ $fact Lagrangian quantization is the wrong guide; there
simply exists \textit{no covariant position operator} whose spectral
projectors fulfill the requirements of covariant localization. Wigner was well
aware of this limitation when he constructed relativistic particles by
representation theory and not by quantization.

In the book on string theory by Polchinski he used this classical relativistic
particle Lagrangian as a "trailer" for presenting a relativistic quantum
theory of strings based on the Nambu-Goto action which replaces the $ds^{2}$
under the square root by the corresponding covariant surface differential.
Hence instead of being helpful, this analogy turns out to be a squid load. The
quantization of the Nambu-Goto Lagrangian according to the correct rules for
quantization in the presence of re-parametrization invariance resembles that
of quantizing the Einstein-Hilbert action; It is certainly non-renormalizable
and has no natural relation to the Poincar\'{e} group which acts on the
embedding Minkowski spacetime \cite{Bahns}. There is another approach to the
square root N-G Lagrangian which is due to Pohlmeyer \cite{Pohl}; it is based
on the observation that the classical system is integrable. So instead of
confronting the problem of quantization of reparametrization-invariant actions
which inevitably leads to renormalization problems, he proposes to quantize
the Poisson relations between the infinitely many conserved "charges". The
problem with this quantization is that one looses the connection with
localization in spacetime and Poincar\'{e} covariance.

On the other hand the Polyakov Lagrangian has a direct relation to chiral
conformal QFT, so one believes to be on conceptually safe grounds. Here the
problem is that the representation of the irreducible oscillator algebra
behind the operator formalism (\ref{con}) which serves for the representation
of the Poincar\'{e} group (and the ensuing intrinsic localization concept
which comes with positive energy representation of the Poincar\'{e} group
\cite{BGL}) is not the same as the one which localizes the chiral model on the
lightray. With other words the Hilbert space representations of the oscillator
algebra are not equivalent. The charge spectrum of the chiral theory is the
whole $\mathbb{R}^{n}$ and the sigma-model fields $\Psi$ in (\ref{con}) are
the charge carriers. On the other hand the spectrum of the representation of
the Poincar\'{e} group is contained in the forward light cone and has mass
gaps. The the spectrum of the zero mode multicomponent charge operator covers
the full spectrum of the charge superselection structure. The treacherous
nature of the analogy between the mass spectrum and the conformal dimensional
spectrum in (\ref{con}) has been overlooked by string theorists.

These analogies become even more seductive if one realizes that a particular
discrete particle representation of the Poincar\'{e} group (the superstring
representation) does appears on the oscillator algebra of a 10 component
supersymmetric current model (unique up to a finite discrete "M-theoretic"
variation). But what has this group theoretic coincidence between a spectrum
of a discrete Poincar\'{e} group representation on the oscillator algebra of a
supersymmetric 10-component abelian current to do with Mandelstam's S-matrix
project? The answer is nothing beyond the appearance of crossing symmetric
analytic functions. Nevertheless the group theoretic content of this relation
is interesting from a historial viewpoint because it is the only known
solution of the 1932 Majorana project to find an irreducible algebra which
carries a purely discrete representation of the Poincar\'{e} group.

In distinction to the string-localization of matter fields interacting with
vectorpotentials in previous section, the representations occurring in the
superstring representation are pointlike generated. This was precisely what
the calculations of the (graded) spacelike commutator of the putative
string-fields by string-theorists in the 90s showed \cite{Martinec}%
\cite{Lowe}. The situation is somewhat confusing as a result of the fact that
the distribution representing the infinite component quantum field is
extremely singular since the localization points of all pointlike components
fall on top of each other. It is an interesting historical question why the
string community agreed with the authors that the localization is stringlike
(a point on an invisible string?). Looking back with some hinsight, the dual
model and string theory are certainly the most curious results from an epoch
in which conceptually unguided calculations combined with sophisticated
mathematics was expected to lead to a unified theory of everything (a TOE).
Historians of science will have a lot of problems to understand the related
Zeitgeist, but the almost 50 years lasting popularity (longer than the
phlogiston theory) will leave them no choice but to try to explain to a
curious public what really went on in the minds of people.

\section{Localization and phase-space degrees of freedom}

In a course on QM one learns that the number of "degrees of freedom" (quantum
states) per unit cell of phase space is finite. Already in the beginning of
the 60s it became clear that this not compatible with the causal localization
in QFT. The first computation revealed that the infinity is not worse than
that of a compact set \cite{H-Sw} which in later work of Buchholz and Wichmann
became sharpened to the cardinality of a \textit{nuclear set} \cite{Haag};
together with modular localization theory it led to the important concept of
modular nuclearity \cite{Haag}.

The physical motivation of these investigations is the desire to understand
the connection between field localization and the presence of particles; in
particular the circumstances under which the causal localization properties of
quantum fields lead to particles with discrete masses including the important
property of \textit{asymptotic completeness\footnote{The equality of the
Hilbert space with a Wigner-Fock particle space.}.} One remarkable result in
the more than eight decades lasting attempts to prove the existence of models
of QFT with interactions and to obtain mathematically controlled
approximations is the before mentioned existence proof for certain strictly
renormalizable integrable models. Such models are characterized in terms of
its factorizing S-matrices which permits a classification in terms of
matrix-valued 2-particle scattering functions (section 6). In that case one
knows the particle structure and one would like to find the net of local
algebras and the their generating quantum fields whose collision theory
reproduces the known particle content. The S-matrix determines the structure
of the wedge alge$\ell$bras. In order to obtain a nontrivial net of compact
localized double cone algebras one can use the aforementioned modular
nuclearity property of phase space degrees of freedom which follows from the
analytic properties of the scattering functions.

On the positive side these models have a realistic short distance behavior as
one expects it from renormalizabity, i.e. they are not superrenormalizable as
polynomial self-interactions between scalar $d_{sd}=0$ (logarithmic divergent
short distance behavior) fields in two dimensions\footnote{The d=1+1
superrenormalizable theory can still be treated within a measure-theoretic
functional quantization setting \cite{Gl-Ja}, no use of modular localization
properties is needed.}. The fact that integrability in QFT can only be
achieved in d=1+1 did not affect their usefulness as a "theoretical
laboratory" of QFT. The existence of these models can be controlled with the
help of "modular nuclearity" \cite{Lech}.

Another important use of these ideas consists in the \textit{exclusion} of
models with unphysical causality properties. Lagrangian quantization seems to
lead inevitably to divergent renormalized perturbative series, and hence it is
not suited for addressing problems of existence of models. It is however
important to maintain the formal causality properties of Lagrangian
quantization in the better mathematically controlled LQP setting of QFT.
Whereas the spacelike Einstein causality property is easily taken care of, the
relevance of the causal completion (causal shadow) property is sometimes
overlooked. One reason is that this quantum counterpart of causal propagation
cannot be formulated in terms of individual fields; its precise formulation
needs the algebraic setting as in section 4.

It is easy to write down generalized free fields which fulfill Einstein
causality but violate the causal completeness property (the local version of
the old time-slice property \cite{H-S}). A recent illustration of a violation
of this important physical property is the conformal covariant generalized
free field which results from a normal free field on a AdS spacetime through
the mathematical AdS$_{n+1}$-CFT$_{n}$ isomorphism \cite{Du-Reh}. The physical
defect of fields which violate the causal completeness property is that they
lead a "poltergeist effect" in the causal shadow region; as one "moves up"
from the spacetime region $\mathcal{O}$ into its causal completions
$\mathcal{O}^{\prime\prime}$ there are causality violating degrees of freedom
apparently coming from nowhere.

The LQP setting reveals that this physical defect is of a general nature and
may be viewed as a manifestation of the holistic nature of spacetime
localization. As the holistic nature of life needs the right amount of
chemicals, the holistic nature of causal localization in spacetime needs the
right cardinality of degrees of freedom which is appropriate for causal
localization. Starting from a physical AdS theory, one obtains an
"overpopulated" CFT model which leads to the mentioned poltergeist phenomenon.
In the opposite direction a "physical healthy" CFT passes to an "anemic" AdS
theory which does not have enough degrees of freedom which are needed for a
nontrivial realization of causality; in the case at hand one has to go to
noncompact spacetime regions in order to find at all any degrees of freedom
\cite{Reh}.

It is interesting to note that this pathology is absent in holographic
projections onto null-surfaces; unlike in isomorphic correspondences,
holographic projections dilute (loss of imbedding information) degree of
freedom by the right amount which fits the lower dimensional surface.

It is interesting to take a closer look at a special misinterpretation which
played an important role in ST. As mentioned before, the irreducible
oscillator algebra of the 10 component chiral current admits two inequivalent
representations, one which is important for the invariance under the conformal
M\"{o}bius group and the pointlike localized fields on a lightray, and the
other which carries the mentioned 10 dimensional superstring positive energy
representation of the Poincar\'{e} group. Both representations are pointlike
generated; this is a property shared by all positive energy representations
with the exception of Wigner's infinite spin representations. But there is a
huge difference in the cardinality of freedom; the oscillator representation
carries the superstring Poincar\'{e} group representation, but certainly not
the superstring field representation which is canonically associated with it
and hence it is not possible to view the one as embedded into the other. The
misplaced terminology "ST" which refers to a stringlocal object in a target
spacetime is the result of anincorrect picture.

At best this terminolgy could refer to an internal oscillator chain (after
taking out the zero mode degree of freedom) "over" a spacetime localization
point which carries the $(m,s)$ representation as well as additional operators
which are not needed for the representation of the Poincar\'{e} group, but
interlink the different levels of the (m,s) tower and in this way secure the
embedding of the reducible supersting representation of the Poincar\'{e} group
into an irreducible algebra. Such a tower of free fields piling up over one
point leads to pointlike singularities which are beyond those of ordinary
(Wightman) QFT even though each individual component is an ordinary free
field. Perhaps this could have been the reason why, despite their correct
calculation, the authors in \cite{Lowe}\cite{Martinec} presented their result
as a confirmation of stringlike spacetime localization by declaring the
localization point to be the center point on an imagined spacetime string.

As previously mentioned the embedding of lower dimensional QFTs into higher
dimensional ones and its Kaluza-Klein inverse are also inconsistent with the
holistic localization principle. Arguments based on quasiclassical
approximations or on "massaging" Lagrangians do not help on issues which
directly relate the cardinality of degrees of freedom with quantum causal
localization. Different from quantum mechanical matter the spacetime dimension
is an inseparable part of what constitutes causally localized quantum matter.
The only known exceptions are holographic projections onto null-surfaces dor
which the cardinality of degrees of freedom is thinned out in the right way
\cite{BMS}.

These insights into the connection between the cardinality of degrees of
freedom and localization immediately disproves the Maldacena conjecture which
claims that both sides of the AdS$_{5}$-CFT$_{4}$ represent \textit{physical}
theories. As a coauthors of a 1962 paper \cite{H-S} which led to the concept
of the causal completion property (which later on was related with the degree
of freedom issue \cite{Haag} it is particularly distressing to look at the
present situation in which globalized communities of particle theorists have
fallen behind previously attained levels of knowledge about important concepts
and where historical ignorance prizes; good for the recipients but bad for the
future of particle physics.

\textbf{Sagredo}: Dear Simplicio, some of our friends tell me that you claim
that the dual model and ST led to a derailment of an important part of
particle theory?

\textbf{Simplicio}: Although my attitude with respect to those attempts
concerning a "theory of everything" has been indeed very critical, I have good
reasons to avoid expressing my critique in this way. What prevents me is the
fact that I share the goals of an S-matrix-based alternative to the
quantization approach. Hence criticizing a certain unfortunate direction which
this has taken in the form of string theory should not be misunderstood as a
dismissal of the aims of the project.

After the successful closure of the dispersion relation project it seemed
natural to look for a setting in which the analytic properties derived from
the relativistic causality of QFT can be extended in such a way that they may
be used for dynamical calculations in particle physics. So instead of starting
with quantized fields and deriving properties of interacting particles
(scattering amplitudes, formfactors), why not start directly with objects
referring to particles and address the problem of whether these results can be
backed up by a more foundational QFT to a later stage. It is customary to
refer to such a particle-based construction as an "on-shell" projects and to
quantum field based approach as "off-shell" since scattering amplitudes and
formfactors are formally related to mass shell restrictions of Fourier
transforms of field correlations$.$ Different from the off-shell project of
QFT for which one will know the physical content of the model-defining field
theoretic interaction only at the end of the calculation, the on-shell
particle-based project is a "top-to-bottom" setting in which the physical
properties are laid out before one starts to work one's way down to the field
theoretic description.

The problem is of course that our conceptual/mathematical understanding works
best on the level of the foundational causal localization principles of
quantum fields, whereas it is difficult to directly convert the apparent
immediteness of observed particles with the help of analytic properties of
scattering amplitudes into concrete predictions. Whereas the foundational
properties of fields lead to analytic properties of off-shell field
correlations, it is extremely hard to extract from them on-shell analytic
properties. Even in perturbation theory where the graphical aspects of
crossing properties are obvious, the proof that there is an analytic on-shell
path which relates a scattering amplitude to its crossed counterpart is
anything but simple. Stanley Mandelstam, one of the protagonists of an
on-shell project, knew that on-shell analytic properties beyond those which
were needed for the derivation of the particle analog of the Kramers-Kronig
dispersion relations are hard to get at. His proposal of the Mandelstam double
spectral representation for the elastic scattering amplitude, was guess and
not a derivation from the causality principles. It was Venezianos guess of a
dual model and its later conversion into string theory which led to the
derailment of Mandelstam's project.

Looking back at that epoch with today's hindsight it is clear that there was
no chance for such a project to succeed at that time. An important aspect of
the S-matrix which tightens its link with the causality principle of local
quantum physics was still missing namely the fact that the S-matrix, in
addition of describing the collision of particles, is also a relative modular
invariant of the wedge algebra $\mathcal{A}(W).$ For integrable models of QFT
(a property which unfortunately is limited to d=1+1 and which forces the
S-matrix to be purely elastic) the on-shell project has a unique solution; in
this case one can really start from the classification of S-matrices and
arrive at a unique integrable QFT which is associated to that integrable
S-matrix. Even without integrability there are some ideas, but due to the
complexity of the problem there has been no significant progress.

\textbf{Sagredo}: But how was string theory related to Mandelstam's on-shell
project and what was its impact ?

\textbf{Simplicio}: Mandelstam realized that an on-shell approach to particle
theory idea must start with a profound understanding of the analytic crossing
property of scattering amplitudes of which the elastic part is the simplest.
As a starting point he postulated a two-variable representation which became
known under the name "the Mandelstam representation". Unfortunately no
crossing symmetric solution of this representation was found.

In order to understand the next step one needs to recall a bit of the spirit
of the times. When a seemingly well-defined but nonlinear problem did not
admit any solution this was sometimes taken as a hint that if the problem
admits any soiution at all, this should be rather unique. This was the view
about solutions of the nonlinear Schwinger-Dyson equations and this was not
different in case of the nonlinear bootstrap project. It may nowadys appear
naive, but the idea that Poincar\'{e} invariance, unitarity and the crossing
property lead to a unique S-matrix (a TOE apart from gravity) had a strong
spell on many people. and even prominent physicists as Freeman Dyson supported
it for some time.

When Veneziano, while playing with properties of Euler beta functions, found a
meromorphic crossing symmetric functions with an infinite family of first
order poles, there was a lot of commotion in the phenomenologically motivated
particle theory community. Veneziano's proposal to view it as a model of an
approximation (it was not unitary and had no elastic cut) to a crossing
symmetric scattering amplitude received widespread acceptance and also
Mandelstam's blessing. Nowadays we know that such functions occur in models of
conformal QFT and have no relation to scattering amplitudes. When the use of
the dual model functions in scattering theory was finally given up, the reason
was not the existence of a conceptual flaw but rather the fact that new
experimental results removed the phenomenological basis for the interest in
such models. This was the end of the Mandelstam on-shell project but not that
of the dual model formalism. The new idea was that Veneziano's mathematical
dual model observations were anyhow too sophisticated for strong interaction
phenomenology and one should find a more foundational application. This was
the birth of string theory which pushed the somewhat modified dual model
formalism from its application to strong interactions all the way up to the
Planck scale; in this way it became the millenniums TOE.

\textbf{Sagredo}: But doesn't this mean that string theory rid itself from the
impossible relation to Mandelstam's on-shell project? How does this fit in
with your belief that ST a failed theory?

\textbf{Simplicio}: The 10 dimensional free superstring is a second quantized
version of the so-called superstring representation. This is a positive energy
Wigner representation on the irreducible operator algebra associated with a
certain supersymmetric 10-component abelian chiral current algebra. One has
all the right to be surprised about the existence of such a representation
since it is the only known entirely discrete positive energy representation on
an irreducible algebra; representations on field algebras coming from QFT
inevitably have the continuous contribution from scattering theory. It is the
first and only known solution of Majorana's 1932 problem \cite{Maj} to find an
irreducible algebra which can support an infinite component \textit{discrete}
positive energy Wigner representation (an "infinite component field
equation"). This group theoretic problem was solved by the string theorists
construction of the "superstring representation" on the algebra of the
supersymmetric 10-component abelian chiral current model; this is their achievement.

\textbf{Sagrado}: But what about strings in spacetime ?

\textbf{Simplicio}: The terminology "string theory" is misleading since the
superstring field creates states which decomposes into irreducible pointlike
generated irreducible Wigner components. The only positive energy Wigner
representations which are genuinely stringlocal are the massless infinite spin
representations but they are absent in the superstring representation. Since
the relation between states and field operators in case of linear (free)
fields is unique, the pointlike nature of states passes immediately to the
fields. By projecting states on finite invariant energy subspaces one can
explicitly see that the "string" field is the singular limit of pointlike
ordinary fields.

There is an important philosophical message which these failures reveal.
Independent of how theoretical discoveries are obtained, the aim must always
be to understand them as a realization of physical principles.
String-localization cannot be based on similarities of an infinite $(m,s)$
tower spectrum with that of a quantum mechanical chain of oscillators; causal
localization is a totally intrinsic property of local quantum physics and the
concept of modular localization expresses this fact in its
conceptual/mathematical most concise form.

Of course what we consider to be a foundational principle is subject to future
refinements. The idea of finding a TOE by playing mathematical games is not
the way in which the material world reveals itself to us. Such a theory is its
own principle whereas all our experience shows that the real interesting part
of nature is that it offers a wealth of different realizations of its principles.

The explanation of why the popularity of a TOE reached its peak at the turn of
the millennium will be problem for historians of science. As the phlogiston
theory, string theory lasted too long in order to be overlooked in the history
of physics. Whereas the phlogiston theory was abandoned as a result of
contradictions with measurements, the contradictions of string theory with
existing principles of particle physics were always present for anybody with a
strong conceptual awareness. The final word about its legacy is up to
historians of science.

My dear Sagredo, at this late hour I propose to close our dialog.//

\section{Resum\'{e} and concluding remarks}

QFT provides particle theory with an important conceptual structure: the
causal localization principle. It results from the amalgamation of the
Faraday-Maxwell-Einstein classical causality in Minkowski spacetime with the
operator-algebraic formulation of quantum theory in Hilbert space. Its
conceptual strength is matched by its concise mathematical formulation: the
adaptation of the Tomita-Takesaki theory of operator algebras in the form of
modular localization. One reason for submitting the present work to a
history/philosophy oriented physics journal is the fact that this new
framework of QFT sheds additional light on a famous debate in the history of
QFT namely the dispute between Einstein and Jordan which finally led Jordan to
the discovery of QFT. Its main message concerning the vacuum-polarization
caused statistical mechanics nature of the spacetime-restricted vacuum has
sometimes been misinterpreted in terms of the quantum mechanical particle-wave
duality \cite{Du-Ja}.

The new modular localization-based formulation removes the alleged spacetime
string-localization from string theory and shows that such models are special
examples of infinite component pointlike fields. It reveals the conceptual
origin of the particle crossing property and explains the solvability of
integrable models in terms of the simplicity of generators of
modular-localized wedge algebras. It suggests to construct nonperturbative
QFTs by starting from the modular structure of wedge algebras and obtain
compact localized operator algebras in terms of intersections of wedge algebras.

The enormous conceptual range of modular localization unfolds in numerous
applications. In certain cases this led to clashes with existing results and
their interpretation. This happened in particular with ideas which originated
in string theory as dimensional embeddings and reductions (the use of
Kaluza-Klein ideas outside of (quasi)classical approximations) and Maldacena's
incorrect claim that the mathematical AdS-CFT isomorphism can be used to
relate two causally localized QFTs in different spacetime dimensions.

On the constructive side it led to a deeper conceptual understanding of the
limitations of BRST gauge theory and a how to overcome them in a new Hilbert
space setting of stringlocal fields. This in turn led to a demystification of
the Higgs mechanism and its alleged symmetry-breaking Mexican hat potential in
terms of massive vectormesons coupled to Hermitian (instead of complex) fields
and their induced second order interactions. The new Hilbert space setting of
interacting higher spin fields leads in particular to the new concept of
stringlocal Hermitian "escort fields" which in the case of $s=1~$are in many
aspects Higgs-like, except that they appear as an inexorable part of the
massive vectormesons rather than independent scalar fields to be coupled to
massive vectormesons. This new concept has no counterpart in the pointlike
gauge setting and therefore cannot be adequately described in the terminology
of pointlike fields.

These theoretical results present a new meeting ground of ideas coming from
foundational local quantum physics with problems arising from the
observation-oriented research on the Standard Model. It does not exclude Higgs
couplings but it denies the existence of a Higgs mechanism of mass creation by
symmetry breaking.

An unsolved problem of at least comparable importance is the derivation of
gluon/quark confinement from the QCD coupling. As explained in the text, the
problem amounts to establish the vanishing of all \textit{correlation
functions which contain stringlocal gluon or quark operators }\footnote{The
exception are $q-\bar{q}$ pairs for which the string-directions have been
chosen in a particular manner.} in the limit of vanishing gluon mass so that
only pointlike correlation functions of composites (hadron and gluonium)
survive. The stringlike nature results from the very restrictive Hilbert space
positivity, which was not available in the Krein space gauge setting. Using
the vectormeson mass $m$ as a natural covariant infrared cutoff and the fact
that the $m\rightarrow0$ limes comes with logarithmic divergencies, one
expects to be able to prove this by resumming the leading long distance log
terms in perturbation theory. The computation with stringlocal covariant
fields should be similar but somewhat more complicated than the well known
(extended) YFS calculation for the scattering amplitude in massive QED.

The purpose of this article has been accomplished if it succeeds to draw
attention to the enormous unifying power of modular localization for problems
of QFT and particle physics.

\textbf{Acknowledgements}: Since I am neither a historian nor a philosopher of
science, but rather was led by the late J\"{u}rgen Ehlers to the fascinating
E-J problem (to which I could apply my knowledge about modular localization
theory), my foremost but sadly posthumous thanks go to him. I also acknowledge
some more recent advice from John Stachel. I am indebted to Jens Mund for
making his yet unpublished systematic results on the modifications of the
Epstein-Glaser iteration in the presence of string-crossings available to me.
Last not least I thank Raymond Stora for his encouraging interest in the SLF
Hilbert space setting and its relation to the BRST gauge formulation. It was
his conviction that gauge theory is collection of successful computational
recipes which is still in need of a major conceptual reformulation (in which
the very restrictive power of Hilbert space positivity plays a dominant role)
which encouraged me to participate in such a project despite my advanced age.

\end{document}